\documentclass[11pt,a4paper]{article}
\usepackage[a4paper, left=1in, right=1in, top=1in, bottom=1in, includehead, includefoot, foot = 0pt, head=0pt]{geometry} 
\parskip=\medskipamount 

\usepackage{booktabs} 
\usepackage{enumitem} 

\usepackage{natbib}

\bibpunct{(}{)}{;}{a}{,}{,}
\newcommand{\biblist}{
\bibliographystyle{apalike-doi}
\bibliography{References}  
}
\usepackage[hyphens,spaces,obeyspaces]{url}
\usepackage[pdftex,colorlinks,citecolor=blue,urlcolor=blue,linkcolor=red]{hyperref} 
\DeclareUrlCommand\doi{\def\UrlLeft##1\UrlRight{doi:\hspace{0.1cm} \href{http://dx.doi.org/##1}{##1}}\urlstyle{rm}}

\usepackage{graphicx} 
\usepackage[small,bf,hang]{caption}	

\usepackage{tabularx}
\usepackage[export]{adjustbox}

\usepackage{amssymb} 
\usepackage{amsmath} 
\usepackage{accents}
\usepackage{bbm} 
\usepackage{amsbsy} 
\usepackage{multirow} 
\usepackage[dvipsnames]{xcolor} 
\usepackage{mathtools}
\usepackage{eurosym}

\usepackage[ruled, lined, longend]{algorithm2e}
\usepackage{float}
\usepackage{rotating}
\usepackage[caption=false]{subfig}
\usepackage{pdflscape}
\usepackage[defaultcolor=red]{changes}

\usepackage{bbm}
\usepackage{changepage}

\usepackage{subcaption}

\usepackage{threeparttable}
\newsavebox{\covbox}
\newsavebox{\srcbox}
\usepackage{makecell}
\usepackage{xcolor}
\definecolor{ggred}{HTML}{F8766D}
\definecolor{ggteal}{HTML}{00BFC4}

\definecolor{dailyweather}{HTML}{2C7FB8}
\definecolor{intraday}{HTML}{7FCDBB}
\definecolor{dayrelated}{HTML}{6A51A3}
\definecolor{objectrelated}{HTML}{A52A2A}

\usepackage{authblk} 

\usepackage{fancyhdr}
\bibpunct[, ]{(}{)}{;}{a}{,}{,}

\usepackage{tikz,pgfplots}
\usetikzlibrary{fit, positioning,chains,shapes.arrows,fit}
\usetikzlibrary{shapes,arrows, arrows.meta}
\usetikzlibrary{decorations.pathreplacing}
\usetikzlibrary{calc} 
\usetikzlibrary{matrix} 
\usepackage{pgf}
\usetikzlibrary{decorations.text}
\tikzset{cross/.style={cross out, draw=black, minimum size=2*(#1-\pgflinewidth), inner sep=0pt, outer sep=0pt},
cross/.default={1pt}}
\usetikzlibrary{shapes.misc}
\usepackage{environ}
\makeatletter
\newsavebox{\measure@tikzpicture}
\NewEnviron{scaletikzpicturetowidth}[1]{%
  \def\tikz@width{#1}%
  \begin{lrbox}{\measure@tikzpicture}%
  \BODY
  \end{lrbox}%
  \pgfmathparse{#1/\wd\measure@tikzpicture}%
  \BODY
}
\makeatother
\tikzdeclarecoordinatesystem{page}{
    \parsecomma#1\endparsecomma
    \pgfpointanchor{current page}{north east}
    \pgf@xc=\pgf@x%
    \pgf@yc=\pgf@y%
    \pgfpointanchor{current page}{south west}
    \pgf@xb=\pgf@x%
    \pgf@yb=\pgf@y%
    \pgfmathparse{(\pgf@xc-\pgf@xb)/2.*\page@x+(\pgf@xc+\pgf@xb)/2.}
    \expandafter\pgf@x\expandafter=\pgfmathresult pt
    \pgfmathparse{(\pgf@yc-\pgf@yb)/2.*\page@y+(\pgf@yc+\pgf@yb)/2.}
    \expandafter\pgf@y\expandafter=\pgfmathresult pt
}
\makeatother
\usetikzlibrary{arrows,automata}
\usepackage{ragged2e}
\definecolor{arrowcolor}{RGB}{201,216,232}
\definecolor{circlecolor}{RGB}{79,129,189}
\colorlet{textcolor}{white}
\colorlet{bordercolor}{white}

\pgfdeclarelayer{background}
\pgfsetlayers{background,main}

\definecolor{airforceblue}{rgb}{0.36, 0.54, 0.66}
\definecolor{forestgreen}{rgb}{0.13, 0.55, 0.13}\definecolor{fulvous}{rgb}{0.86, 0.52, 0.0}
\definecolor{gray}{rgb}{0.5, 0.5, 0.5}
\definecolor{bistre}{rgb}{0.24, 0.17, 0.12}\definecolor{bostonuniversityred}{rgb}{0.8, 0.0, 0.0}
\definecolor{purpleheart}{rgb}{0.41, 0.21, 0.61}
\definecolor{lightsalmonpink}{rgb}{1.0, 0.6, 0.6}\definecolor{arrowcolor}{rgb}{0.92, 0.92, 0.92}
\tikzset{
inner/.style={
  on chain,
  circle,
  inner sep=4pt,
  fill=circlecolor,
  line width=1.5pt,
  draw=bordercolor,
  text width=1.2em,
  align=center,
  text height=1.25ex,
  text depth=0ex
},
on grid
}

\newcommand\drawarrow{

\node[on chain] (f) {};
\begin{pgfonlayer}{background}
\node[
  inner sep=10pt,
  single arrow,
  single arrow head extend=0.6cm,
  draw=none,
  fill=arrowcolor,
  fit= (c1) (f)
] (arrow) {};
\fill[white] 
  (arrow.before tail) -- (c1|-arrow.west) -- (arrow.after tail) -- cycle;
\end{pgfonlayer}
}

\numberwithin{equation}{section}

\begin{document}
\title{Modelling the occurrence and reporting dynamics of weather-related home insurance claims}

\author[1,*]{Paul Wilsens}
\author[1,2]{Katrien Antonio}
\author[1]{Gerda Claeskens}
\author[3]{Koos Gubbels}
\affil[1]{Faculty of Economics and Business, KU Leuven, Belgium.}
\affil[2]{Faculty of Economics and Business, University of Amsterdam, the Netherlands.}
\affil[3]{TS Economics and Management, Tilburg University, the Netherlands.}
\affil[*]{Corresponding author: \href{mailto:paul.wilsens@kuleuven.be}{paul.wilsens@kuleuven.be}}
\date{\vspace{-1.25cm}} 
\maketitle
\thispagestyle{empty}

\begin{abstract}
\noindent 

We study the occurrence and reporting dynamics of weather-related claims in private home insurance, combining detailed portfolio and claim information with publicly available high-resolution weather data. Existing approaches typically start from an identified severe weather event, such as a hailstorm, and model the dynamics of the claims caused by that event. In practice, however, claims arise under moderate as well as severe conditions, and the claims handling process runs continuously. An insurer therefore requires insight into the occurrence and reporting of claims on every calendar day. 
Modelling all calendar days jointly poses two challenges: recently occurred claims are only partially reported, and days with severe weather are rare while accounting for a large share of the claims. We propose a multi-regime EM-XGBoost framework that addresses both: an expectation-maximisation setup accounts for the incurred-but-not-reported (IBNR) claims, while a data-driven regime structure lets the occurrence and reporting dynamics differ across weather conditions. In a case study on a Dutch home insurance portfolio, we compare single- and two-regime specifications for two spatial resolutions: the country as a whole, and its individual regions. We find that the regional two-regime model outperforms the regional single-regime model as well as both national models, indicating that the benefit of a multi-regime setting depends on a sufficiently granular spatial resolution. We also show that daily covariates constructed from hourly weather observations, capturing the duration and the intra-day timing of severe weather, improve both the occurrence and the reporting model.

\end{abstract}

\vspace{-0.25cm}
\vspace{-0.25cm}
\section*{Statements and declarations}
\vspace{-0.15cm}
\paragraph{Data and code availability statement:} The weather data used in this article are publicly available from the sources listed in Table~\ref{tab:sources}. The portfolio and claims data were provided by a Dutch insurer under a confidentiality agreement and were analysed on protected servers; they cannot be made publicly available. Code implementing the multi-regime EM-XGBoost framework is available on \url{https://github.com/PaulWilsens/weather-IBNR-homeinsurance}.
\vspace{-0.75cm}
\paragraph{Funding statement and acknowledgements:} The authors gratefully acknowledge funding from the FWO and Fonds De La Recherche
Scientifique - FNRS (F.R.S.-FNRS) under the Excellence of Science (EOS) program, project
ASTeRISK Research Foundation Flanders [grant number 40007517]. 

\vspace{-0.25cm}
\paragraph{Conflict of interest disclosure:} The authors declare no conflict of interest.

\pagebreak


\section{Introduction} \label{sec:intro}

Weather events, from everyday rainfall and moderate wind to extreme floods and storms, are a major driver of property damage losses. Reports indicate that their contribution to overall property damage-related losses has increased over recent decades and is projected to rise further \citep{SwissReInstitute2024Sigma1,GallagherRe2025NatCat2024}. Insurers are exposed to this risk primarily through home insurance, which covers damage to the building and its contents caused by perils such as storm, hail and water. Insuring weather risks can be challenging: since a single event often affects an entire region, the resulting claims are spatially correlated, requiring careful consideration of portfolio composition. Additionally, the reporting of weather-related claims by policyholders is often delayed for several reasons, such as prolonged weather events, complicated claim report forms, policyholder characteristics or damage that is not immediately visible. In this article, we aim to increase the understanding and modelling of the occurrence and reporting dynamics of weather-related home insurance claims by combining portfolio data with high-resolution weather data, enabling better decision-making for insurers.

To model the occurrence and reporting of insurance claims, one must account for claims that are incurred but not yet reported (IBNR). Each claim is characterised by two dates: the day on which the damage occurs and the day on which the claim is reported to the insurer. Since reporting is often delayed, the claim counts for recent occurrence days are only partially observed at any evaluation date. Understanding the occurrence and reporting dynamics of these claim counts allows an insurer to accurately estimate reserves, and improves operational efficiency via capacity planning. The latter is especially relevant for weather-related claims: if a hailstorm occurs late in the evening, the insurer has to anticipate a large number of reports the next morning. Estimating these outstanding counts requires modelling both the occurrence of claims and the delay with which they are reported. The chain ladder method \citep{mack1993distribution} estimates the outstanding counts from historical development factors, using only the occurrence and reporting periods of the claims aggregated in a run-off triangle, without any covariate information. Related literature leverages expectation-maximisation (EM) algorithms to jointly estimate the occurrence and reporting processes of claim counts. \cite{verbelen2022modeling} proposed a flexible regression framework accommodating any parametric structure for both processes, with the event counts aggregated at a temporal resolution of the modeller's choosing. They treat the right-truncation of the reporting delays as a missing data problem, so that estimation proceeds via an EM algorithm. As part of their occurrence and development model, \cite{crevecoeur2023bridging} extended the EM framework of \cite{verbelen2022modeling} to the individual level via the inclusion of policyholder covariates. Recently, \cite{wilsens2025machine} incorporated XGBoost and neural network models in the maximisation step, improving predictive performance in high-dimensional settings with complex non-linear effects. The EM strand of literature is directly relevant to this article's setting, as it accommodates home insurance claims at any temporal and spatial resolution, in the presence of high-dimensional covariate data.

In recent years, weather data have become publicly available at high resolution, via initiatives such as the \texttt{Copernicus}\footnote{\url{https://cds.climate.copernicus.eu/}} programme or national weather institutes. The occurrence and reporting frameworks discussed above, however, do not incorporate this information. Weather covariates are informative for both processes: the intensity of a weather event drives how many claims occur on a given day and in a given area, while the nature and timing of the event affect how quickly policyholders notice the damage and file a claim. A first strand of literature models claim dynamics conditional on a delineated weather event, mostly the severity of the resulting claims, and in some cases their frequency or reporting delay. \cite{shi2022assessing} proposed a dependent marked point process for hail damage claims in a US portfolio, in which each radar-identified hailstorm initiates a counting process for the reporting of the ensuing claims, with the claim amounts treated as marks. A copula links the reporting time to the claim amount, and storm-level hail covariates drive the reporting intensity. \cite{gao2022leveraging} model the frequency and geographical distribution of hail damage claims as a spatial point pattern, where each hailstorm produces one pattern of claim locations and the historical storms are treated as repeated realisations of a common process. Their mixed-effects Poisson point process combines a storm-specific random effect, capturing that some storms generate far more claims than others, with policy-level and spatially varying weather covariates. \cite{gao2025risk} proposed a copula-based regression model for the severity of US hail damage claims, separating the dependence between two claims caused by the same storm into a spatial part, decaying with the distance between the properties, and a storm-specific part shared by all claims from that storm. \cite{shi2024leveraging} predicted the loss distribution of reported hail claims in a US portfolio from the hourly weather sequence over the course of the loss event, showing that the intra-day evolution of the weather carries predictive information for claim severity beyond static daily summaries. Recently, \cite{gao2026zero} proposed a zero-inflated mixed-effects spatial point process for claims arising from the same storm, addressing the spatial misalignment between granular weather and exposure predictors and claim counts reported at a coarser, grouped level. A second strand focuses on the occurrence of weather events themselves, rather than modelling occurrence dynamics conditional on identified events. \cite{bourget2024data} combined machine learning and regional climate models to assess climate change impacts on pluvial flood occurrence over Canada and the United States at a 10-25 km resolution, modelling the occurrence of flood events rather than the resulting claims. \cite{fung2025investigating} jointly modelled storm occurrences and the corresponding claim frequencies in Greek motor insurance. They used a finite mixture of compound negative binomial regressions to recover the distribution of actual storm events from records in which storms without claims are unobserved. Weather-related information can also be leveraged for non-damage related insurance applications. \cite{robben2026short} illustrated the association between short-term (weekly) mortality rates and environmental variables, specifically via air pollution and weather covariates. \cite{robben2025granular} proposed a regime-switching framework that allows for deviations from a baseline seasonal mortality when there are heatwaves or outbreaks of respiratory diseases such as Covid-19 or Influenza.

In practice, an insurer needs insight into the occurrence and reporting dynamics of weather-related claims on all calendar days, not only during severe storms but also under moderate conditions. Moderate days generate fewer claims, but are numerous, accounting for a substantial share of the annual claim inflow. Moreover, the designation of a day as a weather event (e.g.~a hailstorm) happens after the fact and relies on subjective criteria, while in practice claims occur and the claims handling process runs continuously, indifferent to whether a day was labelled as an event. The first strand of literature discussed above takes a weather event as externally identified and models the occurrence or severity of the resulting claims, conditional on that event, with the claims grouped per storm. Within this strand, occurrence and reporting are not modelled jointly, and reporting is addressed only by \cite{shi2022assessing}, per storm and conditional on its identification. Moreover, linking the delay to the event rather than to the occurrence day obscures the day-level reporting pattern, especially when the weather persists over several days. In this article, we instead put forward a framework that jointly models the number of weather-related home insurance claims and how these counts are distributed over the reporting delays, across all calendar days and postal code areas. The framework is event-agnostic: no event definition is needed, and the model describes the claim dynamics of an arbitrary day
rather than those of an identified storm. Hence, we obtain estimates for the outstanding claim counts per day and area rather than per storm event. Specifically, we propose a multi-regime EM-XGBoost framework, extending \cite{wilsens2025machine}, which accommodates the highly correlated weather information while accounting for the IBNR claims. Modelling all calendar days jointly means that days with severe weather, which are rare but generate a large share of the claims, are pooled with the many moderate days. The multi-regime structure, inspired by \cite{robben2025granular}, addresses this: weather regimes learned from the data allow the occurrence and reporting dynamics to differ across weather conditions. Unlike \cite{robben2025granular}, where the regimes are latent states of a Markov chain with covariate-driven
transition probabilities, we identify the regimes beforehand by clustering the weather covariates, so that each day-area observation is assigned to exactly one regime, see Section~\ref{sec:regimetheory}. We demonstrate our proposed approach on a Dutch home insurance portfolio, where we fit regime-specific occurrence models alongside a common reporting model, and compare single- and two-regime models at the national and regional level, in the presence of a substantial number of weather-, object- and day-specific covariates. We find that a multi-regime specification requires a sufficiently high spatial resolution. Intra-day weather information, i.e.~measurements at a sub-daily temporal resolution, has been shown to improve the prediction of claim severity for reported claims \citep{shi2024leveraging}. We extend this idea to the occurrence and reporting dynamics of claim counts, by exploiting hourly observations to construct daily covariates capturing the duration and the intra-day timing of severe weather.

Section~\ref{sec:data} introduces the Dutch home insurance portfolio and the public weather data sources, along with the notation and problem setup. Section~\ref{sec:feature} describes the feature engineering steps linking the weather data to the portfolio to construct the observed dataset. Section~\ref{sec:model} outlines the EM-XGBoost framework and its multi-regime extension. Section~\ref{sec:casestudy} presents the case study, comparing single- and two-regime models at the national and regional level. Section~\ref{sec:discussion} concludes this article.

\section{Home insurance portfolio and weather data} \label{sec:data}

In this section, we first introduce a notational framework for modelling the occurrence and reporting dynamics of weather-related claims in a home insurance portfolio, combining object-, day- and weather-specific information. We then introduce the Dutch home insurance portfolio and the associated claims that we analyse in Section~\ref{sec:casestudy}. In Section~\ref{sec:weatherinfo}, we discuss the public sources from which we collect the weather information, along with the resolutions and measurement types they provide.

\subsection{Notation and problem setup} \label{sec:notation}

We largely follow the notation introduced in \cite{crevecoeur2023bridging} and \cite{wilsens2025machine}. Let $\mathcal{D}=(N_{i,1},\ldots,N_{i,d}, e_{i},\boldsymbol{x}_i)_{i=1}^{n}$ denote the complete data consisting of $n$ observations where each observation corresponds to a unique combination of an occurrence day and an area in which claims can occur, i.e.~$\text{day}(i)$ and $\text{area}(i)$. We fix the temporal resolution to calendar days throughout this article, since this is the most granular level at which the claim information and most weather covariates are available, as discussed in the remainder of this section. By construction, the number of observations $n$ depends on the chosen level of spatial (and temporal) resolution. The vector $\boldsymbol{x}_{i}=(\boldsymbol{x}_{\text{object}(i)},\boldsymbol{x}_{\text{day}(i)},\boldsymbol{x}_{\text{weather}(i)})$ denotes the covariate information: the object-specific information $\boldsymbol{x}_{\text{object}(i)}$, the day-specific information $\boldsymbol{x}_{\text{day}(i)}$ and the weather-related information $\boldsymbol{x}_{\text{weather}(i)}$. The exposure $e_{i}$ denotes the number of insured objects in $\text{area}(i)$ on $\text{day}(i)$. Additionally, we have that
\begin{itemize}
    \item[\tiny{$\bullet$}] $N_{i,j}$ is the number of claims that occurred in $\text{area}(i)$ on $\text{day}(i)$, which were reported $j-1$ days after occurrence, and
    \item[\tiny{$\bullet$}] $N_i = \sum^{d}_{j=1} N_{i,j}$ denotes the total number of claims that occurred in $\text{area}(i)$ on $\text{day}(i)$, where $d$ denotes a maximum reporting delay of $d-1$ days.
\end{itemize}
At a given time $\tau$, we have no information about claims that have occurred but have not yet been reported. We denote the observed data $D=(N_{i,1},\ldots,N_{i,\tau_{i}},e_{i},\boldsymbol{x}_i)_{i=1}^{n}$ with $\tau_{i}=\min(d,\tau-\text{day}(i)+1)$ the number of days between $\text{day}(i)$ and $\tau$ during which we can observe the reporting of claims. For partially observed observations, i.e.~with $\text{day}(i)$ close to $\tau$, we estimate the number of claims that have already occurred but are not yet reported, i.e.~the IBNR claim count $\sum_{j=\tau_i+1}^{d} N_{i,j}$. In periods of volatile weather, this can help insurers improve capacity planning for their claim processing teams. Second, for future observations, i.e.~occurrence days beyond $\tau$ for which no claims have been reported yet, we aim to predict the total number of claims $N_i$ as well as the corresponding reporting dynamics, enabling portfolio management, e.g.~via scenario generation or early warning systems. The rate at which claims occur, and how quickly they are reported, depends strongly on the weather of the day in question: many moderate days produce a relatively low number of claims, whereas a single severe storm can generate a large number of claims at once, as illustrated for the Dutch portfolio in Figure~\ref{fig:generaloverview}. Reporting also depends on when the damage occurs: fewer claims are reported on the occurrence day itself when this is a holiday or when the damage occurs late in the evening, as shown in Figures~\ref{fig:reportingempirical_holiday}
and~\ref{fig:reportingempirical}.

\subsection{Portfolio information}

\begin{figure}[h] 
\centering
\begin{adjustbox}{width=0.99\textwidth}
    \subfloat[Number of objects]{

\begin{adjustbox}{width=0.33\textwidth}
    \includegraphics[width = \textwidth]{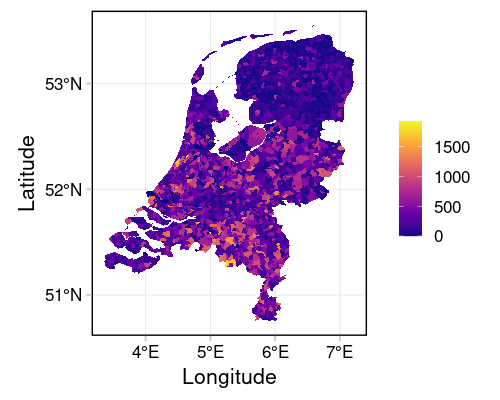}
\end{adjustbox}

}
\subfloat[Proportion detached houses]{

\begin{adjustbox}{width=0.33\textwidth}
\includegraphics[width = \textwidth]{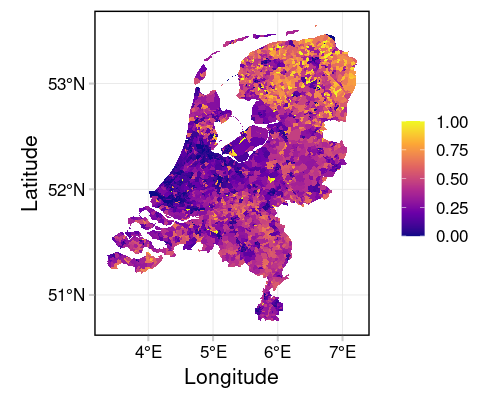}

\end{adjustbox}

} 
\subfloat[Proportion built $<$ 1950]{

\begin{adjustbox}{width=0.33\textwidth}
\includegraphics[width = \textwidth]{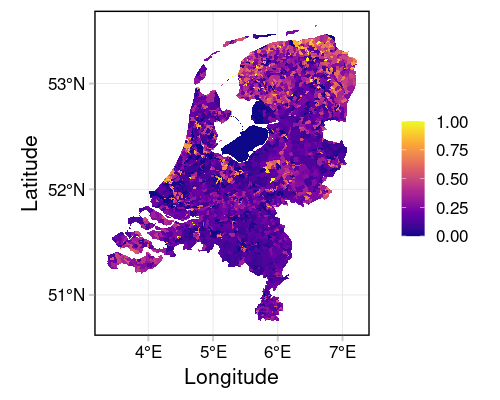}

\end{adjustbox}

}
\end{adjustbox}
\caption{ Portfolio composition on 1 January 2017, visualised for each PC4 postal code area. The left panel shows the number of insured objects with an active policy, the middle panel the proportion of detached houses and the right panel the proportion of objects built before 1950.} 
\label{fig:exposure1jan2017}
\end{figure}

We consider a portfolio of private home insurance policies, provided by a Dutch insurer, of around one million active policies between January 2017 and December 2022. For each policy, we know the start and end date of the contract as well as the corresponding insured object. All insured objects are located within the Netherlands. Figure~\ref{fig:exposure1jan2017} visualises the number of insured objects with an active policy on 1 January 2017, per PC4 postal code area.\footnote{Dutch postal codes consist of four digits followed by two letters. Truncating to the first two or four digits yields a nested hierarchy of areas: 90 PC2 areas and just over 4000 PC4 areas covering the Netherlands, see Figure~\ref{fig:PCmaps}.} The areas around Amsterdam and Tilburg have the highest concentration of insured objects in the portfolio, compared to considerably lower exposure in the rural areas in the north of the Netherlands. Next to the PC4 postal code area of each insured object, we have access to several object characteristics: the roof type, the type of object and the building year. The middle and right panels of Figure~\ref{fig:exposure1jan2017} illustrate the portfolio composition on 1 January 2017: detached houses and older buildings are more prevalent in the rural north, whereas urban areas contain relatively more recent, non-detached objects.

\subsection{Claims information}
\begin{figure}[H] 
\centering
\begin{adjustbox}{width=0.95\textwidth}
    \subfloat[Occurrence structure]{\includegraphics[width = 0.50\textwidth]{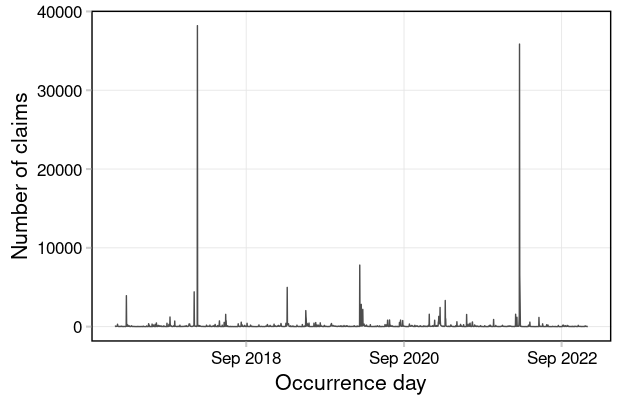}} 
\subfloat[Reporting structure]{\includegraphics[width = 0.50\textwidth]{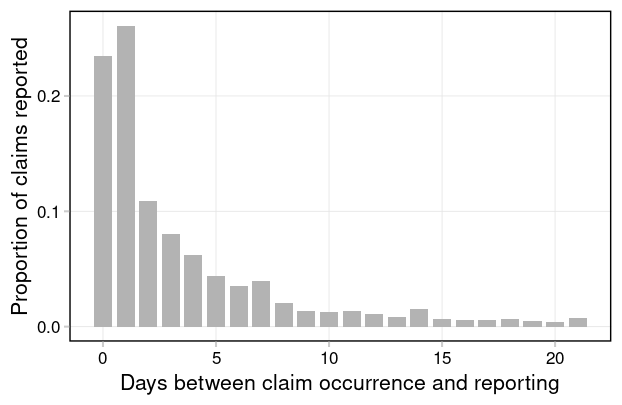}}
\end{adjustbox}
\caption{Overview of the occurrence and reporting structure for the observed period, i.e.~between January 2017 and December 2022. The left-hand side shows the number of claims occurred per day within the observed period. The right-hand side shows the proportion of claims reported with a delay of $0,\ldots,21$ days, respectively, across all days where claims occurred.} 
\label{fig:generaloverview}
\end{figure}

Between 1 January 2017 and 31 December 2022, we have information about 267\ 618 insurance claims filed under a standard home insurance coverage, labelled by the insurer with a weather-related peril. For each claim, we know the day when the damage occurred as well as the day on which the claim was reported to the insurer, with a maximal reporting delay up to three weeks. Figure~\ref{fig:generaloverview} shows the overall occurrence (left panel) and reporting (right panel) dynamics. Two days, i.e.~18 January 2018 and 18 February 2022, account for more than a quarter of all claims in our data. Both dates correspond to well-documented windstorms.\footnote{Thursday 18 January 2018 is referred to as storm `\href{https://www.knmi.nl/over-het-knmi/nieuws/trends-in-stormen-zoals-friederike}{Friederike}', with extreme windspeeds resulting in a nationwide transport shutdown. Friday 18 February 2022 is referred to as storm \href{https://www.knmi.nl/over-het-knmi/nieuws/drielingstorm-dudley-eunice-en-franklin}{`Eunice'} and is part of the triplet windstorm Dudley-Eunice-Franklin in February 2022, which caused massive damage in the Netherlands.} The right panel of Figure~\ref{fig:generaloverview} indicates that most claims are reported the day itself or the day after, with fewer claims reported over time as the number of days since occurrence increases.

\begin{figure}[h] 
\centering
\begin{adjustbox}{width=0.95\textwidth}
    \subfloat[Claims occurred on a non-holiday.]{\includegraphics[width = 0.50\textwidth]{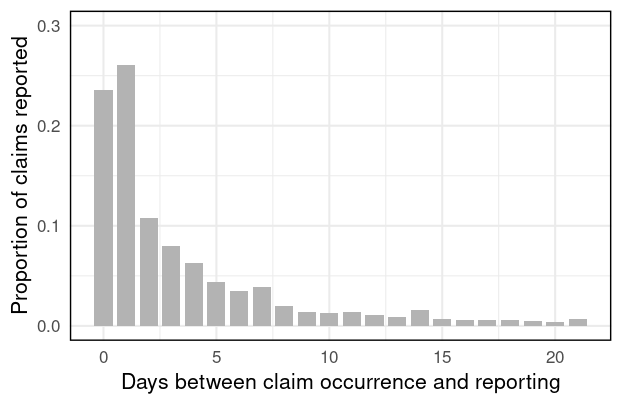}} 
\subfloat[Claims occurred on a Dutch holiday.]{\includegraphics[width = 0.50\textwidth]{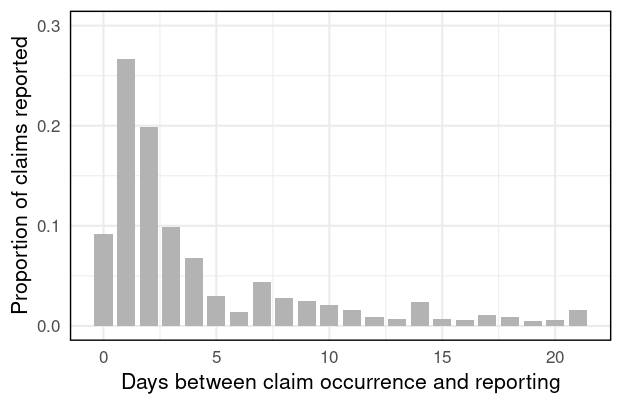}}
\end{adjustbox}
\caption{Visualisation of empirical reporting delay distributions. The left panel shows the 
    proportion of claims reported at a delay of $0,\ldots,21$ days, across all 
    claims that did not occur on a holiday within the observed period. The right panel 
    shows the same distribution for the claims that did occur on a Dutch holiday.} 
\label{fig:reportingempirical_holiday}
\end{figure}

For each calendar day on which a claim can occur or be reported, we collect day-specific information, similar to \cite{verbelen2022modeling}. Since all insured objects are situated in the Netherlands, we consider whether a given day corresponds to a Dutch national holiday, as well as the day immediately before or after a holiday. In addition, we account for the day of the week and whether the day falls in a weekend. Lastly, we also consider whether it is the first or last day of the month. The left and right panels of Figure~\ref{fig:reportingempirical_holiday} show the empirical reporting distributions for claims that occurred on a non-holiday or a Dutch holiday, respectively. We observe that considerably fewer claims are reported on the day of claim occurrence, if that day happens to be a holiday.

\subsection{Weather information} \label{sec:weatherinfo}
We collect weather data for the Netherlands across 2017-2022 from multiple public sources. Different public sources are available at different spatial and temporal resolutions. Additionally, the type of measurements differs: reanalysis data use a numerical weather model to interpolate historical observations onto a grid, whereas observational data (station- or radar-based) are derived directly from measurements. To cover a range of weather covariates for multiple weather data types and spatial/temporal resolutions, we collect data from three public sources, summarised in Table~\ref{tab:sources}.

\begin{table}[H]
\centering
\begin{adjustbox}{width=0.85\textwidth}
    \begin{threeparttable}
\begin{tabular}{@{}>{\raggedright\arraybackslash}p{0.24\textwidth} p{0.34\textwidth} >{\raggedright\arraybackslash}p{0.30\textwidth}@{}}
\toprule
\textbf{Source} & \textbf{Description} & \textbf{References} \\
\midrule
ECMWF Reanalysis v5 \newline \textbf{(ERA5)}
      & Post-processed daily statistics combining model data with observations from across the globe, available via the Copernicus programme\textsuperscript{a}
      & \citet{hersbach2020era5} \newline
        C3S \citeyearpar{copernicus2024era5daily} \\
\addlinespace
European Observational dataset \newline \textbf{(E-OBS)}
      & Daily gridded meteorological data for Europe derived from in-situ observations, available via the Copernicus programme\textsuperscript{a}
      & \citet{cornes2018ensemble} \newline
        C3S \citeyearpar{copernicus2020eobs} \\
\addlinespace
Koninklijk Nederlands Meteorologisch Instituut \newline\textbf{(KNMI)}
      & Various daily and hourly meteorological datasets, both radar-based and station-based observational data, available via the KNMI data store\textsuperscript{b}
      & Hourly station data \citeyearpar{knmi_stationgegevens}\newline
        Radar hail warning \citeyearpar{knmi_radar_hail_warning_24h_v1_0} \newline
        Radar precipitation (gauge-adjusted) \citeyearpar{knmi_rad_nl25_rac_mfbs_1h_2_0} \\
\bottomrule
\end{tabular}
\begin{tablenotes}[flushleft]
\footnotesize
\item[a] Copernicus Climate Data Store, available at \url{https://cds.climate.copernicus.eu/}.
\item[b] KNMI Data Platform, available at \url{https://dataplatform.knmi.nl/} and \url{https://www.daggegevens.knmi.nl/klimatologie/daggegevens}.
\end{tablenotes}
\end{threeparttable}
\end{adjustbox}
\caption{Overview of the three public weather data sources used in this article. For each source, the table indicates the name and the abbreviation used throughout, the description and corresponding references. For KNMI, the references list the three individual datasets collected.}
\label{tab:sources}
\end{table}

Table~\ref{tab:weatheroverview} indicates the weather covariates collected from each of the three public sources, i.e.~information about wind speed, precipitation, hail, temperature and surface pressure. Additionally, it shows the type of weather data, the coverage, temporal resolution and the granularity of the gridded data. ERA5 has the widest coverage as it is globally available but has the least granular grid. E-OBS data are more granular with (land-only) availability within Europe. KNMI datasets are specific to the Netherlands and generally available at very high granularity, except station-based data, which are limited to the number of weather stations. 

\begin{table}[H]
\centering
\begin{adjustbox}{width=0.90\textwidth}
    \begin{threeparttable}
    \begin{tabular}{lcccc}
\toprule
& Unit & \textbf{ERA5} & \textbf{E-OBS} & \textbf{KNMI} \\
\midrule
Mean wind speed     & m/s & $\checkmark$ & $\checkmark$ & $\checkmark$ \\
Maximum wind speed  & m/s & $\checkmark$ &  & $\checkmark$ \\
Surface pressure   & hPa & $\checkmark$ & $\checkmark$ &  \\
Precipitation      & mm & $\checkmark$ & $\checkmark$ & $\checkmark$ \\
Hail probability   & $[0,1]$  &  &  & $\checkmark$ \\
Temperature  & °C &  & $\checkmark$ &  \\
\midrule
Grid  & & $0.25^\circ \times 0.25^\circ$& $0.1^\circ \times 0.1^\circ$& mixed\textsuperscript{a} \\
Coverage &  & worldwide & Europe (land-only) & the Netherlands \\
Type &  & reanalysis & observational & observational\\
Temporal resolution &  & daily & daily & daily and hourly \\
\bottomrule
\end{tabular}
\begin{tablenotes}[flushleft]
\footnotesize
\item[a] The precipitation and hail data are
radar-based on a $1 \times 1$~km grid, the wind
data are station-based.
\end{tablenotes}
\end{threeparttable}
\end{adjustbox}
\caption{Overview of the collected weather data for ERA5, E-OBS and KNMI, as well as their data characteristics. The upper part indicates the weather measurements collected for each source, as well as the measurement units. The lower part summarises the main characteristics of the public data sources, including grid size, coverage, measurement type and temporal resolution.}
\label{tab:weatheroverview}
\end{table}

\begin{figure}[H] 
\centering
\begin{adjustbox}{width=0.95\textwidth}
    \subfloat[ERA5]{

\begin{adjustbox}{width=0.33\textwidth}
    \includegraphics[width = \textwidth]{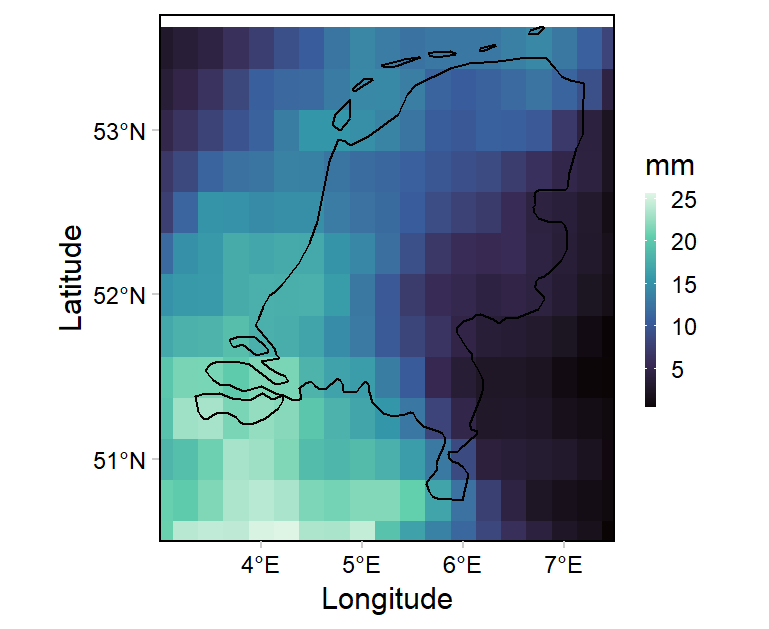}
\end{adjustbox}

}
\subfloat[E-OBS]{

\begin{adjustbox}{width=0.33\textwidth}
\includegraphics[width = \textwidth]{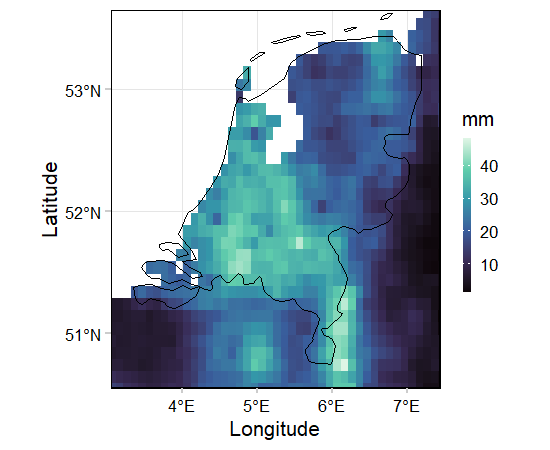}

\end{adjustbox}

} 
\subfloat[KNMI]{

\begin{adjustbox}{width=0.33\textwidth}
\includegraphics[width = \textwidth]{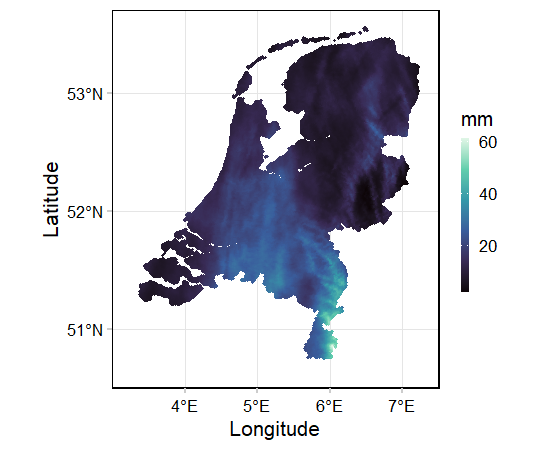}

\end{adjustbox}

}
\end{adjustbox}
\caption{Visualisation of the raw daily gridded precipitation data of 29 April 2018 across the Netherlands, for ERA5 (left panel), E-OBS (middle panel) and KNMI (right panel). White areas indicate that there is no coverage for a specific public source. }
\label{fig:grids}
\end{figure}

Figure~\ref{fig:grids} illustrates the difference in grid coverage described in Table~\ref{tab:weatheroverview} when comparing the different public weather data sources as well as the spatial resolution at which the data are recorded. Additionally, we observe that the magnitude of the measurements can substantially differ between sources, e.g.~due to the difference between radar- and station-based measurements, making it difficult to directly compare reported values.

\begin{figure}[H] 
\centering
\begin{adjustbox}{width=0.98\textwidth}
    \subfloat[Precipitation]{

\begin{adjustbox}{width=0.32\textwidth}
    \includegraphics[width = \textwidth]{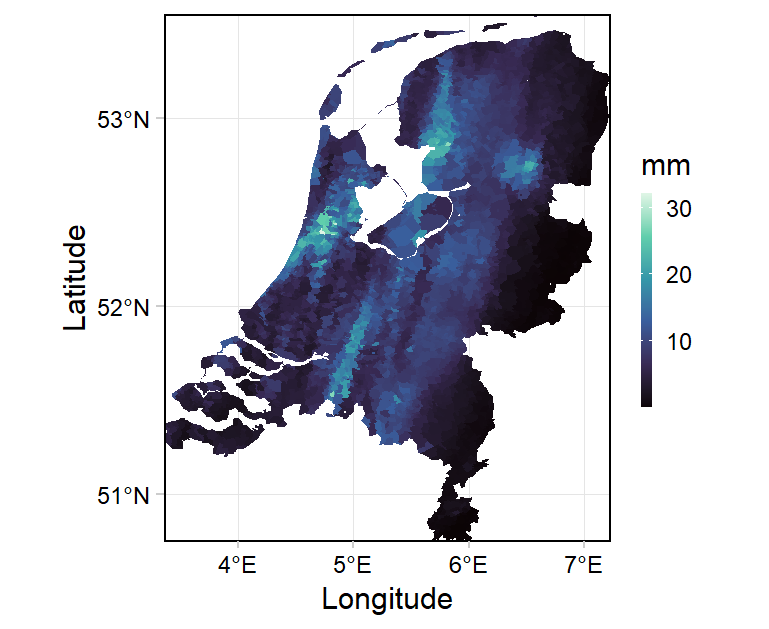}
\end{adjustbox}

}
\subfloat[Hail probability]{

\begin{adjustbox}{width=0.32\textwidth}
\includegraphics[width = \textwidth]{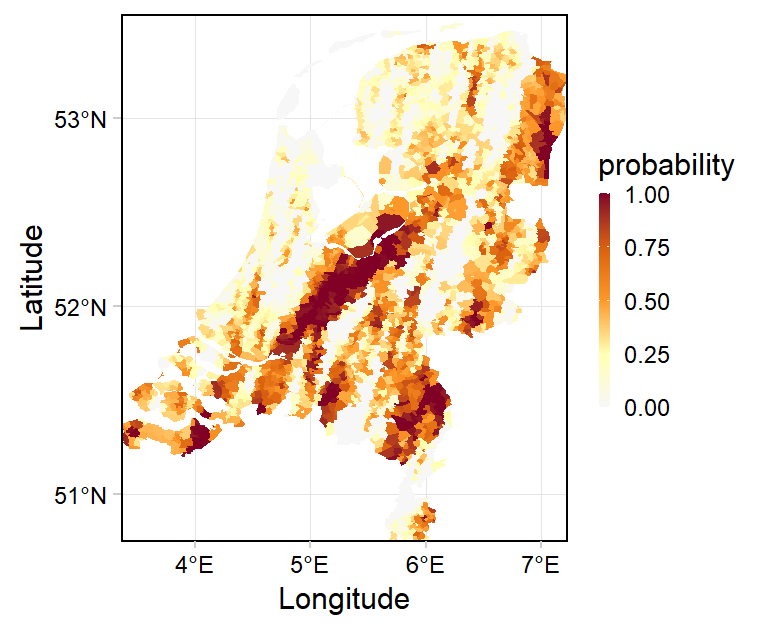}

\end{adjustbox}

} 
\subfloat[Claim occurrence]{

\begin{adjustbox}{width=0.32\textwidth}
\includegraphics[width = \textwidth]{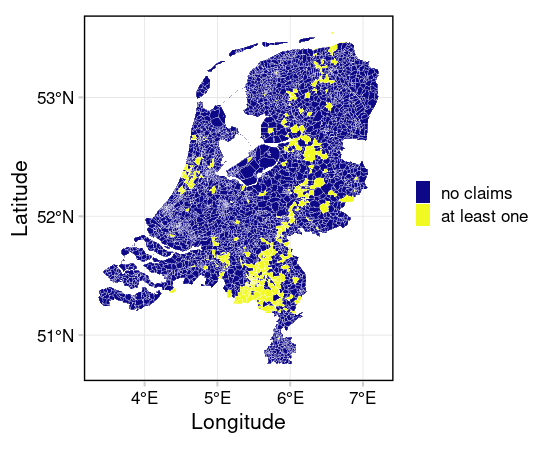}

\end{adjustbox}

}
\end{adjustbox}
\caption{Visualisation of weather and claim occurrence data for 4 June 2019 across the Netherlands. The weather data are daily KNMI measurements: the left and middle panels show the precipitation measurements (in mm) and the hail probability (between 0 and 1), respectively. The right panel shows the PC4 areas in which at least one claim occurred. Data are shown at the PC4 postal code area level.}
\label{fig:4june2019}
\end{figure}

Figure~\ref{fig:4june2019} shows the weather on a specific day, i.e.~4 June 2019, on which there was a rain- and hailstorm. The left and middle panels visualise the precipitation (measured in mm) and the hail probability (expressed as a number between 0 and 1), respectively. The right panel shows the PC4 areas in which at least one claim occurred on that day. In the left panel, the band appears as a light-coloured strip of high precipitation, and in the middle panel as a dark red strip of elevated hail probability, both running from the south of the Netherlands to the north-east and indicating where the storm passed. Claims arise broadly along that same band, illustrating how the claim pattern resembles the weather pattern.

\begin{figure}[H] 
\centering
\begin{adjustbox}{width=0.95\textwidth}
    \subfloat[Claims occurred on a Tuesday.]{\includegraphics[width = 0.50\textwidth]{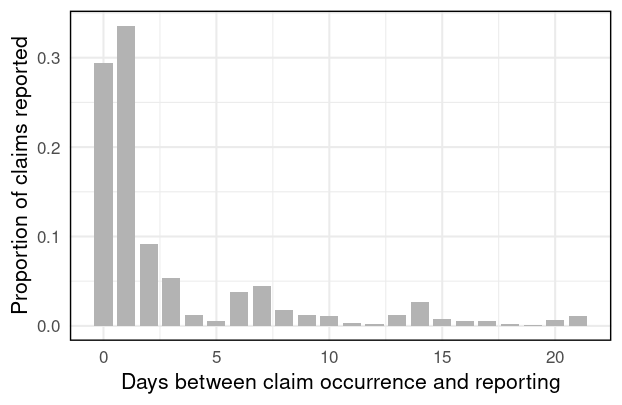}} 
\subfloat[Claims occurred on Tuesday 4 June 2019.]{\includegraphics[width = 0.50\textwidth]{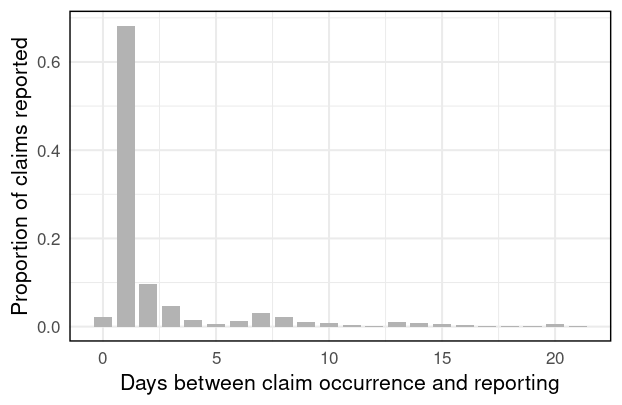}}
\end{adjustbox}
\caption{Visualisation of empirical reporting delay distributions. The left panel shows the 
    proportion of claims reported at a delay of $0,\ldots,21$ days, across all 
    claims that occurred on a Tuesday within the observed period. The right panel 
    shows the same distribution for the claims that occurred on Tuesday 
    4 June 2019, the rain- and hailstorm day visualised in 
    Figure~\ref{fig:4june2019}.} 
\label{fig:reportingempirical}
\end{figure}

Figure~\ref{fig:reportingempirical} shows the empirical reporting pattern for all claims that occurred on a Tuesday (left panel) and for the claims that occurred on Tuesday 4 June 2019 (right panel). In general, around one third of claims occurring on a Tuesday are reported on the occurrence day itself. For 4 June 2019, however, hardly any claims were reported on the day of occurrence, with reporting instead concentrated on the day thereafter. The explanation lies in the timing of the event: the rain- and hailstorm on that day struck mainly during the late evening, leaving policyholders little time to discover damage and report a claim before the end of the day. Daily weather observations cannot distinguish an evening storm from a morning storm. This motivates the construction of covariates from more granular (e.g.~hourly) weather observations, capturing when during the day the most severe weather occurred, which we discuss in Section~\ref{sec:feature}.
\section{Feature engineering and aggregation} \label{sec:feature}

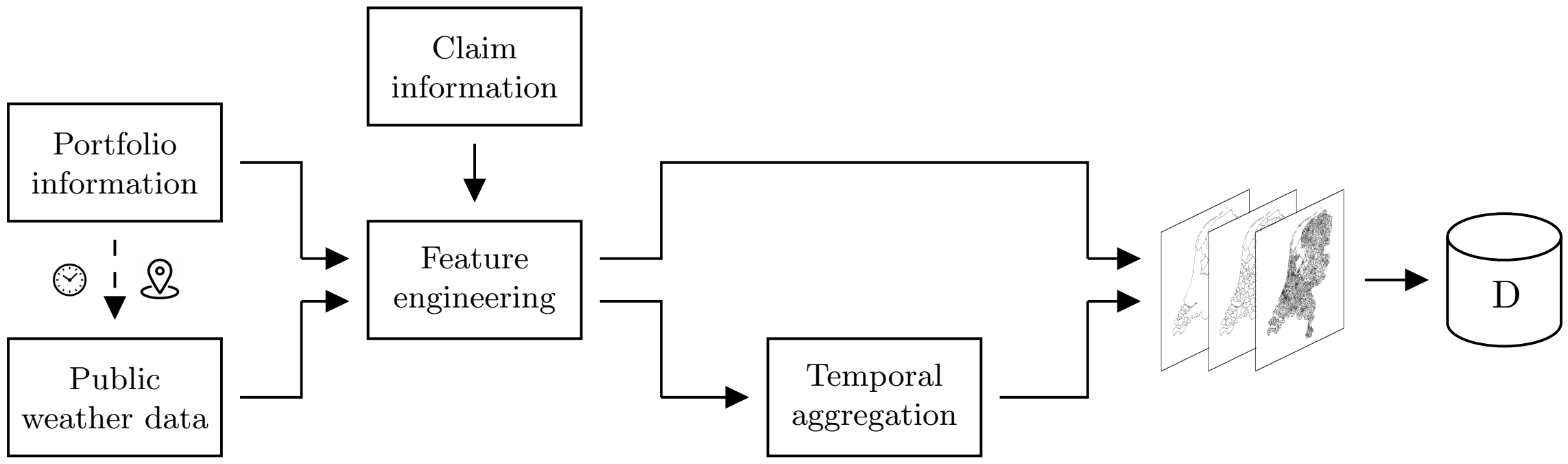
\begin{figure}[H]
\centering
\begin{adjustbox}{width=0.95\textwidth}

\tikzset{every picture/.style={line width=0.75pt}} 

\begin{tikzpicture}[x=0.75pt,y=0.75pt,yscale=-1,xscale=1]

\draw   (25,26) -- (105,26) -- (105,70) -- (25,70) -- cycle ;
\draw   (25,114) -- (105,114) -- (105,158) -- (25,158) -- cycle ;
\draw   (160,-10) -- (240,-10) -- (240,34) -- (160,34) -- cycle ;
\draw    (200,41) -- (200,60) ;
\draw [shift={(200,63)}, rotate = 270] [fill={rgb, 255:red, 0; green, 0; blue, 0 }][line width=0.08]  [draw opacity=0] (8.93,-4.29) -- (0,0) -- (8.93,4.29) -- cycle ;

\draw  [dash pattern={on 4.5pt off 4.5pt}]  (65,77) -- (65,104) ;
\draw [shift={(65,107)}, rotate = 270] [fill={rgb, 255:red, 0; green, 0; blue, 0 }  ][line width=0.08]  [draw opacity=0] (8.93,-4.29) -- (0,0) -- (8.93,4.29) -- cycle    ;
\draw (48,92) node  {\includegraphics[width=10.46pt,height=9.44pt]{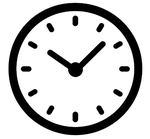}};
\draw (82,92) node  {\includegraphics[width=12.59pt,height=12.59pt]{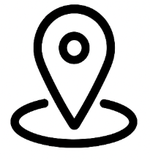}};

\draw    (112,48) -- (135,48) -- (135,84) ;
\draw    (135,84) -- (150,84) ;
\draw [shift={(153,84)}, rotate = 180] [fill={rgb, 255:red, 0; green, 0; blue, 0 }  ][line width=0.08]  [draw opacity=0] (8.93,-4.29) -- (0,0) -- (8.93,4.29) -- cycle    ;
\draw    (112,136) -- (135,136) -- (135,100) ;
\draw    (135,100) -- (150,100) ;
\draw [shift={(153,100)}, rotate = 180] [fill={rgb, 255:red, 0; green, 0; blue, 0 }  ][line width=0.08]  [draw opacity=0] (8.93,-4.29) -- (0,0) -- (8.93,4.29) -- cycle    ;

\draw   (160,70) -- (240,70) -- (240,114) -- (160,114) -- cycle ;

\draw    (247,84) -- (270,84) -- (270,48) -- (430,48) -- (430,84) ;
\draw    (430,84) -- (447,84) ;
\draw [shift={(450,84)}, rotate = 180] [fill={rgb, 255:red, 0; green, 0; blue, 0 }  ][line width=0.08]  [draw opacity=0] (8.93,-4.29) -- (0,0) -- (8.93,4.29) -- cycle    ;
\draw    (247,100) -- (270,100) -- (270,136) ;
\draw    (270,136) -- (300,136) ;
\draw [shift={(303,136)}, rotate = 180] [fill={rgb, 255:red, 0; green, 0; blue, 0 }  ][line width=0.08]  [draw opacity=0] (8.93,-4.29) -- (0,0) -- (8.93,4.29) -- cycle    ;
\draw   (310,114) -- (390,114) -- (390,158) -- (310,158) -- cycle ;
\draw    (397,136) -- (430,136) -- (430,100) ;
\draw    (430,100) -- (447,100) ;
\draw [shift={(450,100)}, rotate = 180] [fill={rgb, 255:red, 0; green, 0; blue, 0 }  ][line width=0.08]  [draw opacity=0] (8.93,-4.29) -- (0,0) -- (8.93,4.29) -- cycle    ;

\draw (492,92) node  {\includegraphics[width=52.5pt,height=52.5pt]{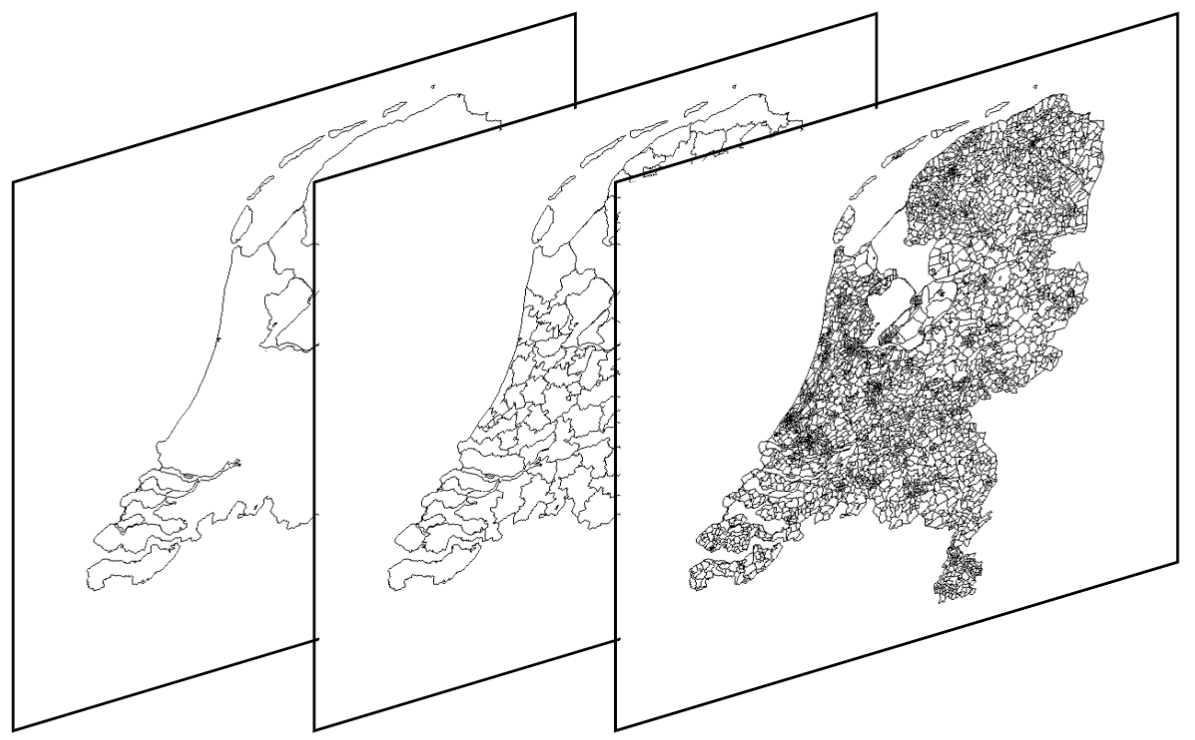}};

\draw    (534,92) -- (555,92) ;
\draw [shift={(558,92)}, rotate = 180] [fill={rgb, 255:red, 0; green, 0; blue, 0 }  ][line width=0.08]  [draw opacity=0] (8.93,-4.29) -- (0,0) -- (8.93,4.29) -- cycle    ;
\draw   (607.61,75.84) -- (607.61,108.16) .. controls (607.61,112.97) and (598.07,116.86) .. (586.3,116.86) .. controls (574.53,116.86) and (564.99,112.97) .. (564.99,108.16) -- (564.99,75.84)(607.61,75.84) .. controls (607.61,80.65) and (598.07,84.54) .. (586.3,84.54) .. controls (574.53,84.54) and (564.99,80.65) .. (564.99,75.84) .. controls (564.99,71.04) and (574.53,67.14) .. (586.3,67.14) .. controls (598.07,67.14) and (607.61,71.04) .. (607.61,75.84) -- cycle ;

\draw (65,48) node [font=\footnotesize, align=center] {Portfolio\\information};
\draw (65,136) node [font=\footnotesize, align=center] {Public\\ weather data};
\draw (200,92) node [font=\footnotesize, align=center] {Feature\\engineering};
\draw (350,136) node [font=\footnotesize, align=center] {Temporal\\aggregation};
\draw (580.1,90.87) node [anchor=north west][inner sep=0.75pt] [align=left] {D};
\draw (200,12) node [font=\footnotesize, align=center] {Claim\\information};

\end{tikzpicture}
\end{adjustbox}
\caption{Visualisation of the construction of the observed dataset $D$. The portfolio information provides the location of the insured objects and the period of interest during which claims are recorded, which are used to collect the public weather data (dashed arrow). The data sources undergo feature engineering; the hourly weather features are temporally aggregated from the hourly to the daily level. Spatial aggregation to a chosen resolution (e.g.~PC2 postal code area) then yields the observed dataset $D$, with one observation per area and day.}
\label{fig:datapipeline}
\end{figure}

Figure~\ref{fig:datapipeline} visualises the steps required to construct the observed dataset $D$ from the portfolio data, claims data and the public weather data. The portfolio information provides the location of each insured object and the period of interest during which claims are recorded, determining for which area and calendar days public weather data are collected. Aggregating the number of insured objects per area and day yields the exposure $e_i$, and the claims data provide the claim counts $N_{i,j}$. The data sources subsequently undergo feature engineering. The object- and day-specific covariates are by construction available at daily resolution. Since some of the (KNMI) weather features are at an hourly resolution, we temporally aggregate these to the daily level. Next, the covariate information is spatially aggregated to a chosen resolution, as motivated for our case study in Section~\ref{sec:casestudy}, yielding the observed dataset $D=(N_{i,1},\ldots,N_{i,\tau_{i}},e_{i},\boldsymbol{x}_i)_{i=1}^{n}$ with one observation per area and 
day. We note that the need for temporal and spatial aggregation depends on the resolution at which the collected data are available, as well as the spatial and temporal resolution chosen by the modeller. In the remainder of this section, we discuss each of these steps in more detail. Table~\ref{tab:covariates} in Appendix~\ref{app:casestudy} lists all constructed covariates and their descriptions.

\subsection{Aggregation of portfolio information}

The object-specific characteristics, e.g.~roof type (thatched or not), are available at the individual object level. Since object locations are only known at the PC4 postal code area level, we model at coarser spatial resolutions than the individual insured object level. This requires aggregating the object-specific characteristics, i.e.~roof type, type of object and building year, per area. At coarser spatial resolutions, each observation $i$ represents a unique regional subportfolio of objects located in $\text{area}(i)$ with an active policy on $\text{day}(i)$, given that the temporal resolution is fixed at the daily level. For a given spatial resolution, we aggregate the object-specific information to reflect the overall composition of that regional subportfolio. Let $\mathcal{O}_i$ denote the set of insured objects in $\text{area}(i)$ with an active policy on $\text{day}(i)$. We can then express the exposure as $e_i = |\mathcal{O}_i|$. For a categorical characteristic with $C$ categories, we denote the proportion of objects in each category as
\begin{align*}
    x_{\text{object}(i)}^{(c)} = \frac{1}{e_i} \sum_{o \in \mathcal{O}_i} \mathbbm{1}\{\text{object } o \text{ belongs to category } c\}, \:\:\: \text{for} \:\:\: c = 1, \ldots, C,
\end{align*}
so that a single categorical characteristic yields $C$
covariates, one proportion per category, describing the
composition of the regional subportfolio corresponding to
observation $i$. In the case study in Section~\ref{sec:casestudy}, we only have access to categorical object-specific covariates. When given continuous characteristics, e.g.~surface area, one can compute summary statistics such as the mean, median or maximum of the set of values recorded for the objects in the regional subportfolio.

\subsection{Aggregation of weather information}

\paragraph{Spatial aggregation of weather information}

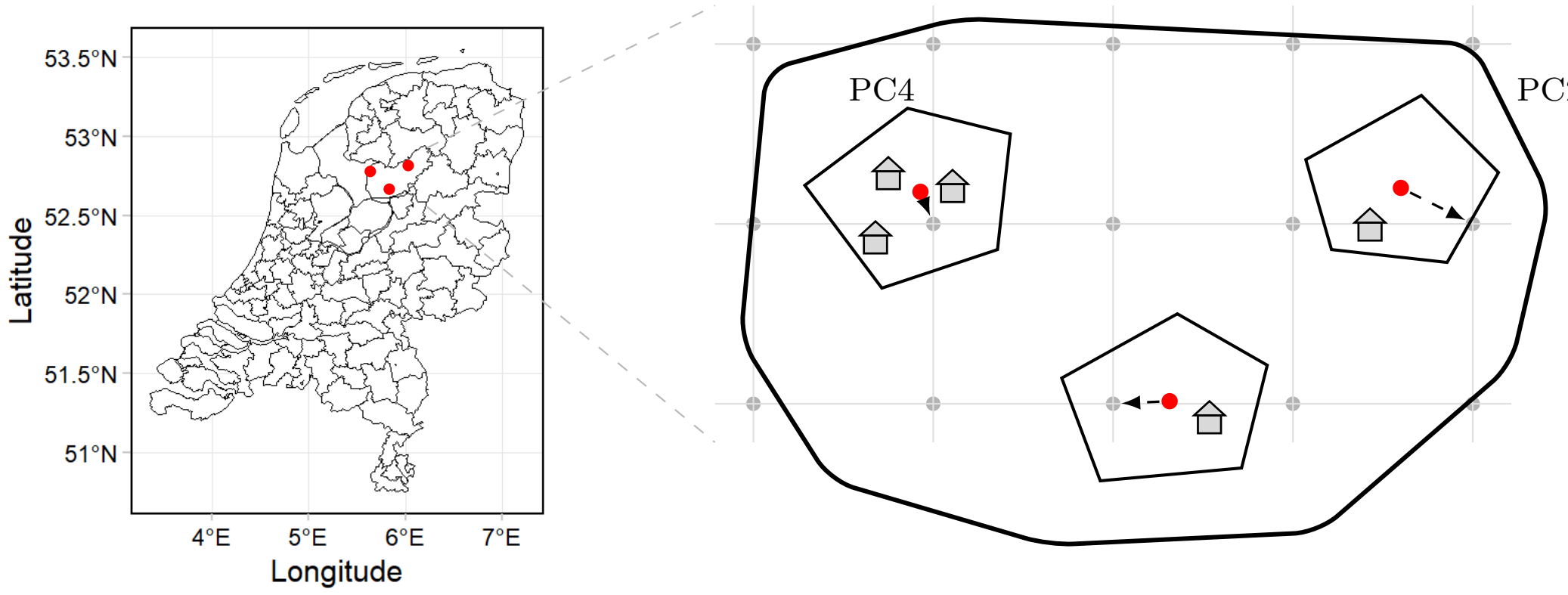
\begin{figure}[H]
\centering
\begin{adjustbox}{width=0.9\textwidth}
    \begin{tikzpicture}[scale=1.0, every node/.style={font=\scriptsize}]
\node[inner sep=0pt, anchor=south west] (nlmap) at (0,0)
  {\includegraphics[width=5.4cm]{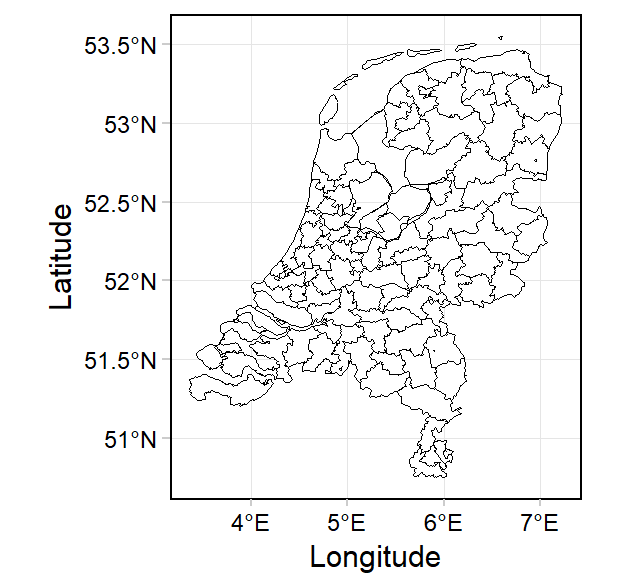}};
\begin{scope}[x={($(nlmap.south east)-(nlmap.south west)$)},
              y={($(nlmap.north west)-(nlmap.south west)$)},
              shift={(nlmap.south west)}]
  \fill[red] (0.65,0.73) circle (1.4pt);
  \fill[red] (0.68,0.70) circle (1.4pt);
  \fill[red] (0.71,0.74) circle (1.4pt);
  \coordinate (zoomTop) at (0.74,0.77);
  \coordinate (zoomBot) at (0.74,0.67);
\end{scope}
\begin{scope}[shift={(6.8,1.70)}, scale=1.1]
\newcommand{\house}[2]{
  \draw[line width=0.5pt, fill=gray!30]
    (#1-0.09,#2-0.08) rectangle (#1+0.09,#2+0.06);
  \draw[line width=0.5pt, fill=gray!30]
    (#1-0.12,#2+0.06) -- (#1,#2+0.17) -- (#1+0.12,#2+0.06) -- cycle;
}
\foreach \x in {0,1.4,2.8,4.2,5.6}
  \foreach \y in {0,1.4,2.8}
    \fill[gray!60] (\x,\y) circle (1.6pt);
\foreach \x in {0,1.4,2.8,4.2,5.6}
  \draw[gray!25, line width=0.4pt] (\x,-0.3) -- (\x,3.1);
\foreach \y in {0,1.4,2.8}
  \draw[gray!25, line width=0.4pt] (-0.3,\y) -- (5.9,\y);
\draw[line width=1.1pt, rounded corners=6pt]
  (-0.1,0.5) -- (0.1,2.6) -- (1.6,3.0) -- (3.6,2.9) -- (5.6,2.8)
  -- (6.2,1.6) -- (5.9,0.3) -- (4.4,-1.0) -- (2.3,-1.1) -- (0.6,-0.6)
  -- cycle;
\node[font=\scriptsize] at (6.2,2.45) {PC2};
\node[font=\scriptsize] at (1,2.45) {PC4};
\draw[line width=0.75pt] (0.4,1.7) -- (1.2,2.3) -- (2.0,2.1) -- (1.9,1.2)
  -- (1.0,0.9) -- cycle;
\house{1.05}{1.75}
\house{1.55}{1.65}
\house{0.95}{1.25}
\fill[red] (1.30,1.65) circle (1.8pt);
\draw[dashed, line width=0.6pt, -latex, shorten <=2.5pt, shorten >=2pt]
  (1.30,1.65) -- (1.4,1.4);
\draw[line width=0.75pt] (2.4,0.2) -- (3.3,0.7) -- (4.0,0.3) -- (3.8,-0.5)
  -- (2.7,-0.6) -- cycle;
\house{3.55}{-0.15}
\fill[red] (3.24,0.02) circle (1.8pt);
\draw[dashed, line width=0.6pt, -latex, shorten <=2.5pt, shorten >=2pt]
  (3.24,0.02) -- (2.8,0);
\draw[line width=0.75pt] (4.3,1.9) -- (5.2,2.4) -- (5.8,1.8) -- (5.4,1.1)
  -- (4.5,1.2) -- cycle;
\house{4.80}{1.35}
\fill[red] (5.04,1.68) circle (1.8pt);
\draw[dashed, line width=0.6pt, -latex, shorten <=2.5pt, shorten >=2pt]
  (5.04,1.68) -- (5.6,1.4);
\coordinate (gridTL) at (-0.3,3.1);
\coordinate (gridBL) at (-0.3,-0.3);
\end{scope}
\draw[gray!55, line width=0.4pt, dashed] (zoomTop) -- (gridTL);
\draw[gray!55, line width=0.4pt, dashed] (zoomBot) -- (gridBL);
\end{tikzpicture}
\end{adjustbox}
\caption{Stylised example of the linking of gridded weather data to PC4 postal code areas, shown for a single weather grid and three PC4 areas located within the same PC2 area. The left side situates the three PC4 areas within the Netherlands, shown at the PC2 level, with the red dots denoting their centroids. The right side zooms in on these areas: the grey dots represent the gridpoints of a public weather data source, the red dots the centroids of the three PC4 areas and each house an insured object with an active policy. For each PC4 area, the weather information assigned to that area is taken from the gridpoint closest in geodesic distance to its centroid, as indicated by the dashed arrows.}
\label{fig:gridlinking}
\end{figure}

For each insured object, we know in which PC4 postal code area it is located. As discussed in Section~\ref{sec:data}, public weather data are collected on a grid, which differs across sources. We therefore link the weather information to each PC4 postal code area, the level at which object locations are known. Specifically, for each weather grid, we assign to a PC4 area the gridpoint that minimises the geodesic distance to its longitude-latitude centroid, computed on the ellipsoidal Earth model WGS84, which takes into account that Earth is not a perfect sphere \citep{karney2013algorithms}. Figure~\ref{fig:gridlinking} illustrates this for three stylised PC4 areas. To model the data at coarser spatial resolutions than the PC4 level, where for a given observation $i$, $\text{area}(i)$ consists of multiple PC4 areas, we need to spatially aggregate this information. An unweighted mean would treat all PC4 areas within $\text{area}(i)$ equally, regardless of how many insured objects they contain. Since the claims we model arise from the insured objects themselves, we instead compute a weighted mean, weighting each PC4 area by its number of insured objects, so that PC4 areas holding a larger part of the regional (e.g.~PC2) subportfolio contribute proportionally more to the aggregated weather value. Returning to Figure~\ref{fig:gridlinking}, the PC2 area shown there contains three PC4 areas consisting of three, one and one insured objects respectively, so the value of each weather covariate assigned to the PC2 area is the weighted mean of the three PC4-level values, with weights $3/5$, $1/5$ and $1/5$, respectively.

\begin{figure}[H]
\centering
\begin{adjustbox}{width=0.80\textwidth}
\begin{tabular}{@{}cc@{}}
\subfloat[Maximum wind speed, 3 January 2018]{%
  \includegraphics[width=0.42\textwidth]{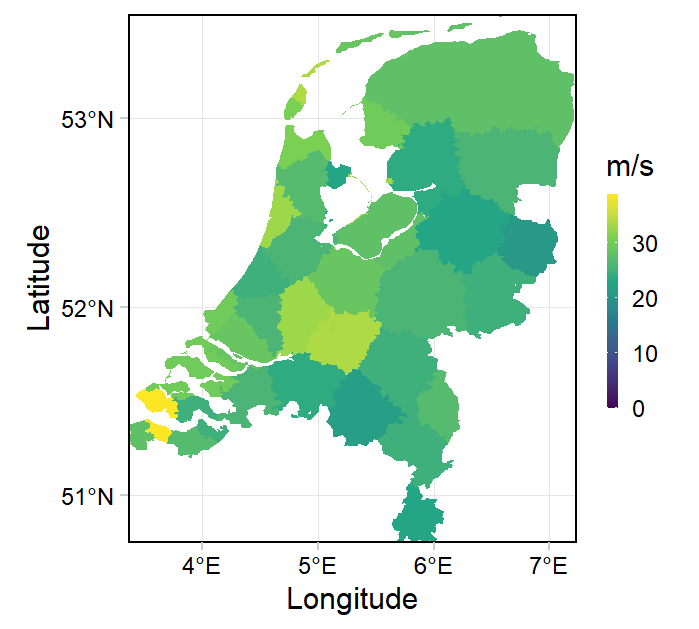}} &
\subfloat[Maximum wind speed, 9 February 2020]{%
  \includegraphics[width=0.42\textwidth]{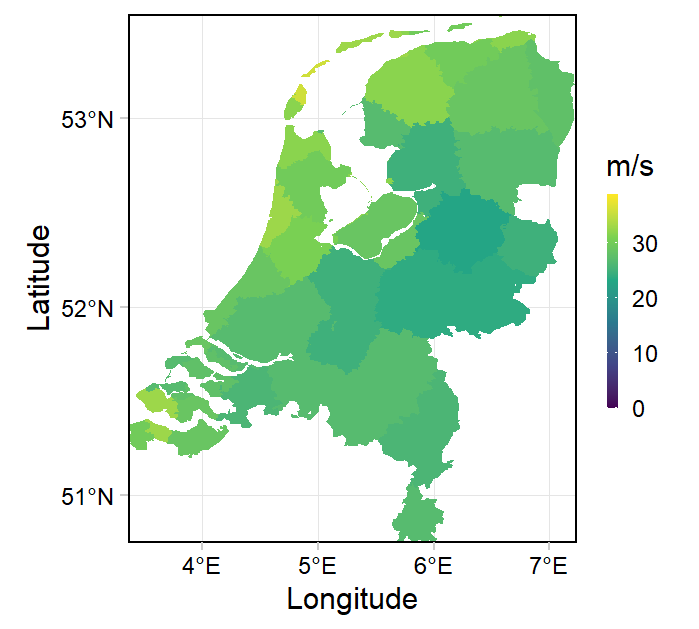}} \\
\subfloat[Hours above 25~m/s, 3 January 2018]{%
  \includegraphics[width=0.42\textwidth]{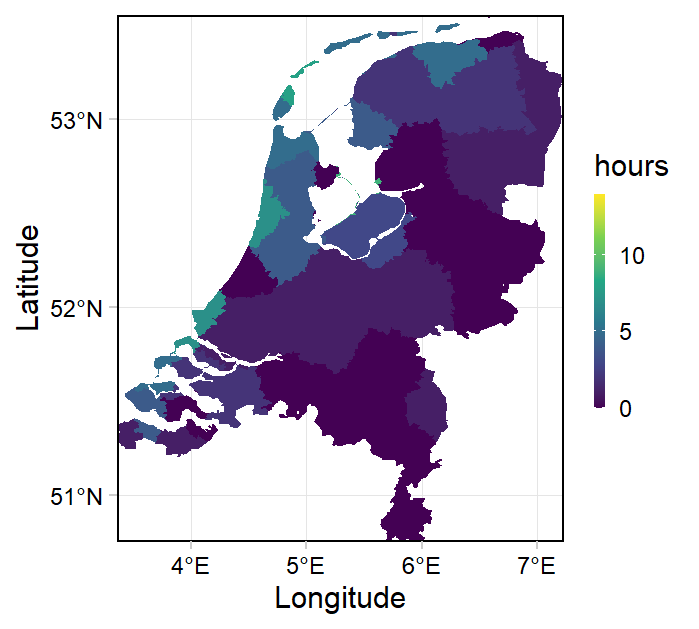}} &
\subfloat[Hours above 25~m/s, 9 February 2020]{%
  \includegraphics[width=0.42\textwidth]{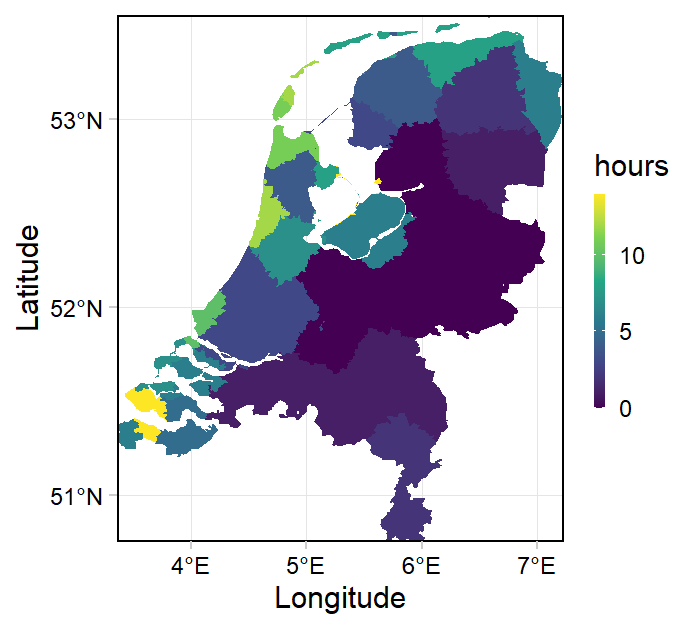}} \\
\end{tabular}
\end{adjustbox}
\caption{KNMI wind variables, linked to the PC4 areas as in Figure~\ref{fig:gridlinking} and aggregated from hourly to daily values. The upper panels show the maximum daily wind speed in m/s, the lower panels the number of hours during which the maximum hourly wind speed exceeded 25~m/s. The left panels correspond to 3 January 2018, the right panels to 9 February 2020.}
\label{fig:windintensity_example}
\end{figure}

\begin{figure}[H]
\centering
\begin{adjustbox}{width=0.99\textwidth}
  \subfloat[Proportion of claims per precipitation block]{
\begin{adjustbox}{width=0.33\textwidth}
    \includegraphics[width=\textwidth]{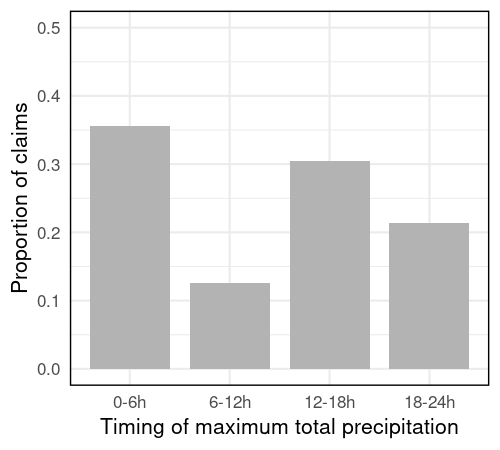}
\end{adjustbox}
}
\subfloat[Morning (0-6 hour block)]{
\begin{adjustbox}{width=0.33\textwidth}
    \includegraphics[width=\textwidth]{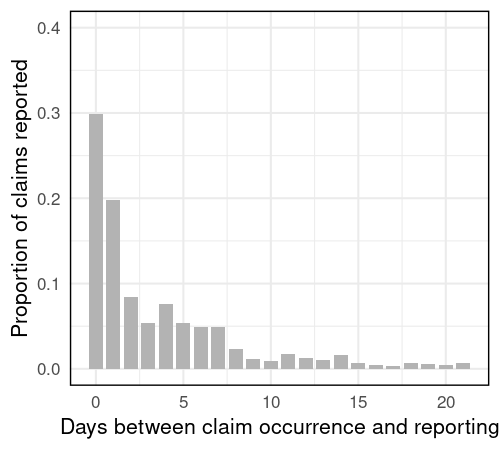}
\end{adjustbox}
}
\subfloat[Evening (18-24 hour block)]{
\begin{adjustbox}{width=0.33\textwidth}
    \includegraphics[width=\textwidth]{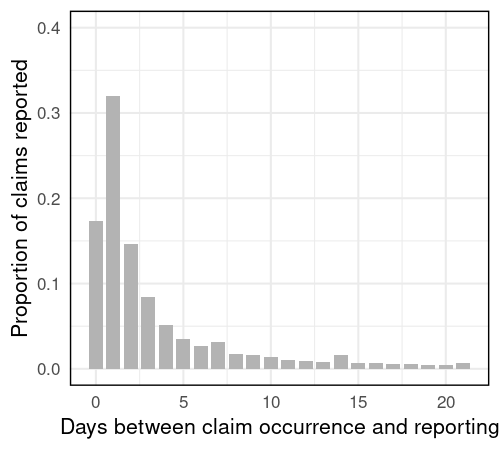}
\end{adjustbox}
}
\end{adjustbox}
\caption{Descriptive information for the intra-day precipitation timing variable, constructed from hourly KNMI precipitation measurements. The left panel shows, across all claims in the observed period, the proportion of claims for which the largest 6-hour precipitation sum fell in each block. The middle and right panels show the empirical reporting delay distribution, i.e.~the proportion of claims reported at a delay of $0,\ldots,21$ days, for the claims occurring on days with most precipitation between 0-6h and between 18-24h, respectively.}
\label{fig:timingdelay}
\end{figure}

\paragraph{Temporal aggregation of intra-day weather information}
Both the precipitation radar data and the station wind observations, collected from the KNMI, are available at hourly intervals. Since we model claim dynamics at a daily temporal resolution, we construct daily aggregates from these hourly values. In general, wind-related covariates such as daily mean wind speed and daily maximum wind speed do not capture how long severe conditions persisted during the day. Two days with the same mean or maximum wind speed can differ considerably in how long the damaging conditions lasted. Therefore, we construct a set of wind-related duration variables, denoting the number of hours during which the maximum hourly wind speed was above thresholds of 20, 25 and 30 m/s. In the literature, wind thresholds are often used to define wind-related weather events and to identify the
range in which wind damage becomes likely, see e.g.~\cite{bonazzi2012spatial} or \cite{fonseca2025empirical}. Figure~\ref{fig:windintensity_example} motivates this construction, showing the 25 m/s threshold variable for 3 January 2018 and 9 February 2020 (bottom panels), on which respectively 4413 and 7810 claims occurred. Despite the top panels showing similar maximum wind speeds for both days, there were considerably more claims on 9 February 2020. Our intra-day 25 m/s threshold variable shows a possible explanation: the wind speeds most likely to cause damage had a substantially longer duration on 9 February 2020 compared to 3 January 2018.
A second set of constructed wind- and precipitation-based covariates relates to the timing of weather severity, by splitting the day into four six-hour blocks and identifying the block in which the wind speed or the precipitation
sum was highest. Figure~\ref{fig:timingdelay} visualises the reporting dynamics for a precipitation-based intra-day variable indicating the 6-hour block during which the sum of hourly precipitation measurements was highest. The middle and right panels show the empirical reporting distribution for claims that occurred on a day where the hourly precipitation sum was highest for the 0-6h (early morning) and 18-24h (evening) periods. We observe that high precipitation in the evening shifts the empirical reporting distribution to the right, in line with Figure~\ref{fig:reportingempirical} in Section~\ref{sec:data}. Note that this can also be informative for the occurrence process, since damage occurring late in the day may be wrongly attributed to the following day. 

\section{Occurrence and reporting model} \label{sec:model}
In this section, we discuss the models used for analysing the occurrence and reporting dynamics of the private home insurance portfolio in the case study in Section~\ref{sec:casestudy}. We first recall the EM-XGBoost framework for reporting delays with high-dimensional data from \cite{wilsens2025machine}, which accommodates the many correlated weather covariates in our setting without requiring manual variable selection. Section~\ref{sec:regimetheory} presents our multi-regime extension to their framework. This extension allows us to fit separate occurrence and/or reporting models per regime, e.g.~under different weather conditions, instead of imposing one common set of effects on all observations. A multi-regime setup has practical advantages over, for example, including a regime label as a covariate, since the feature importance can then be inspected per regime instead of requiring an interaction analysis.

\subsection{Single-regime EM-XGBoost framework} \label{sec:assumptions}
We follow the distributional assumptions of the EM literature discussed in Section~\ref{sec:intro} \citep{verbelen2022modeling,crevecoeur2023bridging,wilsens2025machine}: a Poisson assumption for the number of claims and a multinomial assumption for their distribution over the reporting delays. This allows the complete log-likelihood (see below) to be separated into an occurrence and a reporting term, so that both can be estimated within the maximisation step. For the number of claims for observation $i$, we assume
\begin{align*}
     N_i \sim \text{Poisson}(e_{i} \cdot \lambda_{i}) \:\:\: \text{with} \:\:\: \lambda_{i} := \lambda(\boldsymbol{x}_i) \ge 0.
\end{align*}
For the number of claims reported with a delay of $j-1$ days, conditional on $N_i$, we have
\begin{align*}
N_{i,j}|N_i \sim \text{Multinomial}(p_j(\boldsymbol{x}_i)) \:\:\: \text{with} \:\:\: p_{i,j} := p_j(\boldsymbol{x}_i), j=1,\ldots,d \:\:\: \text{and} \:\:\: \sum_{j=1}^{d} p_j(\boldsymbol{x}_i)=1.
\end{align*}
Following the assumptions stated above, we denote the observed log-likelihood, which is computed over the reported claims, as
\begin{align*}
    LL(\lambda_i, p_{i,j};D)=\sum_{i=1}^n \sum_{j=1}^{\tau_i}\bigg[-e_{i}\cdot\lambda_i \cdot p_{i,j}+N_{i,j} \cdot \log \left(e_{i}\cdot\lambda_i\right)+N_{i,j} \cdot \log \left(p_{i,j}\right)-\log \left(N_{i,j} !\right)\bigg].
\end{align*}
As stated in \cite{verbelen2022modeling}, the interaction term $-e_{i}\cdot\lambda_i \cdot p_{i,j}$ complicates direct optimisation. The log-likelihood for the complete data $\mathcal{D}$ can be written as
\begin{align*}
    LL_{c}(\lambda_i, p_{i,j};\mathcal{D})=\sum_{i=1}^n\bigg[-e_{i}\cdot\lambda_i+N_i \cdot \log \left(e_{i}\cdot\lambda_i\right)+\sum_{j=1}^{d}\left\{N_{i, j} \cdot \log \left(p_{i,j}\right)-\log \left(N_{i,j} !\right)\right\}\bigg].
\end{align*}
The complete log-likelihood $LL_{c}(\lambda_i, p_{i,j};\mathcal{D})$ has the advantage that it can be decoupled into an occurrence and a reporting part. However, it is also not directly optimisable, since the claim counts $N_{i,\tau_{i}+1},\ldots,N_{i,d}$ with $i=1,\ldots,n$ are unobservable. By applying an EM algorithm, this issue can be circumvented via the optimisation of 
\begin{align*}
    Q^{\text{occ}}(\lambda_{i}^{(k)};\mathcal{D}^{(k)}) 
    &= \sum_{i=1}^n\left[ -e_{i}\cdot\lambda_{i}^{(k)}+N^{(k)}_i \cdot \log \left(e_{i}\cdot\lambda_{i}^{(k)}\right) \right], \:\:\: \text{and}\\
    Q^{\text{rep}}(p_{i,j}^{(k)};\mathcal{D}^{(k)}) &= \sum_{i=1}^n \sum_{j=1}^{d} N^{(k)}_{i, j} \cdot \log \left(p_{i,j}^{(k)}\right), \:\:\: \text{subject to} \:\:\: \sum_{j=1}^d p_{i,j}^{(k)} = 1 \:\:\: \text{for} \:\:\: i=1,\ldots,n,
\end{align*}
which are the expectations of the occurrence and reporting parts of the complete log-likelihood, respectively, given an expectation of the complete information $\mathcal{D}^{(k)}=(N^{(k)}_{i,1},\ldots,N^{(k)}_{i,d}, e_{i},\boldsymbol{x}_i)_{i=1}^{n}$ in the $k$th EM iteration. 

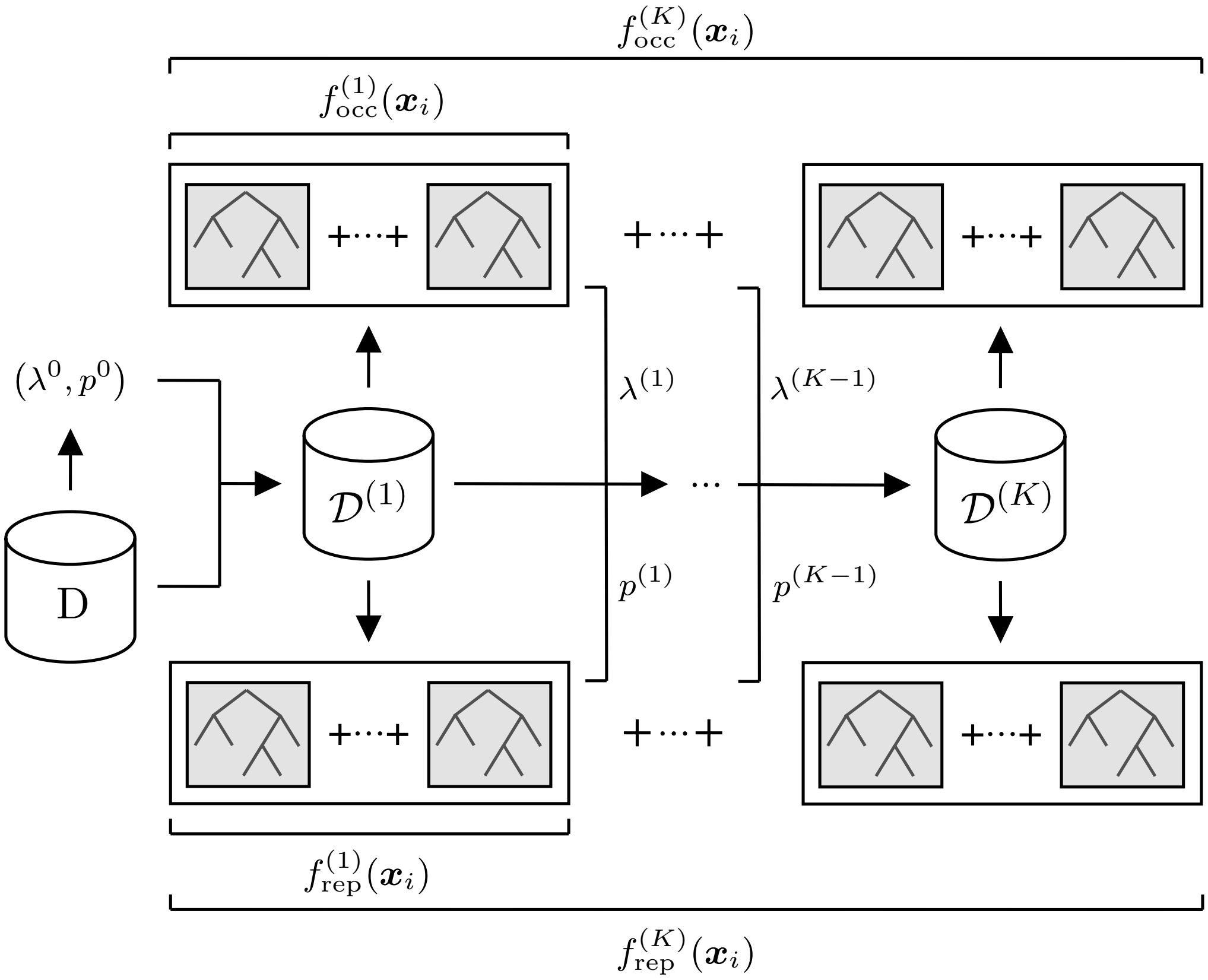
\begin{figure}[H]
\centering
\begin{adjustbox}{width=0.85\textwidth}

\tikzset{every picture/.style={line width=0.75pt}} 

\begin{tikzpicture}[x=0.75pt,y=0.75pt,yscale=-1,xscale=1]

\draw    (438.58,109.7) -- (445.59,109.69) ;
\draw   (290.01,192.52) -- (290.01,224.84) .. controls (290.01,229.65) and (280.47,233.54) .. (268.7,233.54) .. controls (256.93,233.54) and (247.39,229.65) .. (247.39,224.84) -- (247.39,192.52)(290.01,192.52) .. controls (290.01,197.33) and (280.47,201.22) .. (268.7,201.22) .. controls (256.93,201.22) and (247.39,197.33) .. (247.39,192.52) .. controls (247.39,187.72) and (256.93,183.82) .. (268.7,183.82) .. controls (280.47,183.82) and (290.01,187.72) .. (290.01,192.52) -- cycle ;
\draw    (318,208.55) -- (297.5,208.5) ;
\draw    (268.74,159.32) -- (268.7,176.82) ;
\draw [shift={(268.75,156.32)}, rotate = 90.14] [fill={rgb, 255:red, 0; green, 0; blue, 0 }  ][line width=0.08]  [draw opacity=0] (8.93,-4.29) -- (0,0) -- (8.93,4.29) -- cycle    ;
\draw    (318,140.5) -- (297.5,140.45) ;
\draw    (318,208.55) -- (318,140.5) ;
\draw    (335.5,174.57) -- (318,174.53) ;
\draw [shift={(338.5,174.58)}, rotate = 180.14] [fill={rgb, 255:red, 0; green, 0; blue, 0 }  ][line width=0.08]  [draw opacity=0] (8.93,-4.29) -- (0,0) -- (8.93,4.29) -- cycle    ;
\draw   (388.51,158.52) -- (388.51,190.84) .. controls (388.51,195.65) and (378.97,199.54) .. (367.2,199.54) .. controls (355.43,199.54) and (345.89,195.65) .. (345.89,190.84) -- (345.89,158.52)(388.51,158.52) .. controls (388.51,163.33) and (378.97,167.22) .. (367.2,167.22) .. controls (355.43,167.22) and (345.89,163.33) .. (345.89,158.52) .. controls (345.89,153.72) and (355.43,149.82) .. (367.2,149.82) .. controls (378.97,149.82) and (388.51,153.72) .. (388.51,158.52) -- cycle ;
\draw    (326.63,78.28) -- (315.35,86.85) ;
\draw    (326.63,78.28) -- (337.91,86.85) ;

\draw    (337.81,86.74) -- (331.61,96.8) ;
\draw    (337.81,86.74) -- (344,96.8) ;

\draw    (315.81,86.46) -- (309.62,96.52) ;
\draw    (315.81,86.46) -- (322,96.52) ;

\draw    (331.79,96.47) -- (325.6,106.53) ;
\draw    (331.79,96.47) -- (337.98,106.53) ;

\draw  [fill={rgb, 255:red, 193; green, 190; blue, 190 }  ,fill opacity=0.43 ] (307.03,75.76) -- (347.36,75.76) -- (347.36,110.11) -- (307.03,110.11) -- cycle ;
\draw   (353.98,92.91) -- (361.03,92.91)(357.51,89.43) -- (357.51,96.39) ;
\draw    (406.43,78.28) -- (395.15,86.85) ;
\draw    (406.43,78.28) -- (417.71,86.85) ;

\draw    (417.61,86.74) -- (411.42,96.8) ;
\draw    (417.61,86.74) -- (423.8,96.8) ;

\draw    (395.61,86.46) -- (389.42,96.52) ;
\draw    (395.61,86.46) -- (401.81,96.52) ;

\draw    (411.59,96.47) -- (405.4,106.53) ;
\draw    (411.59,96.47) -- (417.79,106.53) ;

\draw  [fill={rgb, 255:red, 193; green, 190; blue, 190 }  ,fill opacity=0.43 ] (386.84,75.76) -- (427.16,75.76) -- (427.16,110.11) -- (386.84,110.11) -- cycle ;
\draw   (373.02,92.91) -- (380.07,92.91)(376.54,89.43) -- (376.54,96.39) ;
\draw   (301.53,69.1) -- (433.08,69.1) -- (433.08,115.76) -- (301.53,115.76) -- cycle ;
\draw    (367.24,125.32) -- (367.2,142.82) ;
\draw [shift={(367.25,122.32)}, rotate = 90.14] [fill={rgb, 255:red, 0; green, 0; blue, 0 }  ][line width=0.08]  [draw opacity=0] (8.93,-4.29) -- (0,0) -- (8.93,4.29) -- cycle    ;
\draw    (367.25,224.04) -- (367.2,206.54) ;
\draw [shift={(367.25,227.04)}, rotate = 269.86] [fill={rgb, 255:red, 0; green, 0; blue, 0 }  ][line width=0.08]  [draw opacity=0] (8.93,-4.29) -- (0,0) -- (8.93,4.29) -- cycle    ;
\draw    (326.88,242.79) -- (315.6,251.36) ;
\draw    (326.88,242.79) -- (338.16,251.36) ;

\draw    (338.06,251.25) -- (331.86,261.31) ;
\draw    (338.06,251.25) -- (344.25,261.31) ;

\draw    (316.06,250.97) -- (309.87,261.03) ;
\draw    (316.06,250.97) -- (322.25,261.03) ;

\draw    (332.04,260.98) -- (325.85,271.04) ;
\draw    (332.04,260.98) -- (338.23,271.04) ;

\draw  [fill={rgb, 255:red, 193; green, 190; blue, 190 }  ,fill opacity=0.43 ] (307.28,240.27) -- (347.61,240.27) -- (347.61,274.62) -- (307.28,274.62) -- cycle ;
\draw   (354.23,257.42) -- (361.28,257.42)(357.76,253.95) -- (357.76,260.9) ;
\draw    (406.68,242.79) -- (395.4,251.36) ;
\draw    (406.68,242.79) -- (417.96,251.36) ;

\draw    (417.86,251.25) -- (411.67,261.31) ;
\draw    (417.86,251.25) -- (424.05,261.31) ;

\draw    (395.86,250.97) -- (389.67,261.03) ;
\draw    (395.86,250.97) -- (402.06,261.03) ;

\draw    (411.84,260.98) -- (405.65,271.04) ;
\draw    (411.84,260.98) -- (418.04,271.04) ;

\draw  [fill={rgb, 255:red, 193; green, 190; blue, 190 }  ,fill opacity=0.43 ] (387.09,240.27) -- (427.41,240.27) -- (427.41,274.62) -- (387.09,274.62) -- cycle ;
\draw   (373.27,257.42) -- (380.32,257.42)(376.79,253.95) -- (376.79,260.9) ;
\draw   (301.78,233.61) -- (433.33,233.61) -- (433.33,280.27) -- (301.78,280.27) -- cycle ;
\draw   (451.55,256.89) -- (460.36,256.89)(455.95,252.42) -- (455.95,261.35) ;
\draw   (475.24,256.89) -- (484.04,256.89)(479.64,252.42) -- (479.64,261.35) ;
\draw    (535.88,242.94) -- (524.6,251.5) ;
\draw    (535.88,242.94) -- (547.16,251.5) ;

\draw    (547.06,251.4) -- (540.86,261.46) ;
\draw    (547.06,251.4) -- (553.25,261.46) ;

\draw    (525.06,251.11) -- (518.87,261.18) ;
\draw    (525.06,251.11) -- (531.25,261.18) ;

\draw    (541.04,261.12) -- (534.85,271.19) ;
\draw    (541.04,261.12) -- (547.23,271.19) ;

\draw  [fill={rgb, 255:red, 193; green, 190; blue, 190 }  ,fill opacity=0.43 ] (516.28,240.41) -- (556.61,240.41) -- (556.61,274.77) -- (516.28,274.77) -- cycle ;
\draw   (563.23,257.57) -- (570.28,257.57)(566.76,254.09) -- (566.76,261.05) ;
\draw    (615.68,242.94) -- (604.4,251.5) ;
\draw    (615.68,242.94) -- (626.96,251.5) ;

\draw    (626.86,251.4) -- (620.67,261.46) ;
\draw    (626.86,251.4) -- (633.05,261.46) ;

\draw    (604.86,251.11) -- (598.67,261.18) ;
\draw    (604.86,251.11) -- (611.06,261.18) ;

\draw    (620.84,261.12) -- (614.65,271.19) ;
\draw    (620.84,261.12) -- (627.04,271.19) ;

\draw  [fill={rgb, 255:red, 193; green, 190; blue, 190 }  ,fill opacity=0.43 ] (596.09,240.41) -- (636.41,240.41) -- (636.41,274.77) -- (596.09,274.77) -- cycle ;
\draw   (582.27,257.57) -- (589.32,257.57)(585.79,254.09) -- (585.79,261.05) ;
\draw   (510.78,233.75) -- (642.33,233.75) -- (642.33,280.42) -- (510.78,280.42) -- cycle ;
\draw   (451.71,92.35) -- (460.52,92.35)(456.11,87.88) -- (456.11,96.81) ;
\draw   (475.4,92.35) -- (484.2,92.35)(479.8,87.88) -- (479.8,96.81) ;
\draw    (536.04,78.4) -- (524.76,86.97) ;
\draw    (536.04,78.4) -- (547.32,86.97) ;

\draw    (547.21,86.86) -- (541.02,96.92) ;
\draw    (547.21,86.86) -- (553.41,96.92) ;

\draw    (525.22,86.57) -- (519.02,96.64) ;
\draw    (525.22,86.57) -- (531.41,96.64) ;

\draw    (541.2,96.58) -- (535,106.65) ;
\draw    (541.2,96.58) -- (547.39,106.65) ;

\draw  [fill={rgb, 255:red, 193; green, 190; blue, 190 }  ,fill opacity=0.43 ] (516.44,75.87) -- (556.77,75.87) -- (556.77,110.23) -- (516.44,110.23) -- cycle ;
\draw   (563.39,93.03) -- (570.44,93.03)(566.92,89.55) -- (566.92,96.51) ;
\draw    (615.84,78.4) -- (604.56,86.97) ;
\draw    (615.84,78.4) -- (627.12,86.97) ;

\draw    (627.02,86.86) -- (620.83,96.92) ;
\draw    (627.02,86.86) -- (633.21,96.92) ;

\draw    (605.02,86.57) -- (598.83,96.64) ;
\draw    (605.02,86.57) -- (611.22,96.64) ;

\draw    (621,96.58) -- (614.81,106.65) ;
\draw    (621,96.58) -- (627.2,106.65) ;

\draw  [fill={rgb, 255:red, 193; green, 190; blue, 190 }  ,fill opacity=0.43 ] (596.24,75.87) -- (636.57,75.87) -- (636.57,110.23) -- (596.24,110.23) -- cycle ;
\draw   (582.43,93.03) -- (589.48,93.03)(585.95,89.55) -- (585.95,96.51) ;
\draw   (510.94,69.21) -- (642.49,69.21) -- (642.49,115.88) -- (510.94,115.88) -- cycle ;
\draw    (445.59,109.69) -- (445.84,239.68) ;
\draw    (438.83,239.68) -- (445.84,239.68) ;
\draw    (395.5,174.57) -- (445.71,174.69) ;
\draw    (463.21,174.73) -- (445.71,174.69) ;
\draw [shift={(466.21,174.74)}, rotate = 180.14] [fill={rgb, 255:red, 0; green, 0; blue, 0 }  ][line width=0.08]  [draw opacity=0] (8.93,-4.29) -- (0,0) -- (8.93,4.29) -- cycle    ;
\draw    (489.08,109.95) -- (496.09,109.94) ;
\draw    (496.09,109.94) -- (496.34,239.93) ;
\draw    (489.33,239.93) -- (496.34,239.93) ;
\draw    (489.21,174.94) -- (496.21,174.94) ;
\draw   (597.26,158.77) -- (597.26,191.09) .. controls (597.26,195.9) and (587.72,199.79) .. (575.95,199.79) .. controls (564.18,199.79) and (554.64,195.9) .. (554.64,191.09) -- (554.64,158.77)(597.26,158.77) .. controls (597.26,163.58) and (587.72,167.47) .. (575.95,167.47) .. controls (564.18,167.47) and (554.64,163.58) .. (554.64,158.77) .. controls (554.64,153.97) and (564.18,150.07) .. (575.95,150.07) .. controls (587.72,150.07) and (597.26,153.97) .. (597.26,158.77) -- cycle ;
\draw    (575.99,125.57) -- (575.95,143.07) ;
\draw [shift={(576,122.57)}, rotate = 90.14] [fill={rgb, 255:red, 0; green, 0; blue, 0 }  ][line width=0.08]  [draw opacity=0] (8.93,-4.29) -- (0,0) -- (8.93,4.29) -- cycle    ;
\draw    (576,224.29) -- (575.95,206.79) ;
\draw [shift={(576,227.29)}, rotate = 269.86] [fill={rgb, 255:red, 0; green, 0; blue, 0 }  ][line width=0.08]  [draw opacity=0] (8.93,-4.29) -- (0,0) -- (8.93,4.29) -- cycle    ;
\draw    (496.21,174.94) -- (543.42,175.04) ;
\draw [shift={(546.42,175.05)}, rotate = 180.13] [fill={rgb, 255:red, 0; green, 0; blue, 0 }  ][line width=0.08]  [draw opacity=0] (8.93,-4.29) -- (0,0) -- (8.93,4.29) -- cycle    ;
\draw    (301.53,59.1) -- (433.08,59.1) ;
\draw    (301.53,64.1) -- (301.53,59.1) ;
\draw    (433.08,64.1) -- (433.08,59.1) ;
\draw    (301.78,290.27) -- (433.33,290.27) ;
\draw    (301.78,290.27) -- (301.78,285.27) ;
\draw    (433.33,290.27) -- (433.33,285.27) ;
\draw    (301.53,34.1) -- (642.49,34.11) ;
\draw    (301.53,39.1) -- (301.53,34.1) ;
\draw    (642.49,39.11) -- (642.49,34.11) ;
\draw    (301.78,315.27) -- (642.74,315.29) ;
\draw    (301.78,315.27) -- (301.78,310.27) ;
\draw    (642.74,315.29) -- (642.74,310.29) ;

\draw (262.5,207.55) node [anchor=north west][inner sep=0.75pt]   [align=left] {D};
\draw (247.66,132.47) node [anchor=north west][inner sep=0.75pt]  [font=\scriptsize] [align=left] {$\displaystyle \left( \lambda ^{0} ,p^{0}\right)$};
\draw (352.5,171.88) node [anchor=north west][inner sep=0.75pt]   [align=left] {$\displaystyle \mathcal{D}^{( 1)}$};
\draw (360,90.5) node [anchor=north west][inner sep=0.75pt]  [font=\small] [align=left] {...};
\draw (360.24,254.9) node [anchor=north west][inner sep=0.75pt]  [font=\small] [align=left] {...};
\draw (461,254.9) node [anchor=north west][inner sep=0.75pt]  [font=\small] [align=left] {...};
\draw (569.24,255.1) node [anchor=north west][inner sep=0.75pt]  [font=\small] [align=left] {...};
\draw (461,90) node [anchor=north west][inner sep=0.75pt]  [font=\small] [align=left] {...};
\draw (569.4,90.5) node [anchor=north west][inner sep=0.75pt]  [font=\small] [align=left] {...};
\draw (471.61,172.8) node [anchor=north west][inner sep=0.75pt]  [font=\small] [align=left] {...};
\draw (561.25,172.13) node [anchor=north west][inner sep=0.75pt]   [align=left] {$\displaystyle \mathcal{D}^{( K)}$};
\draw (498.16,134.72) node [anchor=north west][inner sep=0.75pt]  [font=\scriptsize] [align=left] {$\displaystyle \lambda ^{( K-1)}$};
\draw (499.41,199.72) node [anchor=north west][inner sep=0.75pt]  [font=\scriptsize] [align=left] {$\displaystyle p^{( K-1)}$};
\draw (448.16,134.72) node [anchor=north west][inner sep=0.75pt]  [font=\scriptsize] [align=left] {$\displaystyle \lambda ^{( 1)}$};
\draw (448.41,199.47) node [anchor=north west][inner sep=0.75pt]  [font=\scriptsize] [align=left] {$\displaystyle p^{( 1)}$};
\draw (348.89,38.15) node [anchor=north west][inner sep=0.75pt]  [font=\footnotesize] [align=left] {$\displaystyle f_{\text{occ}}^{(1)}(\boldsymbol{x}_{i})$};
\draw (343.89,293.9) node [anchor=north west][inner sep=0.75pt]  [font=\footnotesize] [align=left] {$\displaystyle f_{\text{rep}}^{( 1)}(\boldsymbol{x}_{i})$};
\draw (447.39,319.9) node [anchor=north west][inner sep=0.75pt]  [font=\footnotesize] [align=left] {$\displaystyle f_{\text{rep}}^{( K)}(\boldsymbol{x}_{i})$};
\draw (447.39,15.15) node [anchor=north west][inner sep=0.75pt]  [font=\footnotesize] [align=left] {$\displaystyle f_{\text{occ}}^{( K)}(\boldsymbol{x}_{i})$};

\end{tikzpicture}
\end{adjustbox}
\caption{Visualisation of the flow of the EM-XGBoost steps \citep{wilsens2025machine}, starting from an observed dataset $D$. The upper part shows the construction of the additive occurrence models $f_{\text{occ}}^{( 1)}(\boldsymbol{x}_{i}),\ldots,f_{\text{occ}}^{(K)}(\boldsymbol{x}_{i})$ resulting in the predicted occurrence intensities $\lambda^{(1)}_{i},\ldots,\lambda^{(K)}_{i}$, respectively. The lower part visualises the construction of the additive reporting models $f_{\text{rep}}^{( 1)}(\boldsymbol{x}_{i}),\ldots,f_{\text{rep}}^{(K)}(\boldsymbol{x}_{i})$ resulting in the predicted reporting probabilities $p_{i,j}^{(1)},\ldots,p_{i,j}^{(K)}$, respectively. The middle part shows the expectation step, i.e.~the construction of $\mathcal{D}^{(1)},\ldots,\mathcal{D}^{(K)}$ via the predicted occurrence intensities and reporting probabilities.}
\label{fig:EMframework}
\end{figure}

\cite{verbelen2022modeling} and \cite{crevecoeur2023bridging} implemented the maximisation step with generalised linear models. This requires manual covariate selection and an explicit specification of the functional form, which becomes impractical with the large number of weather covariates we collect. \cite{wilsens2025machine} proposed using XGBoost models within the maximisation step, allowing the inclusion of many covariates without prior selection and capture non-linear effects. This approach is well-suited to our setting, since the weather covariates described in Section~\ref{sec:data} are highly correlated. Figure~\ref{fig:EMframework} visualises the flow of an EM-XGBoost framework, applied to an observed dataset $D$. We illustrate the consecutive steps of the EM algorithm, alternating between so-called expectation and maximisation steps. In the $k$th expectation step, we construct an expectation for the unobserved, complete dataset $\mathcal{D}$, denoted as $\mathcal{D}^{(k)}=(N^{(k)}_{i,1},\ldots,N^{(k)}_{i,d}, e_{i},\boldsymbol{x}_i)_{i=1}^{n}$, using the previous estimates for the occurrence intensities and reporting probabilities, i.e.~$\lambda^{(k-1)}_{i}$ and $p_{i,j}^{(k-1)}$. In the $k$th maximisation step, we learn the occurrence model $f^{(k)}_{\text{occ}}(\boldsymbol{x}_{i})$ and reporting model $f^{(k)}_{\text{rep}}(\boldsymbol{x}_{i})=(f^{(k)[1]}_{\text{rep}}(\boldsymbol{x}_{i}),\ldots,f^{(k)[d]}_{\text{rep}}(\boldsymbol{x}_{i}))$ by adding additional regression trees to the previous models $f^{(k-1)}_{\text{occ}}(\boldsymbol{x}_{i})$ and $f^{(k-1)}_{\text{rep}}(\boldsymbol{x}_{i})$, respectively. This way, we incorporate the iterative updates of the expectation of the unobserved data into the training process of the occurrence and reporting model. We subsequently set 
\begin{align*}
    \widehat{\lambda}^{(k)}(\boldsymbol{x}_i) = \exp{( \widehat{f}^{(k)}_{\text{occ}}(\boldsymbol{x}_i))} \:\: \text{and} \:\: \widehat{p}_{j}^{(k)}(\boldsymbol{x}_i)= \frac{\exp(\widehat{f}_{\text{rep}}^{(k)[j]}(\boldsymbol{x}_{i}))}{\sum_{z=1}^{d} \exp(\widehat{f}_{\text{rep}}^{(k)[z]}(\boldsymbol{x}_{i}))},
\end{align*}
setting the stage for the $k+1$th expectation step. The full details of the EM-XGBoost algorithm \citep{wilsens2025machine} are in Algorithm~\ref{algo:EM} in Appendix~\ref{app:emalgo}.

\subsection{Multi-regime EM-XGBoost} \label{sec:regimetheory}

\paragraph{Regime construction} We propose an extension to the EM-XGBoost framework \citep{wilsens2025machine} outlined above. A single occurrence model, as in Figure~\ref{fig:EMframework}, describes occurrence dynamics across all observations at once. This is demanding when the portfolio is exposed to distinct conditions, which generate claims through different mechanisms. In our setting, days with severe weather are rare but produce a large share of the claims, and the covariates driving claim occurrence under such conditions need not be the same as those that are relevant on moderate days. We use the term regime for a distinct set of conditions under which claims arise, e.g.~weather conditions such as clear versus severe weather. In contrast to regime-switching models such as \cite{robben2025granular}, in which the regime is a latent state estimated jointly with the model, we identify the regimes through clustering of covariate information, so that each cluster serves as an empirical proxy for a regime. This identification is a preprocessing step: the clustering is performed once, before the EM algorithm is run, and assigns every observation to exactly one regime. Suppose $\mathcal{S}$ is a subset of the available covariates. We then define
\begin{align*}
    \{\boldsymbol{x}_{i}^{(\mathcal{S})} \}_{i=1,\ldots,n},
\end{align*}
where the covariate vector $\boldsymbol{x}_{i}^{(\mathcal{S})}$ only contains the covariates in $\mathcal{S}$. The choice of $\mathcal{S}$ is left to the modeller: $\mathcal{S}$ can consist of any subset of the available covariates. Whether the resulting regimes admit a substantive interpretation depends on this choice, the framework itself does not require the regimes to be interpretable. For example, if $\mathcal{S}$ consists of weather-related variables, we can use it to define different weather regimes. Next, we apply a clustering method of choice that results in $L$ non-overlapping clusters $\mathcal{C}_{1},\ldots,\mathcal{C}_{L}$ such that
\begin{align*}
    \mathcal{C}_{l} \subseteq\{\boldsymbol{x}_{i}^{(\mathcal{S})} \}_{i=1,\ldots,n}
    =
    \bigcup_{l=1}^{L} \mathcal{C}_{l}
    \:\:\: \text{and} \:\:\:
    \mathcal{C}_{l_{1}} \cap \mathcal{C}_{l_{2}} = \varnothing,
    \:\:\: \text{for} \:\:\: l_{1} \neq l_{2}.
\end{align*}
That is, we group observations whose covariates
in $\mathcal{S}$ are similar, and let each group
represent a regime. Subsequently, we split the observed data $D$ according to this partitioning into $L$ different datasets $D_{1},\ldots,D_{L}$ with
\begin{align}
    D_{l} = \{(N_{i,1},\ldots,N_{i,d}, e_{i},\boldsymbol{x}_i) \:\: | \:\: i \in \mathcal{I}_{l}\}, \:\:\: \mathcal{I}_{l} = \{i \:\: | \:\: \boldsymbol{x}_{i}^{(\mathcal{S})} \in \mathcal{C}_{l}\}.
\end{align}
Each cluster $\mathcal{C}_{l}$ thus corresponds to a regime $l$, with $D_{l}$ the data assigned to it. Note that the regime
allocation applies to an observation rather than to an area: when the covariates used for the clustering vary from day to day, as they do for weather covariates, the same area can belong to one regime on a given day and to another the next. Once identified, the regimes are kept fixed: a new observation, e.g.~a future day, is assigned to the existing regime whose observations it most resembles in terms of its covariates in $\mathcal{S}$. For the k-means clustering used in the case study, this is the regime with the closest centroid, see Section~\ref{sec:regimeid}. 

\begin{figure}[h]
\centering
\begin{adjustbox}{width=0.85\textwidth}

\tikzset{every picture/.style={line width=0.75pt}} 

\begin{tikzpicture}[x=0.75pt,y=0.75pt,yscale=-1,xscale=1]

\draw    (306.76,315.13) -- (299.75,315.13) ;
\draw   (54.92,192.87) -- (54.92,225.19) .. controls (54.92,229.99) and (45.38,233.89) .. (33.62,233.89) .. controls (21.85,233.89) and (12.31,229.99) .. (12.31,225.19) -- (12.31,192.87)(54.92,192.87) .. controls (54.92,197.68) and (45.38,201.57) .. (33.62,201.57) .. controls (21.85,201.57) and (12.31,197.68) .. (12.31,192.87) .. controls (12.31,188.07) and (21.85,184.17) .. (33.62,184.17) .. controls (45.38,184.17) and (54.92,188.07) .. (54.92,192.87) -- cycle ;
\draw    (79.43,210.41) -- (61.93,210.37) ;
\draw [shift={(82.43,210.42)}, rotate = 180.14] [fill={rgb, 255:red, 0; green, 0; blue, 0 }  ][line width=0.08]  [draw opacity=0] (8.93,-4.29) -- (0,0) -- (8.93,4.29) -- cycle    ;
\draw  [dashed]  (33.62,238.89) -- (33.62,272) -- (123.01,272) -- (123.01,252.7) ;
\draw [shift={(123.01,249.7)}, rotate = 90] [fill={rgb, 255:red, 0; green, 0; blue, 0 }  ][line width=0.08]  [draw opacity=0] (8.93,-4.29) -- (0,0) -- (8.93,4.29) -- cycle    ;
\draw    (187.34,210.46) -- (166.84,210.41) ;
\draw   (248.64,155.26) -- (248.64,187.58) .. controls (248.64,192.39) and (239.1,196.28) .. (227.33,196.28) .. controls (215.56,196.28) and (206.02,192.39) .. (206.02,187.58) -- (206.02,155.26)(248.64,155.26) .. controls (248.64,160.07) and (239.1,163.96) .. (227.33,163.96) .. controls (215.56,163.96) and (206.02,160.07) .. (206.02,155.26) .. controls (206.02,150.46) and (215.56,146.56) .. (227.33,146.56) .. controls (239.1,146.56) and (248.64,150.46) .. (248.64,155.26) -- cycle ;
\draw   (248.64,233.55) -- (248.64,265.86) .. controls (248.64,270.67) and (239.1,274.56) .. (227.33,274.56) .. controls (215.56,274.56) and (206.02,270.67) .. (206.02,265.86) -- (206.02,233.55)(248.64,233.55) .. controls (248.64,238.35) and (239.1,242.25) .. (227.33,242.25) .. controls (215.56,242.25) and (206.02,238.35) .. (206.02,233.55) .. controls (206.02,228.74) and (215.56,224.85) .. (227.33,224.85) .. controls (239.1,224.85) and (248.64,228.74) .. (248.64,233.55) -- cycle ;
\draw    (306.84,218.37) -- (306.84,249.7) ;
\draw    (187.34,210.46) -- (186.88,171.42) -- (196.02,171.43) ;
\draw [shift={(199.02,171.43)}, rotate = 180.02] [fill={rgb, 255:red, 0; green, 0; blue, 0 }  ][line width=0.08]  [draw opacity=0] (8.93,-4.29) -- (0,0) -- (8.93,4.29) -- cycle    ;
\draw    (187.34,210.67) -- (186.88,249.71) -- (196.02,249.71) ;
\draw [shift={(199.02,249.71)}, rotate = 179.98] [fill={rgb, 255:red, 0; green, 0; blue, 0 }  ][line width=0.08]  [draw opacity=0] (8.93,-4.29) -- (0,0) -- (8.93,4.29) -- cycle    ;
\draw    (186.8,74.6) -- (175.52,83.17) ;
\draw    (186.8,74.6) -- (198.08,83.17) ;

\draw    (197.97,83.06) -- (191.78,93.13) ;
\draw    (197.97,83.06) -- (204.17,93.13) ;

\draw    (175.98,82.78) -- (169.78,92.84) ;
\draw    (175.98,82.78) -- (182.17,92.84) ;

\draw    (191.96,92.79) -- (185.76,102.85) ;
\draw    (191.96,92.79) -- (198.15,102.85) ;

\draw  [fill={rgb, 255:red, 193; green, 190; blue, 190 }  ,fill opacity=0.43 ] (167.2,72.08) -- (207.52,72.08) -- (207.52,106.44) -- (167.2,106.44) -- cycle ;
\draw   (214.15,89.23) -- (221.2,89.23)(217.67,85.76) -- (217.67,92.71) ;
\draw    (266.6,74.6) -- (255.32,83.17) ;
\draw    (266.6,74.6) -- (277.88,83.17) ;

\draw    (277.78,83.06) -- (271.58,93.13) ;
\draw    (277.78,83.06) -- (283.97,93.13) ;

\draw    (255.78,82.78) -- (249.59,92.84) ;
\draw    (255.78,82.78) -- (261.97,92.84) ;

\draw    (271.76,92.79) -- (265.57,102.85) ;
\draw    (271.76,92.79) -- (277.95,102.85) ;

\draw  [fill={rgb, 255:red, 193; green, 190; blue, 190 }  ,fill opacity=0.43 ] (247,72.08) -- (287.33,72.08) -- (287.33,106.44) -- (247,106.44) -- cycle ;
\draw   (233.18,89.23) -- (240.23,89.23)(236.71,85.76) -- (236.71,92.71) ;
\draw   (161.7,65.42) -- (293.25,65.42) -- (293.25,112.08) -- (161.7,112.08) -- cycle ;
\draw    (227.37,122.06) -- (227.33,139.56) ;
\draw [shift={(227.38,119.06)}, rotate = 90.14] [fill={rgb, 255:red, 0; green, 0; blue, 0 }  ][line width=0.08]  [draw opacity=0] (8.93,-4.29) -- (0,0) -- (8.93,4.29) -- cycle    ;
\draw    (227.38,299.06) -- (227.33,281.57) ;
\draw [shift={(227.38,302.06)}, rotate = 269.86] [fill={rgb, 255:red, 0; green, 0; blue, 0 }  ][line width=0.08]  [draw opacity=0] (8.93,-4.29) -- (0,0) -- (8.93,4.29) -- cycle    ;
\draw    (186.8,318.24) -- (175.52,326.81) ;
\draw    (186.8,318.24) -- (198.08,326.81) ;

\draw    (197.97,326.7) -- (191.78,336.77) ;
\draw    (197.97,326.7) -- (204.17,336.77) ;

\draw    (175.98,326.42) -- (169.78,336.49) ;
\draw    (175.98,326.42) -- (182.17,336.49) ;

\draw    (191.96,336.43) -- (185.76,346.5) ;
\draw    (191.96,336.43) -- (198.15,346.5) ;

\draw  [fill={rgb, 255:red, 193; green, 190; blue, 190 }  ,fill opacity=0.43 ] (167.2,315.72) -- (207.52,315.72) -- (207.52,350.08) -- (167.2,350.08) -- cycle ;
\draw   (214.15,332.88) -- (221.2,332.88)(217.67,329.4) -- (217.67,336.36) ;
\draw    (266.6,318.24) -- (255.32,326.81) ;
\draw    (266.6,318.24) -- (277.88,326.81) ;

\draw    (277.78,326.7) -- (271.58,336.77) ;
\draw    (277.78,326.7) -- (283.97,336.77) ;

\draw    (255.78,326.42) -- (249.59,336.49) ;
\draw    (255.78,326.42) -- (261.97,336.49) ;

\draw    (271.76,336.43) -- (265.57,346.5) ;
\draw    (271.76,336.43) -- (277.95,346.5) ;

\draw  [fill={rgb, 255:red, 193; green, 190; blue, 190 }  ,fill opacity=0.43 ] (247,315.72) -- (287.33,315.72) -- (287.33,350.08) -- (247,350.08) -- cycle ;
\draw   (233.18,332.88) -- (240.23,332.88)(236.71,329.4) -- (236.71,336.36) ;
\draw   (161.7,309.06) -- (293.25,309.06) -- (293.25,355.73) -- (161.7,355.73) -- cycle ;
\draw    (306.84,249.7) -- (306.76,315.13) ;
\draw    (306.84,203.07) -- (299.84,203.07) ;
\draw    (255.64,249.7) -- (306.84,249.7) ;
\draw    (306.76,106.01) -- (299.75,106.02) ;
\draw    (255.64,171.42) -- (306.9,171.42) ;
\draw    (306.84,218.37) -- (299.84,218.38) ;
\draw    (364.4,171.67) -- (357.4,171.67) ;
\draw    (306.76,106.01) -- (306.84,203.07) ;
\draw    (324.4,171.46) -- (306.9,171.42) ;
\draw [shift={(327.4,171.47)}, rotate = 180.14] [fill={rgb, 255:red, 0; green, 0; blue, 0 }  ][line width=0.08]  [draw opacity=0] (8.93,-4.29) -- (0,0) -- (8.93,4.29) -- cycle    ;
\draw    (324.34,249.75) -- (306.84,249.7) ;
\draw [shift={(327.34,249.75)}, rotate = 180.14] [fill={rgb, 255:red, 0; green, 0; blue, 0 }  ][line width=0.08]  [draw opacity=0] (8.93,-4.29) -- (0,0) -- (8.93,4.29) -- cycle    ;
\draw    (364.34,249.95) -- (357.34,249.95) ;
\draw    (364.26,315.38) -- (357.26,315.38) ;
\draw    (364.34,218.62) -- (364.34,249.95) ;
\draw    (364.34,249.95) -- (364.26,315.38) ;
\draw    (364.34,203.31) -- (357.34,203.32) ;
\draw    (364.34,218.62) -- (357.34,218.62) ;
\draw    (364.26,106.26) -- (364.34,203.31) ;
\draw    (364.26,106.26) -- (357.26,106.26) ;
\draw   (316.71,88.28) -- (325.52,88.28)(321.11,83.82) -- (321.11,92.74) ;
\draw   (340.4,88.28) -- (349.2,88.28)(344.8,83.82) -- (344.8,92.74) ;
\draw   (316.71,332.28) -- (325.52,332.28)(321.11,327.82) -- (321.11,336.74) ;
\draw   (340.4,332.28) -- (349.2,332.28)(344.8,327.82) -- (344.8,336.74) ;
\draw    (364.4,171.67) -- (411.61,171.77) ;
\draw [shift={(414.61,171.78)}, rotate = 180.13] [fill={rgb, 255:red, 0; green, 0; blue, 0 }  ][line width=0.08]  [draw opacity=0] (8.93,-4.29) -- (0,0) -- (8.93,4.29) -- cycle    ;
\draw    (364.34,249.95) -- (411.55,250.06) ;
\draw [shift={(414.55,250.06)}, rotate = 180.13] [fill={rgb, 255:red, 0; green, 0; blue, 0 }  ][line width=0.08]  [draw opacity=0] (8.93,-4.29) -- (0,0) -- (8.93,4.29) -- cycle    ;
\draw   (464.23,155.62) -- (464.23,187.93) .. controls (464.23,192.74) and (454.69,196.63) .. (442.92,196.63) .. controls (431.15,196.63) and (421.61,192.74) .. (421.61,187.93) -- (421.61,155.62)(464.23,155.62) .. controls (464.23,160.42) and (454.69,164.32) .. (442.92,164.32) .. controls (431.15,164.32) and (421.61,160.42) .. (421.61,155.62) .. controls (421.61,150.81) and (431.15,146.92) .. (442.92,146.92) .. controls (454.69,146.92) and (464.23,150.81) .. (464.23,155.62) -- cycle ;
\draw   (464.17,233.9) -- (464.17,266.22) .. controls (464.17,271.02) and (454.63,274.92) .. (442.86,274.92) .. controls (431.1,274.92) and (421.56,271.02) .. (421.56,266.22) -- (421.56,233.9)(464.17,233.9) .. controls (464.17,238.71) and (454.63,242.6) .. (442.86,242.6) .. controls (431.1,242.6) and (421.56,238.71) .. (421.56,233.9) .. controls (421.56,229.1) and (431.1,225.2) .. (442.86,225.2) .. controls (454.63,225.2) and (464.17,229.1) .. (464.17,233.9) -- cycle ;
\draw    (442.96,122.41) -- (442.92,139.91) ;
\draw [shift={(442.97,119.41)}, rotate = 90.14] [fill={rgb, 255:red, 0; green, 0; blue, 0 }  ][line width=0.08]  [draw opacity=0] (8.93,-4.29) -- (0,0) -- (8.93,4.29) -- cycle    ;
\draw    (402.3,74.68) -- (391.02,83.25) ;
\draw    (402.3,74.68) -- (413.58,83.25) ;

\draw    (413.47,83.14) -- (407.28,93.2) ;
\draw    (413.47,83.14) -- (419.67,93.2) ;

\draw    (391.48,82.86) -- (385.28,92.92) ;
\draw    (391.48,82.86) -- (397.67,92.92) ;

\draw    (407.46,92.87) -- (401.26,102.93) ;
\draw    (407.46,92.87) -- (413.65,102.93) ;

\draw  [fill={rgb, 255:red, 193; green, 190; blue, 190 }  ,fill opacity=0.43 ] (382.7,72.16) -- (423.02,72.16) -- (423.02,106.52) -- (382.7,106.52) -- cycle ;
\draw   (429.65,89.31) -- (436.7,89.31)(433.17,85.84) -- (433.17,92.79) ;
\draw    (482.1,74.68) -- (470.82,83.25) ;
\draw    (482.1,74.68) -- (493.38,83.25) ;

\draw    (493.28,83.14) -- (487.08,93.2) ;
\draw    (493.28,83.14) -- (499.47,93.2) ;

\draw    (471.28,82.86) -- (465.09,92.92) ;
\draw    (471.28,82.86) -- (477.47,92.92) ;

\draw    (487.26,92.87) -- (481.07,102.93) ;
\draw    (487.26,92.87) -- (493.45,102.93) ;

\draw  [fill={rgb, 255:red, 193; green, 190; blue, 190 }  ,fill opacity=0.43 ] (462.5,72.16) -- (502.83,72.16) -- (502.83,106.52) -- (462.5,106.52) -- cycle ;
\draw   (448.68,89.31) -- (455.73,89.31)(452.21,85.84) -- (452.21,92.79) ;
\draw   (377.2,65.5) -- (508.75,65.5) -- (508.75,112.16) -- (377.2,112.16) -- cycle ;
\draw    (442.91,299.42) -- (442.87,281.92) ;
\draw [shift={(442.92,302.42)}, rotate = 269.86] [fill={rgb, 255:red, 0; green, 0; blue, 0 }  ][line width=0.08]  [draw opacity=0] (8.93,-4.29) -- (0,0) -- (8.93,4.29) -- cycle    ;
\draw    (402.3,318.49) -- (391.02,327.06) ;
\draw    (402.3,318.49) -- (413.58,327.06) ;

\draw    (413.47,326.95) -- (407.28,337.02) ;
\draw    (413.47,326.95) -- (419.67,337.02) ;

\draw    (391.48,326.67) -- (385.28,336.74) ;
\draw    (391.48,326.67) -- (397.67,336.74) ;

\draw    (407.46,336.68) -- (401.26,346.75) ;
\draw    (407.46,336.68) -- (413.65,346.75) ;

\draw  [fill={rgb, 255:red, 193; green, 190; blue, 190 }  ,fill opacity=0.43 ] (382.7,315.97) -- (423.02,315.97) -- (423.02,350.33) -- (382.7,350.33) -- cycle ;
\draw   (429.65,333.13) -- (436.7,333.13)(433.17,329.65) -- (433.17,336.61) ;
\draw    (482.1,318.49) -- (470.82,327.06) ;
\draw    (482.1,318.49) -- (493.38,327.06) ;

\draw    (493.28,326.95) -- (487.08,337.02) ;
\draw    (493.28,326.95) -- (499.47,337.02) ;

\draw    (471.28,326.67) -- (465.09,336.74) ;
\draw    (471.28,326.67) -- (477.47,336.74) ;

\draw    (487.26,336.68) -- (481.07,346.75) ;
\draw    (487.26,336.68) -- (493.45,346.75) ;

\draw  [fill={rgb, 255:red, 193; green, 190; blue, 190 }  ,fill opacity=0.43 ] (462.5,315.97) -- (502.83,315.97) -- (502.83,350.33) -- (462.5,350.33) -- cycle ;
\draw   (448.68,333.13) -- (455.73,333.13)(452.21,329.65) -- (452.21,336.61) ;
\draw   (377.2,309.31) -- (508.75,309.31) -- (508.75,355.98) -- (377.2,355.98) -- cycle ;
\draw    (161.7,53.42) -- (293.25,53.42) ;
\draw    (161.7,58.42) -- (161.7,53.42) ;
\draw    (293.25,58.42) -- (293.25,53.42) ;
\draw    (161.7,28.42) -- (508.75,28.5) ;
\draw    (161.7,33.42) -- (161.7,28.42) ;
\draw    (508.75,33.5) -- (508.75,28.5) ;
\draw    (161.7,367.73) -- (293.25,367.73) ;
\draw    (161.7,392.73) -- (161.7,387.73) ;
\draw    (293.25,367.73) -- (293.25,362.73) ;
\draw    (161.7,392.73) -- (508.75,392.81) ;
\draw    (161.7,367.73) -- (161.7,362.73) ;
\draw    (508.75,392.81) -- (508.75,387.81) ;
\draw (123.01,210.67) node  {\includegraphics[width=52.5pt,height=52.5pt]{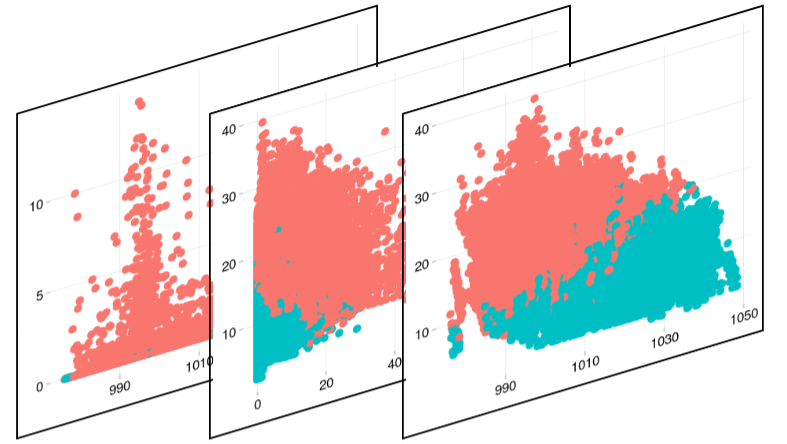}};

\draw (27.42,207.9) node [anchor=north west][inner sep=0.75pt]   [align=left] {D};
\draw (48,275) node [anchor=north west][inner sep=0.75pt]  [font=\scriptsize] [align=left] {$\displaystyle \{\boldsymbol{x}_{i}^{(\mathcal{S})}\}_{i=1}^{n}$};
\draw (220.15,86.8) node [anchor=north west][inner sep=0.75pt]  [font=\small] [align=left] {...};
\draw (220.15,330.42) node [anchor=north west][inner sep=0.75pt]  [font=\small] [align=left] {...};
\draw (280.33,204.61) node [anchor=north west][inner sep=0.75pt]  [font=\scriptsize] [align=left] {$\displaystyle p^{( 1)}$};
\draw (309.83,131.33) node [anchor=north west][inner sep=0.75pt]  [font=\scriptsize] [align=left] {$\displaystyle \lambda _{1}^{( 1)}$};
\draw (310.83,278.33) node [anchor=north west][inner sep=0.75pt]  [font=\scriptsize] [align=left] {$\displaystyle \lambda _{2}^{( 1)}$};
\draw (336.12,169.36) node [anchor=north west][inner sep=0.75pt]  [font=\small] [align=left] {...};
\draw (336.37,247.61) node [anchor=north west][inner sep=0.75pt]  [font=\small] [align=left] {...};
\draw (367.83,131.14) node [anchor=north west][inner sep=0.75pt]  [font=\scriptsize] [align=left] {$\displaystyle \lambda _{1}^{( K-1)}$};
\draw (367.83,278.47) node [anchor=north west][inner sep=0.75pt]  [font=\scriptsize] [align=left] {$\displaystyle \lambda _{2}^{( K-1)}$};
\draw (325.8,85.9) node [anchor=north west][inner sep=0.75pt]  [font=\small] [align=left] {...};
\draw (325.8,330) node [anchor=north west][inner sep=0.75pt]  [font=\small] [align=left] {...};
\draw (212.17,168.68) node [anchor=north west][inner sep=0.75pt]   [align=left] {$\displaystyle \mathcal{D}_{1}^{( 1)}$};
\draw (212.97,246.68) node [anchor=north west][inner sep=0.75pt]   [align=left] {$\displaystyle \mathcal{D}_{2}^{( 1)}$};
\draw (426.92,168.18) node [anchor=north west][inner sep=0.75pt]   [align=left] {$\displaystyle \mathcal{D}_{1}^{( K)}$};
\draw (427.22,247.18) node [anchor=north west][inner sep=0.75pt]   [align=left] {$\displaystyle \mathcal{D}_{2}^{( K)}$};
\draw (435.65,86.8) node [anchor=north west][inner sep=0.75pt]  [font=\small] [align=left] {...};
\draw (435.65,331) node [anchor=north west][inner sep=0.75pt]  [font=\small] [align=left] {...};
\draw (207.55,33.29) node [anchor=north west][inner sep=0.75pt]  [font=\footnotesize] [align=left] {$\displaystyle f_{\text{occ},1}^{( 1)}(\boldsymbol{x}_{i})$};
\draw (311.55,8.29) node [anchor=north west][inner sep=0.75pt]  [font=\footnotesize] [align=left] {$\displaystyle f_{\text{occ},1}^{( K)}(\boldsymbol{x}_{i})$};
\draw (330.33,204.86) node [anchor=north west][inner sep=0.75pt]  [font=\scriptsize] [align=left] {$\displaystyle p^{( K-1)}$};
\draw (201.8,370.83) node [anchor=north west][inner sep=0.75pt]  [font=\footnotesize] [align=left] {$\displaystyle f_{\text{occ},2}^{( 1)}(\boldsymbol{x}_{i})$};
\draw (308.3,394.33) node [anchor=north west][inner sep=0.75pt]  [font=\footnotesize] [align=left] {$\displaystyle f_{\text{occ},2}^{( K)}(\boldsymbol{x}_{i})$};

\end{tikzpicture}

\end{adjustbox}
\caption{Visualisation of the flow of the multi-regime EM-XGBoost steps for a 
    two-regime ($L=2$) example, starting from an observed dataset $D$. Based on 
    the restricted covariate information $\{\boldsymbol{x}^{(\mathcal{S})}_i\}_{i=1}^{n}$, the 
    observations are partitioned, 
    yielding the datasets $D_1$ and $D_2$. The upper and lower 
    part show the construction of the regime-specific additive occurrence models 
    $f^{(1)}_{\text{occ},1}(\boldsymbol{x}_i), \ldots, f^{(K)}_{\text{occ},1}(\boldsymbol{x}_i)$ 
    and $f^{(1)}_{\text{occ},2}(\boldsymbol{x}_i), \ldots, f^{(K)}_{\text{occ},2}(\boldsymbol{x}_i)$, 
    resulting in the predicted occurrence intensities 
    $\lambda^{(1)}_1, \ldots, \lambda^{(K)}_1$ and 
    $\lambda^{(1)}_2, \ldots, \lambda^{(K)}_2$, respectively. The middle part shows 
    the expectation step, i.e.\ the construction of 
    $\mathcal{D}^{(1)}_1, \ldots, \mathcal{D}^{(K)}_1$ and 
    $\mathcal{D}^{(1)}_2, \ldots, \mathcal{D}^{(K)}_2$ via the predicted occurrence 
    intensities and reporting probabilities. The reporting model is shared across 
    regimes, as in the case study in Section~\ref{sec:casestudy}.}
\label{fig:regimeswitchextension}
\end{figure}
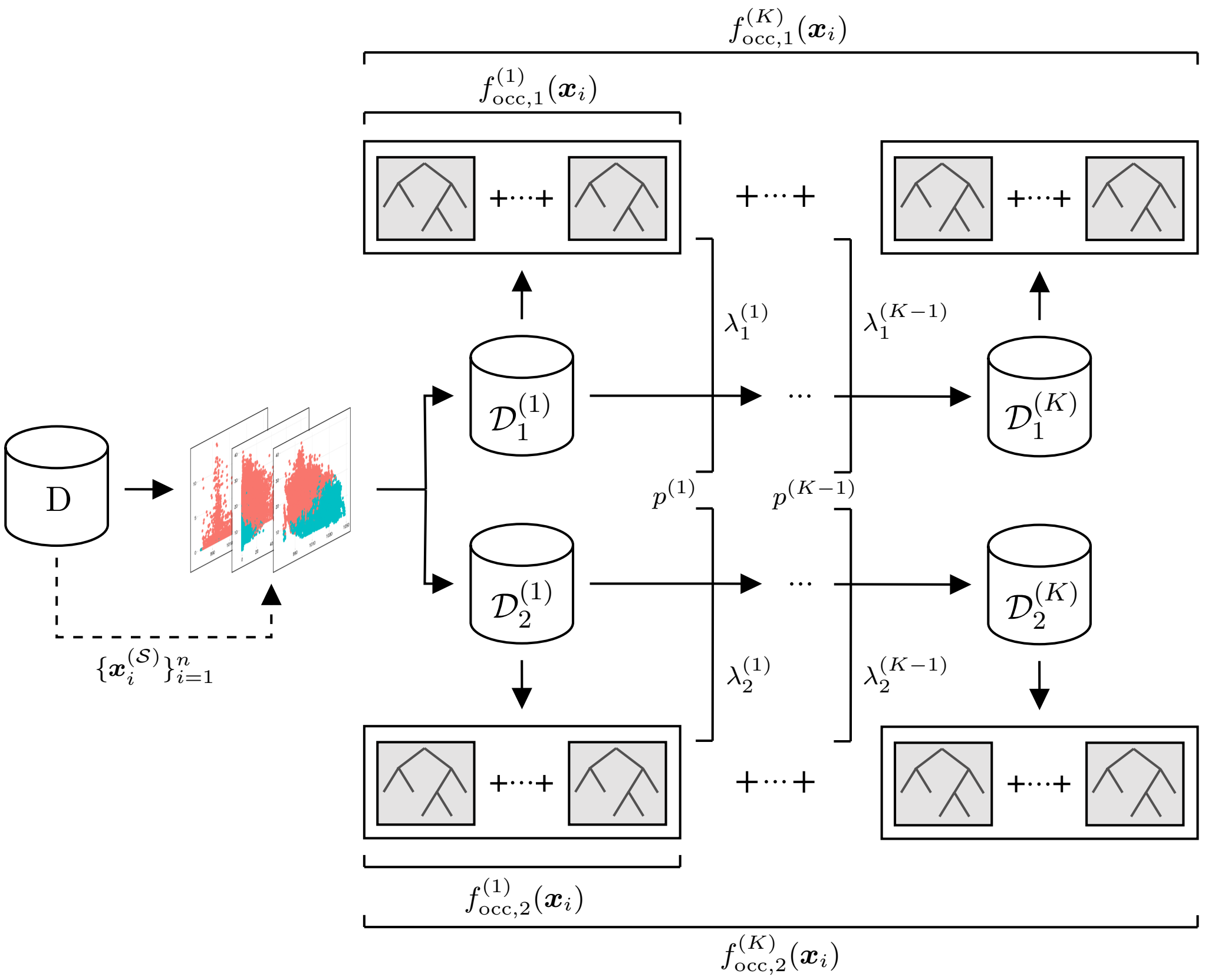

\paragraph{Regimes in EM-XGBoost} Within the EM-XGBoost framework, the regimes allow for separate occurrence models per regime, separate reporting models, or both. Figure~\ref{fig:regimeswitchextension} shows a simplified scheme of the subsequent steps of the multi-regime EM-XGBoost framework for a two-regime ($L=2$) example, with regime-specific occurrence models and a shared reporting model, the chosen configuration in Section~\ref{sec:casestudy}. In the $k$th expectation step, we define our current expectation of the complete data for each regime, i.e.~$\mathcal{D}^{(k)}_{1}$ and $\mathcal{D}^{(k)}_{2}$ for the example in Figure~\ref{fig:regimeswitchextension}. Subsequently, in the $k$th maximisation step, we fit a separate occurrence model for both regimes, denoted as $f_{\text{occ},1}^{(k)}(\boldsymbol{x}_{i})$ and $f_{\text{occ},2}^{(k)}(\boldsymbol{x}_{i})$. The full details can be found in Algorithm~\ref{algo:EMmulti} in Appendix~\ref{app:emalgo}.

\section{Case study} \label{sec:casestudy}

\begin{figure}[h] 
\centering
\begin{adjustbox}{width=0.99\textwidth}
   \subfloat[NL (PC0)]{

\begin{adjustbox}{width=0.33\textwidth}
    \includegraphics[width = \textwidth]{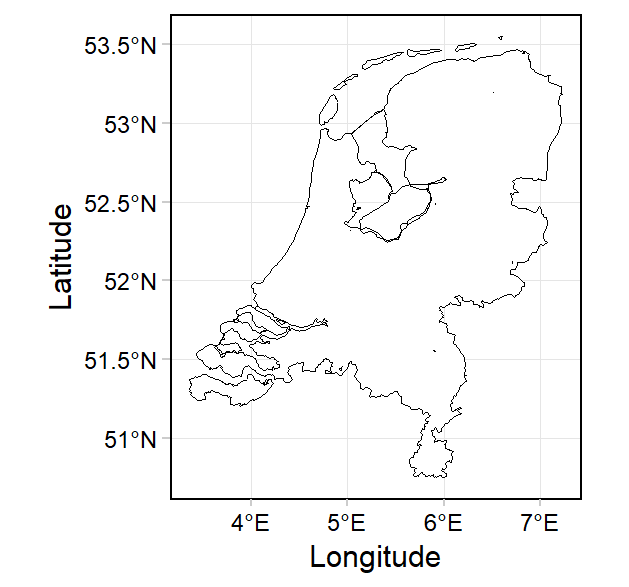}
\end{adjustbox}

}
\subfloat[PC2]{

\begin{adjustbox}{width=0.33\textwidth}
\includegraphics[width = \textwidth]{map_pc2.png}

\end{adjustbox}

} 
\subfloat[PC4]{

\begin{adjustbox}{width=0.33\textwidth}
\includegraphics[width = \textwidth]{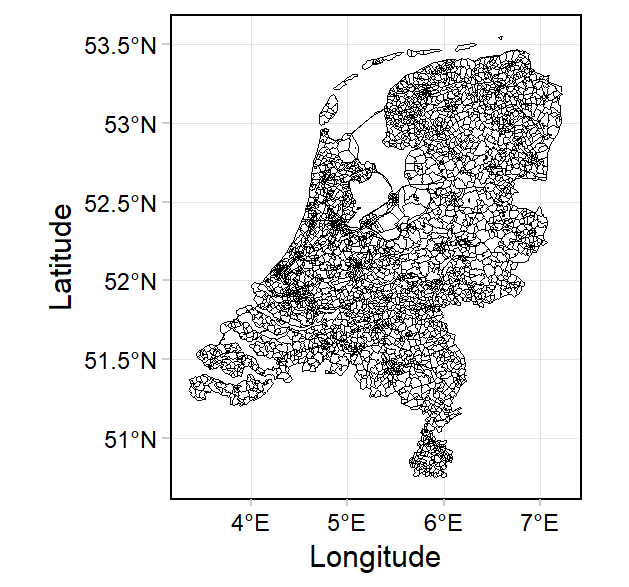}

\end{adjustbox}

} 
\end{adjustbox}
\caption{Illustration of three levels of spatial resolution at which the Netherlands can be represented. The left panel shows the national 
    level (NL), corresponding to the whole of the Netherlands. The middle and 
    right panels show the PC2 and PC4 postal code areas, 
    defined by the first two and first four digits of the Dutch postal code, 
    respectively.}
\label{fig:PCmaps}
\end{figure}

In this section, we apply the proposed framework to the Dutch private home insurance portfolio, introduced in Section~\ref{sec:data}, modelling the occurrence and reporting of claims per calendar day. We assess whether a multi-regime specification improves on a single occurrence model, and examine how this improvement depends on the spatial resolution of the data. Specifically, we compare a single-regime and a two-regime ($L=2$) EM-XGBoost flow, where regimes are identified from weather information. For both the single- and two-regime setting, we consider two levels of spatial resolution: the national (NL) level and the regional (PC2 areas) level. The former treats the Netherlands as a single area, while the latter consists of 90 distinct areas. We compare both levels to assess how much spatial resolution matters: at the national level the weather covariates are averaged over the whole portfolio, so a storm hitting only part of the country cannot be captured.
We fix the number of regimes at two at both spatial levels rather than optimising over $L$. At the national level, each day yields a single observation, so that the severe-weather regime contains only a few days, among which some covariate levels, e.g.~a holiday or a particular weekday, may not occur at all. The regional level could support additional regimes, but we use the same number of regimes at both levels, so that the national and regional models remain directly comparable. The data are spatially aggregated to the chosen spatial resolutions as described in Section~\ref{sec:feature}. Figure~\ref{fig:PCmaps} illustrates the national (NL) and regional (PC2 areas) levels, as well as the PC4 areas. Due to computational limitations\footnote{There are 90 PC2 and around 4000 PC4 postal code areas in the Netherlands. For an observed window of six years, this results in $90\text{x}365\text{x}6=197\ 100$ versus $4000\text{x}365\text{x}6=8\ 760\ 000$ observations, when considering the data at the PC2 or PC4 level, respectively. Since the occurrence and reporting models are fit at every EM iteration, and we analysed the data via protected servers, a parameter grid search at the PC4 level was deemed too time intensive.}, we do not model the data at the PC4 postal code area level.

\subsection{Model calibration} \label{sec:calibration}

In the two-regime setting, we fit regime-specific occurrence models alongside a single reporting model common to both regimes. The expected reporting log-likelihood $Q^{\text{rep}}(p_{i,j}^{(k)};\mathcal{D}^{(k)})$ weights each observation by its claim counts, see Section~\ref{sec:model}. As a result, for a severe-regime reporting model, the fitted reporting distribution would largely be determined by the day-specific covariate values of the windstorms of 18 January 2018 (Friederike) and 18 February 2022 (Eunice). To avoid this, we pool the reporting information across regimes via a shared reporting model. We calibrate the regime-specific occurrence and shared reporting models following Algorithms~\ref{algo:EM} and~\ref{algo:EMmulti} in Appendix~\ref{app:emalgo} for the single- and two-regime settings, respectively. Claims are reported with a maximal reporting delay of three weeks, i.e.~at most $d-1$ days after occurrence with $d=22$. The occurrence
models include the object-specific and weather-related covariates, as well as the day-specific covariates of the occurrence day. The reporting model includes the weather-related covariates and the day-specific covariates of the occurrence day and of each of the $d$ potential reporting days, but not the object-specific covariates. The evaluation date $\tau$ corresponds to 10 December 2022, i.e.~21 days before the end of the observed period. This ensures that we have the complete information available for model evaluation, including the true IBNR claim count, which is otherwise (partially) unobservable for later choices of $\tau$. To select values for the parameters specific to the training process, we randomly split the insured objects into three disjoint groups, covering claims data between 1 January 2017 and 10 December 2022. Next, we apply the feature engineering and aggregation steps described in Section~\ref{sec:feature}, resulting in three separate datasets $D_{\text{train}}$, $D_{\text{val}}$ and $D_{\text{test}}$. $D_{\text{train}}$ consists of 80\% of insured objects and is used for training the occurrence and reporting models. $D_{\text{val}}$ consists of another 10\% of insured objects and is used for early stopping, for model training during a single EM iteration as well as across EM iterations. $D_{\text{test}}$ accounts for the remaining 10\% of objects and is used for out-of-sample evaluation. Note that when we refer to the data including the unreported claim counts, the data is denoted as $\mathcal{D}_{\text{train}}$, $\mathcal{D}_{\text{val}}$ and $\mathcal{D}_{\text{test}}$. We opt for a random grid search as proposed in \cite{bergstra2012random}. Specifically, we select the combination of parameters that results in the highest observed log-likelihood $LL(\lambda_i, p_{i,j};D_{\text{val}})$, computed on the validation data. The grid values and the optimal settings for the national and regional models can be found in Table~\ref{tab:optimalregular} and Table~\ref{tab:optimalregimeswitch} in Appendix~\ref{app:casestudy}, for the single-regime and two-regime case, respectively.

\subsection{Regime identification} \label{sec:regimeid}
\begin{figure}[h] 
\centering
\begin{adjustbox}{width=0.99\textwidth}
  \subfloat[]{

\begin{adjustbox}{width=0.33\textwidth}
    \includegraphics[width = \textwidth]{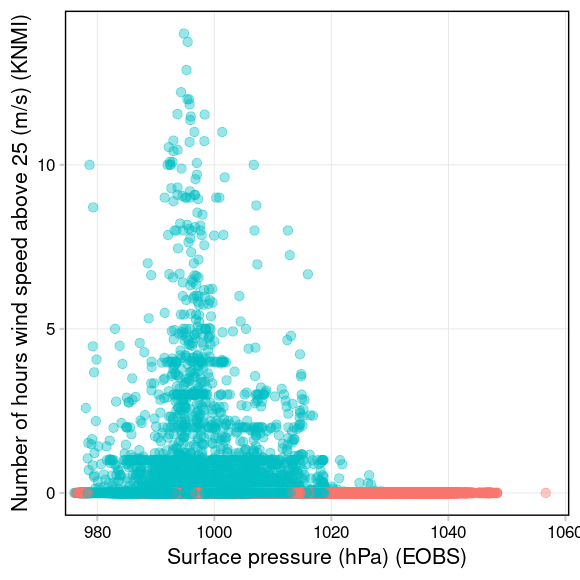}
\end{adjustbox}

}
\subfloat[]{

\begin{adjustbox}{width=0.33\textwidth}
\includegraphics[width = \textwidth]{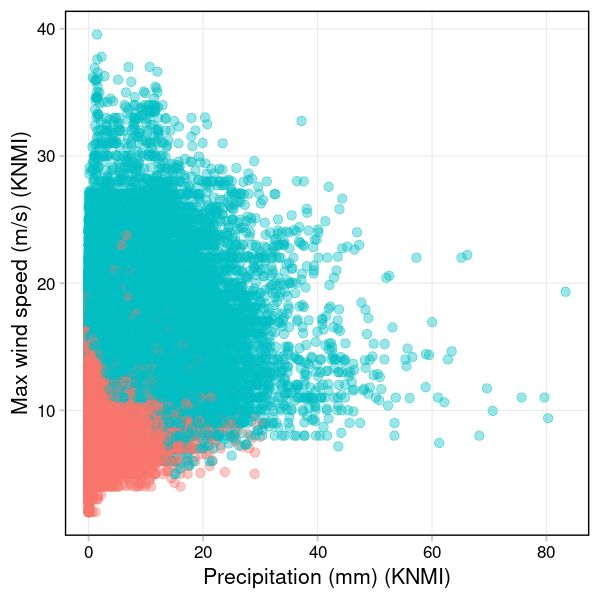}

\end{adjustbox}

} 
\subfloat[]{

\begin{adjustbox}{width=0.33\textwidth}
\includegraphics[width = \textwidth]{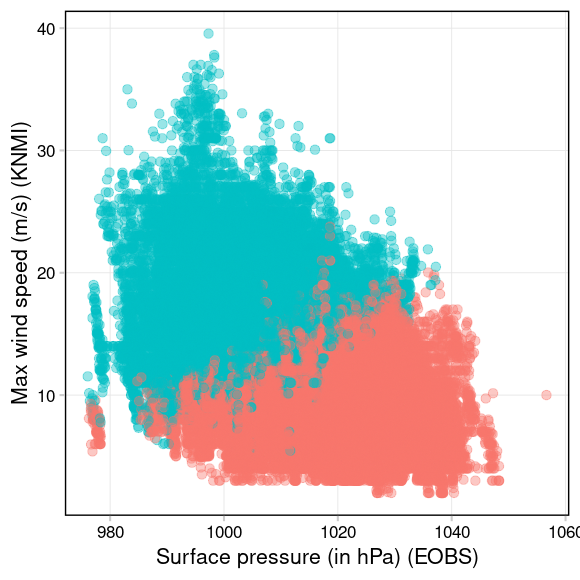}

\end{adjustbox}

}  
\end{adjustbox}
\vspace{0.5em}
{\footnotesize
\textcolor{ggred}{$\bullet$}~Clear-weather regime \hspace{2em}
\textcolor{ggteal}{$\bullet$}~Severe-weather regime}
\caption{Visualisation of the two-regime clustering solution at the PC2 spatial level, shown for three combinations of weather covariates. Each point corresponds to an observation in $D_{\text{train}}$, coloured according to the regime assigned by k-means: the clear-weather regime (red) and the severe-weather regime (teal). The left panel shows the number of hours with a wind speed above 25~m/s (KNMI) against the surface pressure (E-OBS), the middle panel the maximum wind speed (KNMI) against the precipitation (KNMI), and the right panel the maximum wind speed (KNMI) against the surface pressure (E-OBS).}
\label{fig:clustering}
\end{figure}

For the two-regime ($L=2$) models at the national and regional level, we first identify the regimes and then assign a regime label to every observation. Following Section~\ref{sec:regimetheory}, we require a choice of the restricted covariate information $\mathcal{S}$. At a given spatial resolution, we take $\mathcal{S}$ to consist of the weather covariates from the three public sources,
including the intra-day wind duration features, so that the regimes are determined by weather conditions alone. This allows us to compare how claims occur under clear and under severe weather, which is what an insurer needs to
anticipate the inflow on a given day. We identify the regimes on the training data $D_{\text{train}}$, i.e.~the 80\% of insured objects used for training, over all occurrence days between 1 January 2017 and 10 December 2022. All covariates in $\mathcal{S}$ are first standardised, using the means and standard deviations of $D_{\text{train}}$, so that variables measured on a larger scale do not dominate. We then apply k-means clustering \citep{lloyd1982least} on the standardised covariates. This accounts for all weather covariates jointly, rather than, for example, imposing a threshold on a single covariate. This yields a `clear-weather' and a `severe-weather' regime, with corresponding training datasets $D_{\text{train},c}$ and $D_{\text{train},s}$: each day-area observation is assigned to one of the two regimes on the basis of the weather covariates on that particular day in that area. Figure~\ref{fig:clustering} visualises the two-regime clustering solution for different combinations of weather covariates at the PC2 spatial level. Wind-related features dominate the regime label allocation, as observations with high values for those are in the severe regime (denoted with teal dots). The observations in the validation and test data are assigned to the regime whose centroid is closest, after standardising their covariates in $\mathcal{S}$ with the training means and standard deviations, so that no information from these datasets enters the identification step. The EM algorithm then proceeds as illustrated in Figure~\ref{fig:regimeswitchextension}.

\subsection{Initialisation EM algorithm}
The EM algorithm requires starting values $\lambda^{(0)}_i$ and $p^{(0)}_{i,j}$, as shown in Figure~\ref{fig:EMframework}. Instead of initialising all observations with the same, pooled reporting distribution, as in \cite{wilsens2025machine}, we exploit the fact that reporting behaviour differs systematically across days, as illustrated in Figure~\ref{fig:reportingempirical}. Specifically, we partition the training data into cells defined by weekday and an intra-day timing covariate, i.e.~the share of the daily wind intensity occurring in the evening (between 18h and 24h), discretised into four equal bins; see Table~\ref{tab:covariates} in Appendix~\ref{app:casestudy}. Both dimensions capture systematic differences in reporting: delays vary across weekdays, and the intra-day timing of damage determines how much of the day remains afterwards, so that damage late in the evening is largely reported the following day rather than on the occurrence day itself. A pooled distribution would average these differences away, yielding starting values that are off for days whose reporting behaviour deviates from the average. For each cell, we estimate an empirical reporting distribution  using only the fully developed observations, i.e.~where $\tau_i \geq d$ and all $d$ delays are observed, and assign it to every observation in that cell.  Cells that are empty or unobserved fall back to the pooled empirical reporting  distribution across all fully developed observations. The initial daily occurrence intensities are then obtained via the initial reporting probabilities and the observed claim counts, i.e.
\begin{equation*}
    \lambda^{(0)}_i = \frac{\sum_{j=1}^{\tau_i} N_{i,j}}{e_{i} \cdot \sum_{j=1}^{\tau_i} p^{(0)}_{i,j}} \:\:\: \text{for} \:\:\: i=1,\ldots,n_{\text{train}}.
\end{equation*}
Specifically, we divide the number of claims reported up to $\tau_i$ by the exposure $e_i$ and the probability mass that the assigned reporting distribution places on the delays $0, \ldots, \tau_i-1$. This corrects for the claims that are expected to be reported after $\tau$, which would otherwise lead to an underestimation of the occurrence intensities for recent occurrence days. In the two-regime setting, we additionally split the cells by regime label, using the regime-specific pooled reporting distribution as a fallback when a cell is empty or unobserved.

\subsection{Evaluation metrics}
We evaluate the fitted models on the out-of-sample data $\mathcal{D}_{\text{test}}$, via evaluation criteria that check the occurrence and reporting fit, both separately and jointly. As a first metric, jointly assessing both components, we compute the observed log-likelihood $LL(\widehat{\lambda}^{(k)}_i, \widehat{p}^{(k)}_{i,j};D_{\text{test}})$ on the out-of-sample data $D_{\text{test}}$, i.e.~using only the claims that occurred and were reported up to the evaluation date $\tau$. Note that the log-likelihood can be used to compare different models at the same temporal and spatial resolution, but not across different temporal or spatial resolutions. To directly compare the occurrence fit between the regional and national level, we aggregate regional daily predictions to national daily totals. For a given spatial resolution, we denote the occurred and predicted number of claims on day $t$ as $N_{t}=\sum_{i: \text{day}(i)=t }N_{i}$ and $\widehat{N}_{t}=\sum_{i: \text{day}(i)=t }e_{i} \cdot\widehat{\lambda}_{i}$, respectively. We define the mean squared error 
\begin{align*}
    \text{MSE}(\widehat{N}_t;\mathcal{D}_{\text{test}})=\frac{1}{\tau}\sum_{t=1}^{\tau} (N_t - \widehat{N}_t)^{2}
\end{align*}
and the mean absolute error 
\begin{align*}
    \text{MAE}(\widehat{N}_t
;\mathcal{D}_{\text{test}})=\frac{1}{\tau}\sum_{t=1}^{\tau} |N_t - \widehat{N}_t|.
\end{align*}
The former is largely determined by errors on days with many claims, i.e.~severe storms, while the latter better reflects performance on moderate days. For the reporting fit, we assess the ranked probability score 
\begin{align*}
    \text{RPS}(\widehat{p}_{i,j};\mathcal{D}_{\text{test}})&=\frac{1}{d-1} \sum_{j=1}^{d-1} \left( F^{\text{pred}}_j - F^{\text{emp}}_j \right)^2, \:\: \text{where} \\
    F^{\text{pred}}_{j} &=\frac{1}{n}\sum^{j}_{z=1}  \sum^{n}_{i=1}\widehat{p}_{i,z} \:\: \text{and} \:\: F^{\text{emp}}_{j} = \frac{\sum^{j}_{z=1}\sum^{n}_{i=1} N_{i,z}}{\sum^{n}_{i=1}\sum^{d}_{k=1} N_{i,k}}
\end{align*}
denote the predicted and empirical cumulative reporting distributions at a delay of $j-1$ days. Since it compares cumulative reporting distributions, the RPS is directly comparable across spatial resolutions.

\subsection{Results}

\paragraph{Out-of-sample performance}

Table~\ref{tab:evalmetrics} summarises the evaluation metrics defined in the previous section for both the national and regional models, in the single- and two-regime setting, computed on the insured objects in the test data. Note that, as discussed in the previous section, the observed log-likelihood $LL(\widehat{\lambda}^{(k)}_i, \widehat{p}^{(k)}_{i,j};D_{\text{test}})$ is not comparable across different spatial levels. For the remaining criteria, the regional models outperform their national counterparts on both the occurrence and the reporting component. The difference between the regional and national models is considerably larger for the MSE than for the MAE. Since the MSE is dominated by days with many claims, this indicates that the gain from modelling at the regional level is concentrated on days with severe weather, rather than spread evenly across the observed period. Comparing the single- and two-regime specifications, the two-regime model dominates the single-regime model on all evaluation criteria at the regional level. At the national level, results are mixed: the two-regime model attains a better reporting fit, but a lower observed log-likelihood and a worse occurrence fit in terms of both the MSE and the MAE. A multi-regime specification therefore appears most beneficial at a more granular spatial level.

\begin{table}[H]
\centering
\begin{adjustbox}{width=0.8\textwidth}
    \begin{tabular}{lcccc}
    \toprule
    & \multicolumn{2}{c}{\textbf{National} (NL)} & \multicolumn{2}{c}{\textbf{Regional} (PC2)}\\
    \cmidrule(lr){2-3} \cmidrule(lr){4-5}
    & Single-regime & Two-regime & Single-regime & Two-regime\\
    \midrule
    $LL(\widehat{\lambda}^{(k)}_i, \widehat{p}^{(k)}_{i,j};D_{\text{test}})$&\textbf{-20\ 906.70}&-21\ 401.33&-80\ 194.91&\textbf{-79\ 658.72}\\
    [0.5em]
    \textbf{Occurrence model}&&&&\\
    \quad $\text{MSE}(\widehat{N}_t;\mathcal{D}_{\text{test}})$ & \textbf{1234.29} & 1664.14 & 85.18 & \underline{\textbf{50.95}}\\
    \quad $\text{MAE}(\widehat{N}_t;\mathcal{D}_{\text{test}})$ & \textbf{4.7358} & 5.1949 & 3.2599 & \underline{\textbf{2.9464}} \\[0.5em]
    \textbf{Reporting model}&&&&\\
    \quad $\text{RPS}(\widehat{p}_{i,j};\mathcal{D}_{\text{test}})$ & 0.000441 & \textbf{0.000429} & 0.000297 & \underline{\textbf{0.000280}} \\
    \bottomrule
    \end{tabular}
\end{adjustbox}
\caption{Evaluation metrics for the national and regional model, for both the single- and two-regime setting, computed on the test data $\mathcal{D}_{\text{test}}$. The first row shows the observed log-likelihood $LL(\widehat{\lambda}^{(k)}_i, \widehat{p}^{(k)}_{i,j};D_{\text{test}})$. The second and third row show evaluation criteria for the occurrence model, i.e.~the mean squared error $\text{MSE}(\widehat{N}_t;\mathcal{D}_{\text{test}})$ and the mean absolute error $\text{MAE}(\widehat{N}_t;\mathcal{D}_{\text{test}})$, both computed on national daily totals. The fourth row shows the ranked probability score $\text{RPS}(\widehat{p}_{i,j};\mathcal{D}_{\text{test}})$, which evaluates the reporting model. Bold indicates the best value within a spatial level, underlining the best value across both spatial levels. The observed log-likelihood is only comparable within a spatial level and is therefore not underlined.}
\label{tab:evalmetrics}
\end{table}

\begin{figure}[h] 
\centering
\subfloat[National single-regime]{

\begin{adjustbox}{width=0.4\textwidth}
    \includegraphics[width = \textwidth]{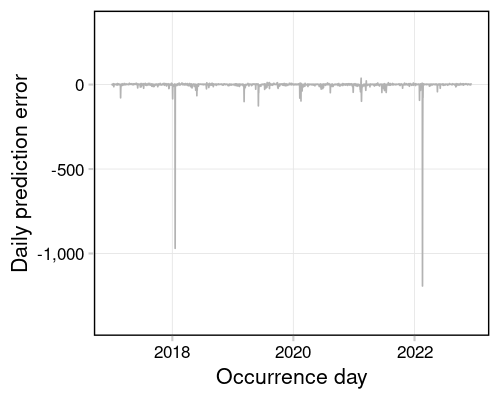}
\end{adjustbox}

}
\subfloat[National two-regime]{

\begin{adjustbox}{width=0.4\textwidth}
\includegraphics[width = \textwidth]{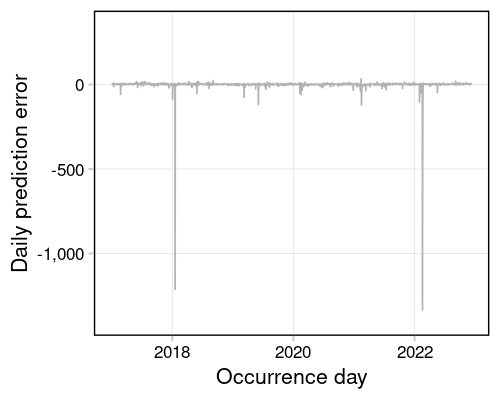}

\end{adjustbox}

} \\
\subfloat[Regional single-regime]{

\begin{adjustbox}{width=0.4\textwidth}
\includegraphics[width = \textwidth]{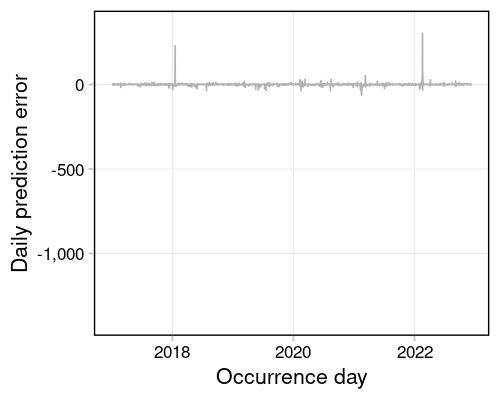}

\end{adjustbox}

}
\subfloat[Regional two-regime]{

\begin{adjustbox}{width=0.4\textwidth}
\includegraphics[width = \textwidth]{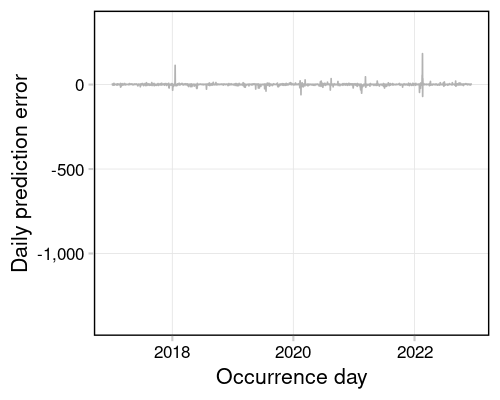}

\end{adjustbox}

}
\caption{Daily prediction error, i.e.~the difference between the predicted and observed number of claims, for each occurrence day in the observed period. The top panels show the national single-regime (top left) and two-regime (top right) model, the bottom panels the regional single-regime (bottom left) and two-regime (bottom right) model. Negative values indicate an underestimation of the number of claims on that day. }
\label{fig:occfit}
\end{figure}

In Figure~\ref{fig:occfit}, we look at the occurrence fit across all occurrence days. Specifically, we consider the daily prediction error $\widehat{N}_{t}-N_t$ for $t=1,\ldots,\tau$, i.e.~the difference between the predicted and observed daily claim counts. We observe that the difference in mean squared error in Table~\ref{tab:evalmetrics} between the national models (top panels in Figure~\ref{fig:occfit}) and the regional models (bottom panels) is largely driven by a significant underestimation of the daily claim counts during the severe windstorms on 18 January 2018 and 18 February 2022, by the national models. The regional two-regime model shows the smallest prediction errors on these days, in line with the results in Table~\ref{tab:evalmetrics}.

\begin{figure}[H] 
\centering
\begin{adjustbox}{width=0.95\textwidth}
    \subfloat[Empirical]{\includegraphics[width = 0.33\textwidth]{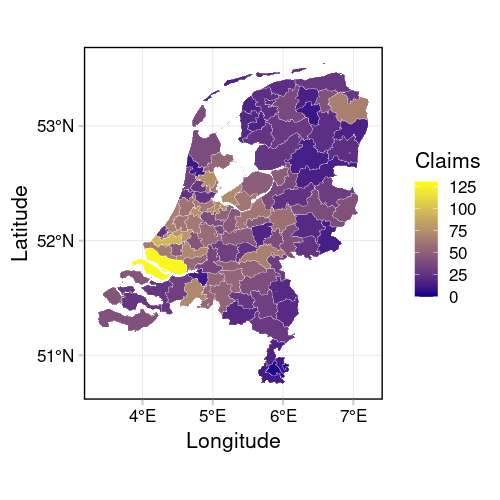}} 
\subfloat[Regional single-regime]{\includegraphics[width = 0.33\textwidth]{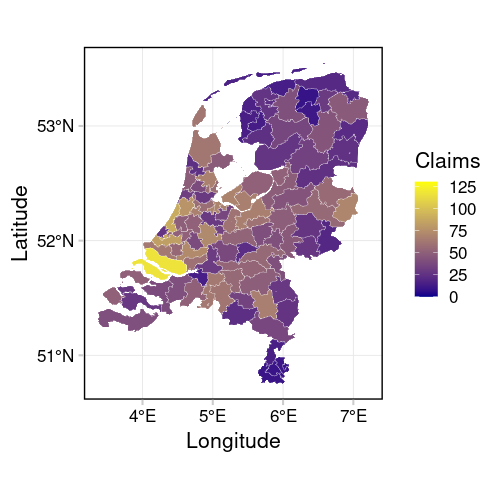}}
\subfloat[Regional two-regime]{\includegraphics[width = 0.33\textwidth]{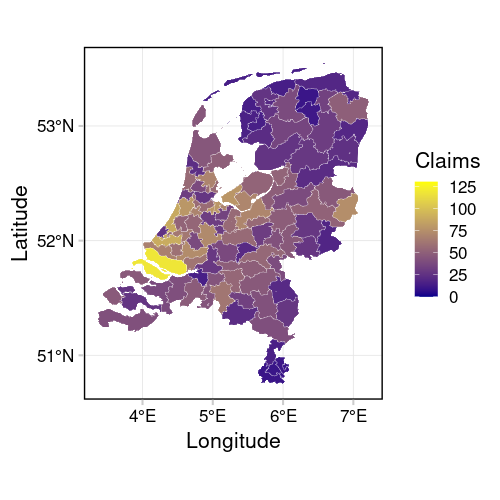}}
\end{adjustbox}
\caption{Visualisation of the observed and predicted number of claims per PC2 
    postal code area on 18 February 2022 (storm Eunice). The left panel 
    shows the observed claim counts, the middle and right panel the out-of-sample predicted 
    claim counts of the regional single-regime and two-regime model, respectively, computed on $\mathcal{D}_{\text{test}}$. } 
\label{fig:occfit_18feb2022}
\end{figure}

Figure~\ref{fig:occfit_18feb2022} zooms in on the out-of-sample spatial occurrence fit across the Netherlands on 18 February 2022 (storm Eunice), one of the two days identified in Figure~\ref{fig:occfit} as driving the difference in mean squared error between the national and regional models. Note that out-of-sample refers here to a different set of insured objects, as discussed in Section~\ref{sec:calibration}. The left panel shows the observed number of claims per PC2 area  alongside the predictions of the single- and two-regime models, in the middle and right panels, respectively. Both the single-regime regional model and the two-regime model capture the spatial pattern across PC2 areas well. The prediction differences between both models are small at the level of an individual area and hence hard to discern on a common colour scale. However, Figure~\ref{fig:occfit} shows a clearly larger overestimation for the single-regime regional model on this day, which is only possible if its per-area errors point in the same direction. The two-regime model thus improves the occurrence fit through a small but systematic correction across areas, rather than through a large correction in a few areas. By construction, the national models cannot capture the spatial occurrence variation across areas. In Appendix~\ref{app:casestudy}, we include an additional spatial occurrence fit plot for 4 June 2019, showing that the models capture the spatial pattern, i.e.~the claims band observed in Figure~\ref{fig:4june2019} in Section~\ref{sec:data}.

\begin{figure}[H] 
\centering
\begin{adjustbox}{width=0.95\textwidth}
    \subfloat[Single-regime]{\includegraphics[width = 0.4\textwidth]{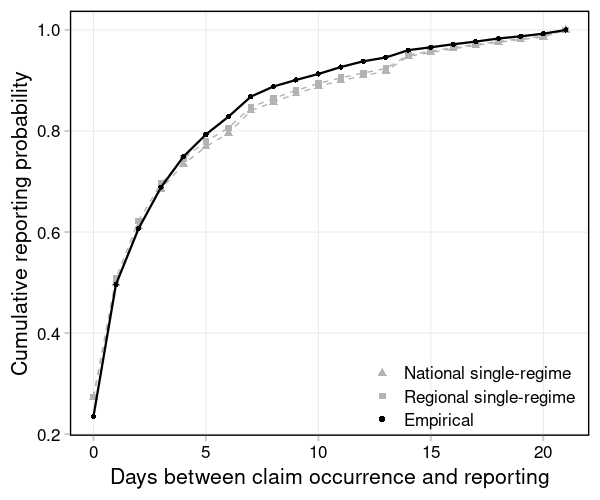}} 
\hspace{1cm}
\subfloat[Two-regime]{\includegraphics[width = 0.4\textwidth]{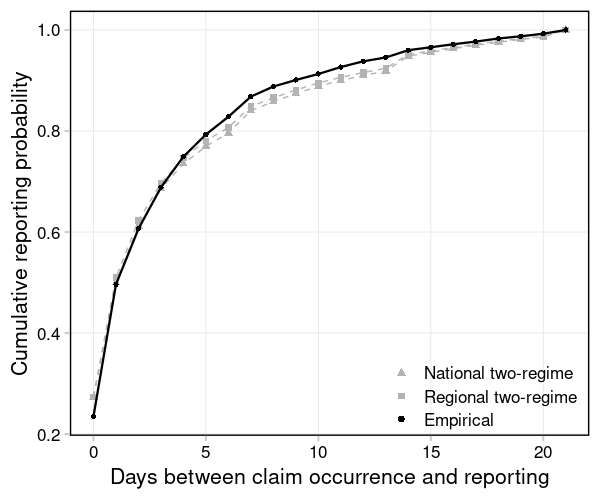}}
\end{adjustbox}
\caption{Out-of-sample cumulative reporting probabilities for a delay of $0,\ldots,21$ days, for the single-regime (left panel) and two-regime 
    (right panel) setting. Each panel compares the fitted cumulative reporting 
    distributions of the national (light grey triangles) and regional (light grey squares) model with the empirical cumulative 
    reporting distribution (black circles), computed on the test data $\mathcal{D}_{\text{test}}$.} 
\label{fig:repfit}
\end{figure}

Figure~\ref{fig:repfit} visualises the out-of-sample cumulative reporting probabilities, for the single-regime (left panel) and two-regime (right panel) models. Across models, we observe an overestimation of the reporting probabilities for short delays (0-2 days) and an underestimation for longer delays. For both the single- and two-regime models, the regional setting provides a slightly better fit than the national setting, with the regional two-regime model having the best overall reporting fit, as indicated in Table~\ref{tab:evalmetrics}.  

\paragraph{Feature importance}

\begin{figure}[H] 
\centering
\begin{adjustbox}{width=0.98\textwidth}
    \subfloat[Occurrence (clear regime)]{\includegraphics[width = 0.33\textwidth]{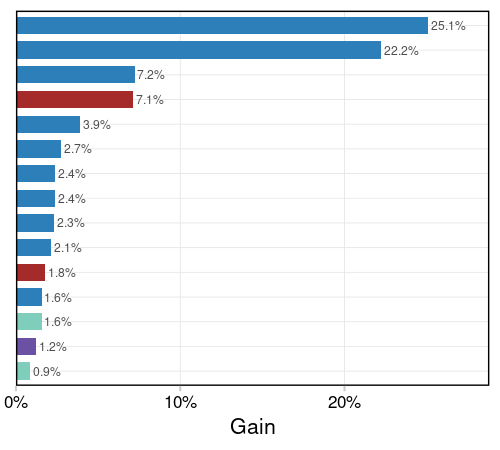}} 
\subfloat[Occurrence (severe regime)]{\includegraphics[width = 0.33\textwidth]{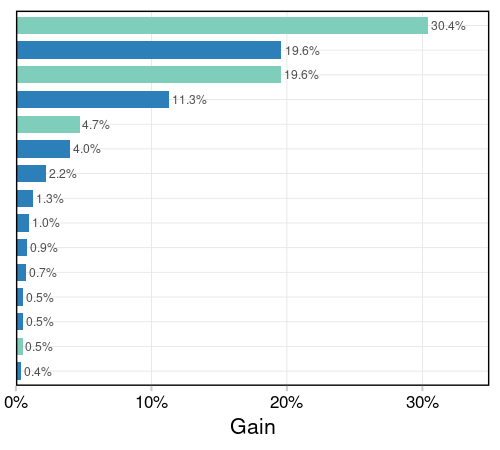}}
\subfloat[Reporting]{\includegraphics[width = 0.33\textwidth]{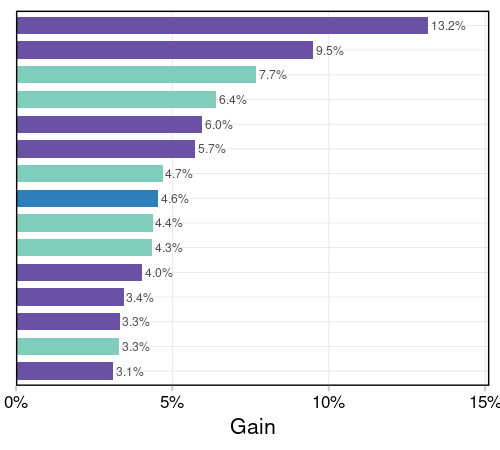}}
\end{adjustbox}
\vspace{0.5em}
{\footnotesize
\textcolor{dailyweather}{$\blacksquare$}~Daily weather \hspace{1.5em}
\textcolor{intraday}{$\blacksquare$}~Intra-day weather \hspace{1.5em}
\textcolor{dayrelated}{$\blacksquare$}~Day-specific \hspace{1.5em}
\textcolor{objectrelated}{$\blacksquare$}~Object-specific}
\caption{Visualisation of the gain-based variable importance for the regional
    two-regime model, showing the 15 covariates with the highest gain per model. Each bar corresponds to a single covariate, coloured according to the covariate group to which it belongs, i.e.\ daily weather, intra-day weather, day-related and object-related covariates. The left and middle panel show the occurrence model for the clear-weather and the severe-weather regime, respectively. The right panel
    shows the reporting model, which is shared across both regimes. } 
\label{fig:varimportance}
\end{figure}

Figure~\ref{fig:varimportance} shows the variable importance\footnote{We use the gain-based importance as implemented in the \texttt{xgboost} package in \textsf{R} \citep{chen2016xgboost}.} for the regional
two-regime model, for both occurrence regimes as well as the (shared) reporting model. For the reporting model, we sum the gain over the $d$ delay-specific models described in Appendix~\ref{app:emalgo}. Covariates are grouped by block as defined in Table~\ref{tab:covariates}, with the weather-related block further consisting of daily and intra-day covariates. For the clear-weather occurrence regime, daily weather covariates dominate, in particular precipitation-related information and temperature. For the severe-weather regime, the intra-day wind duration covariates are the largest contributors, followed by the daily wind covariates. If the regimes differed only in the number of claims they generate, the same covariates would have ranked highest in both models. The change in which covariates matter suggests that the relation between the weather and the occurrence of claims itself differs across regimes. For the shared reporting model, day-specific covariates (mainly weekend-related) and intra-day weather covariates (the timing of the most severe weather) drive the predictions. Object-specific covariates contribute little to the occurrence models, which is most likely a consequence of modelling the data at an aggregated rather than the object level. 

\begin{figure}[h] 
\centering
\includegraphics[width = 0.80\textwidth]{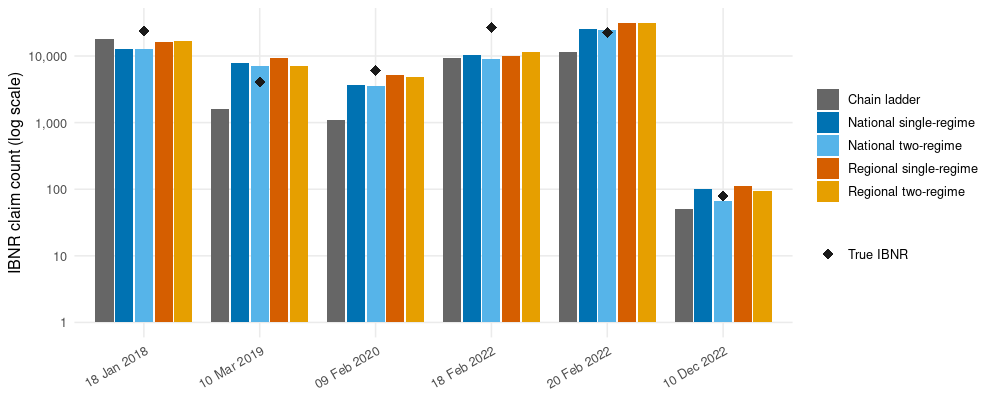}
\caption{Visualisation of the backtesting exercise for the (in-sample) IBNR claim count, for six evaluation dates $\tau$. The bars show the predicted number of outstanding IBNR claims at each evaluation date, for the national and regional model in both the single- and two-regime setting, as well as for a chain ladder benchmark. The black diamonds show the empirically observed number of outstanding IBNR claims. Note that the vertical axis is shown on a logarithmic scale for visualisation purposes.} 
\label{fig:IBNR}
\end{figure}

\paragraph{Backtesting IBNR claim count} In Figure~\ref{fig:IBNR}, we backtest our framework by nowcasting the number of outstanding IBNR claims for different choices of the evaluation date $\tau$. Next to 10 December 2022, we select five storm days on which many claims were recorded. For each chosen evaluation date, all models are refitted using only the claims reported up to that day. We compare our models' nowcasts to a chain ladder benchmark as well as to the empirically observed
values. We observe that the chain ladder method (visualised by the grey bar)
systematically underpredicts the total outstanding IBNR claim count, for all six evaluation dates. However, given 18 January 2018 (storm Friederike) as evaluation date, it outperforms our models as there is only a limited amount of data included, highlighting the need for sufficient historical data to obtain accurate estimates. On 18 February 2022 (storm
Eunice), all models underestimate the number of outstanding IBNR claims, as only a single comparable event, storm Friederike, is available in the training data at that point.

To summarise the results of Figure~\ref{fig:IBNR}, we compute the mean absolute log ratio (MALR). Since the observed number of outstanding IBNR claims ranges from fewer than a hundred to several thousands across the selected evaluation dates, an error measure on the scale of the claim counts, such as the MSE or the MAE, would be determined almost entirely by the storm days. We therefore evaluate the nowcasts on a relative scale, using the log of the accuracy ratio, i.e.~the logarithm of the ratio of the predicted to the observed count \citep{tofallis2015better}. Working on the log scale gives an over- and an underprediction by the same factor the same weight, and yields a dimensionless quantity that is comparable across evaluation dates with different claim magnitudes. For a given set of $M$ evaluation dates $\tau_1,\ldots,\tau_M$, we define

\begin{align*}
    \text{MALR} = \frac{1}{M}\sum_{m=1}^{M}
    \left|\, \log \left(
      \frac{\sum_{i}\sum_{j=\tau_{i,m}+1}^{d} e_{i} \cdot\widehat{\lambda}_i\cdot\widehat{p}_{i,j}}
           {\sum_{i}\sum_{j=\tau_{i,m}+1}^{d} N_{i,j}}
    \right) \right|,
\end{align*}
where $\tau_{i,m} = \min(d, \tau_m - \text{day}(i) + 1)$ denotes the number of observed reporting days for observation $i$ at evaluation date $\tau_m$, and the sums run over the observations with $\text{day}(i) \leq \tau_m$. Table~\ref{tab:ibnr} shows the MALR values, computed for the six evaluation dates in Figure~\ref{fig:IBNR}. The two-regime regional model has the lowest MALR value, in line with the results of Table~\ref{tab:evalmetrics}. The remaining three models perform
similarly, with marginal differences that are difficult to assess given the small number of evaluation dates. 

\begin{table}[H]
  \centering
  \begin{adjustbox}{width=0.90\textwidth}
      \begin{tabular}{lccccc}
    \toprule
    &  & \multicolumn{2}{c}{\textbf{National (NL)}}
      & \multicolumn{2}{c}{\textbf{Regional (PC2)}} \\
     \cmidrule(lr){3-4} \cmidrule(lr){5-6}
    & Chain ladder & Single-regime & Two-regime & Single-regime & Two-regime \\
    \midrule
    MALR & 0.8498 & 0.5171 & \textbf{0.5068} & 0.5056 & \underline{\textbf{0.4124}} \\
    \bottomrule
  \end{tabular}
  \end{adjustbox}
  \caption{Mean absolute log-ratio $(\text{MALR})$ of the IBNR claim count nowcasts across the six evaluation dates shown in Figure~\ref{fig:IBNR}, for the national and regional model in both the single- and two-regime setting, as well as for the chain ladder benchmark. Bold indicates the best value within a spatial level, the overall best value across spatial levels is underlined. }
  \label{tab:ibnr}
\end{table}

\begin{figure}[h]
\centering

\textbf{26 November -- 10 December 2022}\\[0.3em]
\begin{adjustbox}{width=0.98\textwidth}
    \subfloat[]{

\begin{adjustbox}{width=0.32\textwidth}
    \includegraphics[width = \textwidth]{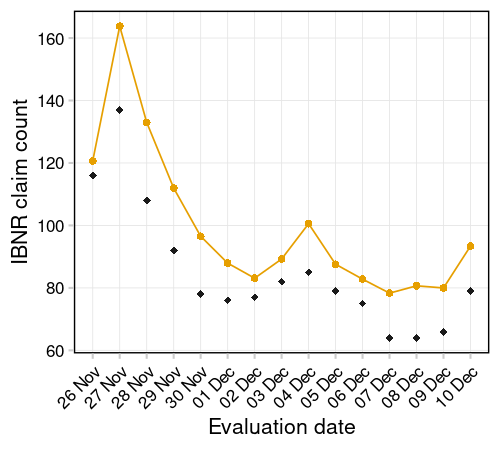}
\end{adjustbox}

}
\subfloat[]{

\begin{adjustbox}{width=0.32\textwidth}
\includegraphics[width = \textwidth]{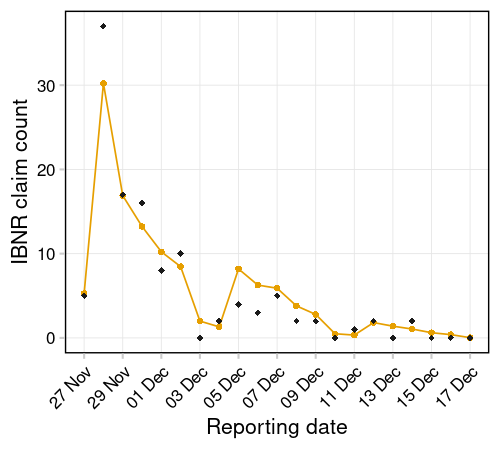}
\end{adjustbox}

}
\subfloat[]{

\begin{adjustbox}{width=0.32\textwidth}
\includegraphics[width = \textwidth]{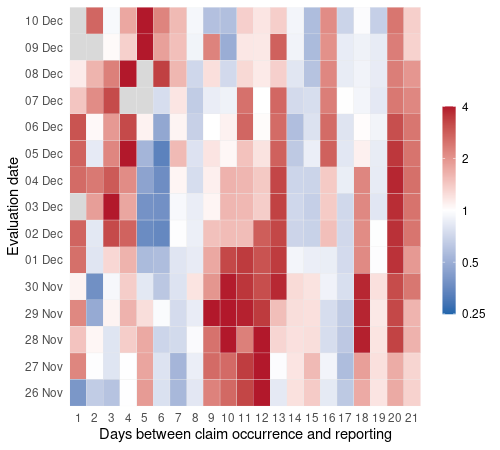}
\end{adjustbox}

}
\end{adjustbox}

\vspace{1em}
\textbf{20 February -- 6 March 2022}\\[0.3em]
\begin{adjustbox}{width=0.98\textwidth}
    \subfloat[]{

\begin{adjustbox}{width=0.32\textwidth}
    \includegraphics[width = \textwidth]{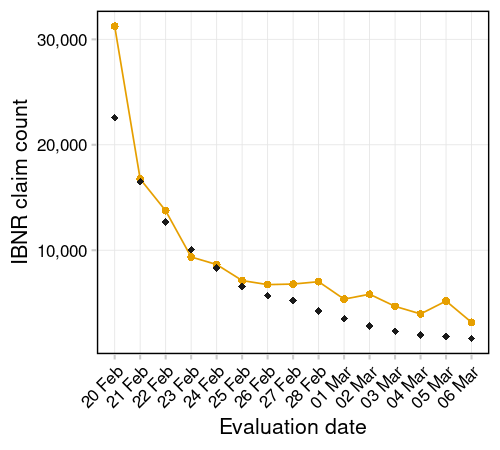}
\end{adjustbox}

}
\subfloat[]{

\begin{adjustbox}{width=0.32\textwidth}
\includegraphics[width = \textwidth]{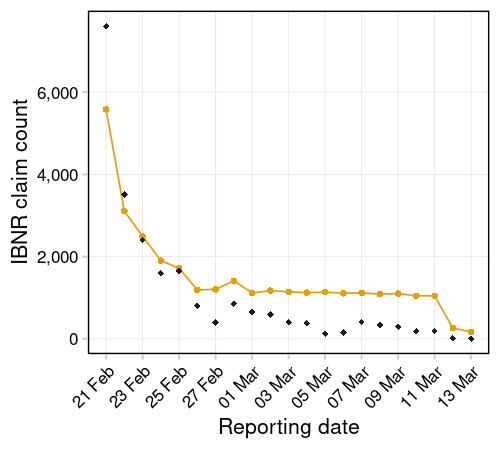}
\end{adjustbox}

}
\subfloat[]{

\begin{adjustbox}{width=0.32\textwidth}
\includegraphics[width = \textwidth]{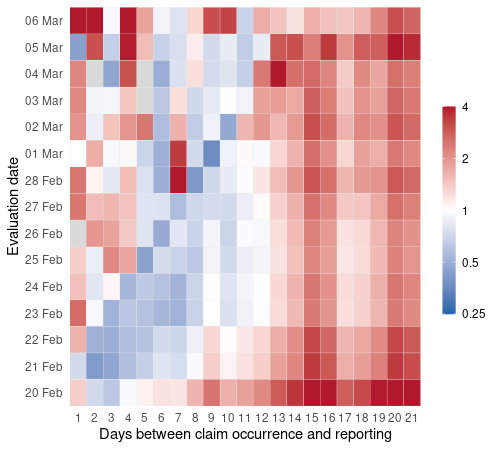}
\end{adjustbox}

}
\end{adjustbox}

\caption{IBNR diagnostics for the regional two-regime model across consecutive evaluation dates. The top row covers 26 November to 10 December 2022, a period with moderate weather. The bottom row covers 20 February to 6 March 2022, a period with many IBNR claims due to storm aftermath. The left panels show the predicted (yellow line) and observed (black diamonds) outstanding IBNR claim counts at each evaluation date. The middle panels show the predicted (yellow line) and observed (black diamonds) outstanding IBNR claim counts at 26 November 2022 (top) and 20 February 2022 (bottom), by the calendar day on which they are reported. The right panels show the logarithmic ratio of the predicted to the observed IBNR claim count per evaluation date and reporting delay, where blue and red cells indicate under- and overprediction, respectively.}
\label{fig:ibnr-diagnostics}
\end{figure}

In Figure~\ref{fig:ibnr-diagnostics}, we explore different IBNR
diagnostics for the two-regime regional model, by refitting it for every day in two periods of consecutive days: from 26 November 2022 to 10 December 2022 and from 20 February 2022 to 6 March 2022. The former is a period of moderate weather with few claims, the latter is the period after several storms including Eunice hit the Netherlands. Each day in these periods serves as an evaluation date $\tau_m$, for which the refitted model only includes the claims reported up to $\tau_m$. As for the MALR, the sums below run over the observations with
$\text{day}(i) \leq \tau_m$, with $\widehat{\lambda}_i$ and
$\widehat{p}_{i,j}$ obtained from the model refitted at $\tau_m$. The left panels show the evolution of the predicted and observed total outstanding IBNR claim count, i.e.
\begin{align*}
  \sum_{i} \sum_{j=\tau_{i,m}+1}^{d} e_i \cdot \widehat{\lambda}_i
  \cdot \widehat{p}_{i,j} \:\:\: \text{and} \:\:\:
  \sum_{i} \sum_{j=\tau_{i,m}+1}^{d} N_{i,j},
\end{align*}
respectively. For the period with moderate weather, we observe
overprediction across the period, which is largely due to an
overprediction of the occurred claims on 27 November. Since this is a Sunday with very little reporting, almost the entire error carries
over into the outstanding IBNR claim count. In general, we observe
that we capture the weekend effect: fewer claims are reported during
the weekend, resulting in a higher number of outstanding IBNR claims.
For the storm aftermath, this effect is less visible since the
additional claims in the weekend are dwarfed by the outstanding claims
from the heavy storms. In the middle panels, we zoom in on the claims
that occurred on or before 26 November 2022 (top middle panel) and 20
February 2022 (bottom middle panel) and were not yet reported at that
date. For each reporting day $\rho > \tau_m$, we show the predicted and
observed number of these claims reported on day $\rho$, i.e.
\begin{align*}
  \sum_{i} \sum_{\substack{j > \tau_{i,m} \\ \text{day}(i)+j-1 = \rho}}
  e_i \cdot \widehat{\lambda}_i \cdot \widehat{p}_{i,j}
  \:\:\: \text{and} \:\:\:
  \sum_{i} \sum_{\substack{j > \tau_{i,m} \\ \text{day}(i)+j-1 = \rho}}
  N_{i,j},
\end{align*}
respectively. We see that we capture the pattern for 26 November well,
showing the drop in claims reported during the weekend. For 20
February, we observe that the model overestimates the number of claims
that occurred on 18 February (storm Eunice) and still need to be
reported. Since this number of claims is very high it dominates the
tail, resulting in a flat pattern from late February onwards and a
drop after 11 March, which is the last possible reporting date for
claims that occurred on 18 February. The right panels decompose the
prediction errors by reporting delay: each row corresponds to one
evaluation date $\tau_m$ and each cell to the logarithmic ratio of the
predicted to the observed IBNR claim count at a reporting delay of
$j-1$ days, i.e.
\begin{align*}
  \log \left(
  \frac{\sum_{i:\, j > \tau_{i,m}} e_i \cdot \widehat{\lambda}_i \cdot
  \widehat{p}_{i,j}}{\sum_{i:\, j > \tau_{i,m}} N_{i,j}} \right),
  \:\:\: \text{for} \:\:\: j = 2, \ldots, d.
\end{align*}
For the moderate period, the deviations do not exhibit a clear
pattern, consistent with the absence of a single dominant occurrence
day. For the storm aftermath, the picture is systematic: the short
delays are well predicted or slightly underpredicted, whereas from
roughly delay 10 onwards the cells are consistently red. This confirms
that the overestimation of the claims occurring on 18 February
propagates into every tail cell of the reporting distribution, rather
than reflecting a misspecified delay pattern.


 



 



 


\section{Discussion}
\label{sec:discussion}
In this article, we study the occurrence and reporting dynamics of weather-related claims in private home insurance, combining detailed portfolio and claim information with publicly available high-resolution weather data. Instead of modelling claim occurrence conditional on externally identified weather events, we model all calendar days and postal code areas jointly. Claims arise under severe as well as moderate conditions: a few days with severe weather each generate a large number of claims, while the many days with moderate weather each generate few claims but jointly account for a substantial share of the total. Modelling all days jointly captures both kinds of days. Specifically, we propose a multi-regime EM-XGBoost framework, in which a data-driven regime structure lets the occurrence and reporting dynamics differ across weather conditions. Via a case study on a Dutch home insurance portfolio, in which we fit regime-specific occurrence models alongside a common reporting model, we show that given a sufficiently high spatial resolution, a multi-regime specification outperforms single-regime models. Additionally, a feature importance analysis indicates that the two occurrence regimes are driven by different covariates, supporting the choice of separate models. We also find that daily covariates constructed from hourly weather observations, capturing the duration and the intra-day timing of severe weather, are informative for both the occurrence and the reporting dynamics. These results have practical implications for portfolio risk management. A better understanding of claim occurrence could support early warning systems, allowing an insurer to warn policyholders ahead of severe weather and thereby potentially reduce the number of claims in its portfolio. For instance, feeding weather forecasts into the fitted occurrence models yields expected claim counts per area for the coming days, which can be used to target warnings at the areas most at risk. Additionally, a better understanding of the reporting dynamics improves the estimation of outstanding claim counts for reserving, and supports capacity planning for claim handling teams, avoiding internal processing delays. An important limitation is the scarcity of severe weather events in the observed period: such events generate a large share of the claims, yet only a handful are available for training, as illustrated by the underestimation of the IBNR claim count for storm Eunice in the backtesting exercise. Additionally, when used for forecasting purposes, timely predictions depend on the availability of accurate weather forecasts. A natural extension to our research is to apply our framework at more granular spatial levels such as the insured object level. Additional data such as house images and flood maps can then for example be used to obtain better predictions (e.g.~window location in combination with wind direction) at the object level, which can be especially useful when modelling claim sizes instead of claim counts.

\biblist

\pagebreak
\appendix

\section{EM algorithms} \label{app:emalgo}
\subsection{Single-regime EM-XGBoost framework} 

\begin{adjustbox}{width=0.85\textwidth}
   \begin{algorithm}[H] \label{algo:EM}
    \caption{EM-XGBoost algorithm, based on \cite{wilsens2025machine}}
\vspace{0.1cm}
\textbf{Data} \hspace*{-3.5cm}\makebox[2.3cm][r]{\textbf{(0)}}\\
Observable datasets $D_{\text{train}}=(N_{i,1},\ldots,N_{i,\tau_{i}},e_{i},\boldsymbol{x}_i)_{i=1}^{n_{\text{train}}}$ and $D_{\text{val}}=(N_{i,1},\ldots,N_{i,\tau_{i}},e_{i},\boldsymbol{x}_i)_{i=1}^{n_{\text{val}}}$.\\
\vspace{0.2cm}

\textbf{Initialisation} \hspace*{-5cm}\makebox[2.3cm][r]{\textbf{(1)}}\\
Choose starting values $p^{(0)}_{i,j}$ for the reporting probabilities, with
$\sum_{j=1}^{d} p^{(0)}_{i,j} = 1$ for $i = 1, \ldots, n_{\text{train}}$, and set
\begin{align*}
    \lambda^{(0)}_i = \frac{\sum_{j=1}^{\tau_i} N_{i,j}}{e_{i} \cdot\sum_{j=1}^{\tau_i}  p^{(0)}_{i,j}}
    \quad \text{for } i = 1, \ldots, n_{\text{train}}.
\end{align*}
A straightforward choice is the pooled empirical reporting distribution
\begin{align*}
    p^{(0)}_{i,j} = \frac{\big(\sum_{z=1}^{n_{\text{train}}} N_{z,j}\big)}{\big(\sum_{z=1}^{n_{\text{train}}} \sum_{j=1}^{\tau_z} N_{z,j}\big)}, 
\end{align*}
which is identical across
observations. More informative, data-driven choices are possible, see for example
Section~\ref{sec:casestudy}.
\vspace{0.2cm}

\For{$k=1,\ldots,K$}{
\vspace{0.2cm}

\textbf{Expectation step} \hspace*{-6.2cm}\makebox[2.3cm][r]{\textbf{(2.1)}} \\

Construct $\mathcal{D}_{\text{train}}^{(k)}=(N^{(k)}_{i,1},\ldots,N^{(k)}_{i,d},\boldsymbol{x}_i)_{i=1}^{n_{\text{train}}}$ and $\mathcal{D}_{\text{val}}^{(k)}=(N^{(k)}_{i,1},\ldots,N^{(k)}_{i,d},\boldsymbol{x}_i)_{i=1}^{n_{\text{val}}}$ by defining
\begin{align*}
    N_{i,j}^{(k)}= \begin{cases}N_{i,j} & j \leq \tau_i \\  e_{i} \cdot\widehat{\lambda}^{(k-1)}(\boldsymbol{x}_i) \cdot \widehat{p}_{j}^{(k-1)}(\boldsymbol{x_i}) & \tau_i<j \leq d\end{cases}.
\end{align*}
\vspace{0.2cm}

\textbf{Maximisation step} \hspace*{-6.5cm}\makebox[2.3cm][r]{\textbf{(2.2)}}\\
\vspace{0.2cm}
\begin{adjustwidth}{0.5cm}{0cm}
\textbf{Occurrence intensities} \\

Apply XGBoost algorithm to learn $f^{(k)}_{\text{occ}}(.)$, full details in \cite{wilsens2025machine}, by maximising $Q^{\text{occ}}(\lambda_{i}^{(k)};\mathcal{D}_{\text{train}}^{(k)})$. Set the predicted occurrence intensities
\begin{align*}
    \widehat{\lambda}^{(k)}(\boldsymbol{x}_i) = \exp{( \widehat{f}^{(k)}_{\text{occ}}(\boldsymbol{x}_i))}.
\end{align*}
\end{adjustwidth}
\vspace{0.2cm}
\begin{adjustwidth}{0.5cm}{0cm}
\textbf{Reporting probabilities} \\

Apply XGBoost algorithm to learn $f^{(k)}_{\text{rep}}(.)=(f^{(k)[1]}_{\text{rep}}(.),\ldots,f^{(k)[d]}_{\text{rep}}(.))$, full details in \newline \cite{wilsens2025machine}, by maximising $Q^{\text{rep}}(p_{i,j}^{(k)};\mathcal{D}_{\text{train}}^{(k)})$. Set the predicted reporting probabilities
\begin{align*}
    \widehat{p}_{j}^{(k)}(\boldsymbol{x}_i) = \frac{\exp(\widehat{f}_{\text{rep}}^{(k)[j]}(\boldsymbol{x}_{i}))}{\sum_{z=1}^{d} \exp(\widehat{f}_{\text{rep}}^{(k)[z]}(\boldsymbol{x}_{i}))}.
\end{align*}
\end{adjustwidth}

\If{$k>\texttt{EM\_patience}$}{
\vspace{0.2cm}
Stop EM algorithm if
\begin{align*}
    LL(\widehat{\lambda}^{(k)}_i, \widehat{p}^{(k)}_{i,j};D_{\text{val}}) = \displaystyle\max_{z = k-\texttt{EM\_patience}, \dots, k} LL(\widehat{\lambda}^{(z)}_i, \widehat{p}^{(z)}_{i,j};D_{\text{val}})
\end{align*}
holds.
}

}
    
\end{algorithm} 
\end{adjustbox}

\subsection{Multi-regime EM-XGBoost}
\begin{adjustbox}{width=0.82\textwidth}
   \begin{algorithm}[H] \label{algo:EMmulti}
    \caption{Multi-regime EM-XGBoost pseudocode}
\vspace{0.1cm}
\textbf{Data} \hspace*{-3.5cm}\makebox[2.3cm][r]{\textbf{(0)}}\\
Observable datasets $D_{\text{train}}=(N_{i,1},\ldots,N_{i,\tau_{i}},e_{i},\boldsymbol{x}_i)_{i=1}^{n_{\text{train}}}$ and $D_{\text{val}}=(N_{i,1},\ldots,N_{i,\tau_{i}},e_{i},\boldsymbol{x}_i)_{i=1}^{n_{\text{val}}}$.\\
\vspace{0.2cm}
\textbf{Regime allocation} \hspace*{-6cm}\makebox[2.3cm][r]{\textbf{(1)}}\\
Define a subset $\mathcal{S}$ of the available covariates. Let $\{\boldsymbol{x}_{i}^{(\mathcal{S})} \}_{i=1,\ldots,n_{\text{train}}}$ be the restricted feature information of $D_{\text{train}}$.\\
\vspace{0.2cm}
Apply a hard clustering technique to the restricted feature information $\{\boldsymbol{x}_{i}^{(\mathcal{S})} \}_{i=1,\ldots,n_{\text{train}}}$ to obtain a partitioning $D_{\text{train},1},\ldots,D_{\text{train},L}$ of the training data $D_{\text{train}}$.\\
\vspace{0.2cm}
Let $\{\boldsymbol{x}_{i}^{(\mathcal{S})} \}_{i=1,\ldots,n_{\text{val}}}$ be the restricted feature information of the validation data $D_{\text{val}}$. Partition the validation data into $D_{\text{val},1},\ldots,D_{\text{val},L}$ by assigning each observation regime $l \in 1,\ldots,L$ for which the regime centroid is closest.\\
\vspace{0.2cm}
\textbf{Initialisation} \hspace*{-5cm}\makebox[2.3cm][r]{\textbf{(2)}}\\
Choose starting values $p^{(0)}_{i,j}$ for the reporting probabilities, with
$\sum_{j=1}^{d} p^{(0)}_{i,j} = 1$ for $i = 1, \ldots, n_{\text{train}}$.\\
\For{$l=1,\ldots,L$}{
\begin{align*}
    \lambda^{(0)}_{l,i} = \frac{\sum_{j=1}^{\tau_i} N_{i,j}}{e_{i} \cdot\sum_{j=1}^{\tau_i} p^{(0)}_{i,j}}
    \quad \text{for } i = 1, \ldots, \texttt{n\_train\_l}.
\end{align*}
}

\For{$k=1,\ldots,K$}{
\vspace{0.2cm}

\textbf{Expectation step} \hspace*{-6.2cm}\makebox[2.3cm][r]{\textbf{(3.1)}} \\

Following step 2.1 in Algorithm~\ref{algo:EM}, construct $\mathcal{D}_{\text{train}}^{(k)}$, $\mathcal{D}_{\text{val}}^{(k)}$ and their corresponding partitionings $\mathcal{D}_{\text{train},1}^{(k)},\ldots,\mathcal{D}_{\text{train},L}^{(k)}$ and $\mathcal{D}_{\text{val},1}^{(k)},\ldots,\mathcal{D}_{\text{val},L}^{(k)}$.

\vspace{0.2cm}

\textbf{Maximisation step} \hspace*{-6.5cm}\makebox[2.3cm][r]{\textbf{(3.2)}}\\
\vspace{0.2cm}
\begin{adjustwidth}{0.5cm}{0cm}
\textbf{Occurrence intensities} \\

\For{$l=1,\ldots,L$}{
Following step 2.2 in Algorithm~\ref{algo:EM}, learn $f^{(k)}_{\text{occ},l}(.)$ by maximising $Q^{\text{occ}}(\lambda_{l,i}^{(k)};\mathcal{D}_{\text{train},l}^{(k)})$ and set 
\begin{align*}
    \widehat{\lambda}_{l}^{(k)}(\boldsymbol{x}_i) = \exp{( \widehat{f}^{(k)}_{\text{occ},l}(\boldsymbol{x}_i))}.
\end{align*}
}
\end{adjustwidth}
\vspace{0.2cm}
\begin{adjustwidth}{0.5cm}{0cm}
\textbf{Reporting probabilities} \\

\For{$l=1,\ldots,L$}{
Following step 2.2 in Algorithm~\ref{algo:EM}, learn $f^{(k)}_{\text{rep},l}(.)=(f^{(k)[1]}_{\text{rep},l}(.),\ldots,f^{(k)[d]}_{\text{rep},l}(.))$, \newline by maximising $Q^{\text{rep}}(p_{l,i,j}^{(k)};\mathcal{D}_{\text{train},l}^{(k)})$. Set the predicted reporting probabilities
\begin{align*}
    \widehat{p}_{l,j}^{(k)}(\boldsymbol{x}_i) = \frac{\exp(\widehat{f}_{\text{rep},l}^{(k)[j]}(\boldsymbol{x}_{i}))}{\sum_{z=1}^{d} \exp(\widehat{f}_{\text{rep},l}^{(k)[z]}(\boldsymbol{x}_{i}))}.
\end{align*}
}
\end{adjustwidth}

\If{$k>\texttt{EM\_patience}$}{
\vspace{0.2cm}
Stop EM algorithm if
\begin{align*}
    LL(\widehat{\lambda}^{(k)}_i, \widehat{p}^{(k)}_{i,j};D_{\text{val}}) = \displaystyle\max_{z = k-\texttt{EM\_patience}, \dots, k} LL(\widehat{\lambda}^{(z)}_i, \widehat{p}^{(z)}_{i,j};D_{\text{val}})
\end{align*}
holds.
}

}

\end{algorithm} 
\end{adjustbox}
\section{Case study} \label{app:casestudy}

\subsection{Covariates}

\begin{table}[htbp]
\centering
\sbox{\covbox}{%
\begin{adjustbox}{max width=\textwidth}
\begin{tabular}{lllc}
\toprule
\textbf{Covariate} & \textbf{Description} & \textbf{Type} & \makecell[l]{\textbf{Number of}\\\textbf{categories\textsuperscript{a}}} \\
\midrule
\multicolumn{4}{l}{\textit{\textbf{Object-specific}}} \\
\midrule
Roof type           & Proportion of insured objects per roof type              & Continuous  & 2 \\
Object type         & Proportion of insured objects per object type            & Continuous  & 12 \\
Building year       & Proportion of insured objects per building year band     & Continuous  & 17 \\
Exposure            & Number of insured objects with an active policy          & Continuous  & / \\
\midrule
\multicolumn{4}{l}{\textit{\textbf{Day-specific}} \textbf{(for delay $0,\ldots,21$ days)}}  \\
\midrule
Holiday             & Indicator for a Dutch national holiday                   & Categorical & 2 \\
Pre-/post-holiday   & Indicator for the day before or after a holiday          & Categorical & 2 \\
Weekday             & Day of the week                                          & Categorical & 7 \\
Weekend             & Indicator for a Saturday or Sunday                       & Categorical & 2 \\
Month boundary      & Indicator for the first or last day of the month         & Categorical & 2 \\
\midrule
\multicolumn{4}{l}{\textit{\textbf{Weather-related}}} \\
\midrule
Mean wind speed     & Daily mean wind speed (m/s); ERA5, E-OBS, KNMI           & Continuous  & / \\
Maximum wind speed  & Daily maximum wind speed (m/s); ERA5, KNMI               & Continuous  & / \\
Total precipitation & Total daily precipitation (mm); ERA5, E-OBS, KNMI        & Continuous  & / \\
Surface pressure    & Daily mean surface pressure at sea level (hPa); ERA5, E-OBS & Continuous & / \\
Temperature         & Daily mean temperature near the surface (\textdegree C); E-OBS & Continuous & / \\
Hail probability    & Daily probability of hail\textsuperscript{b}; KNMI                 & Continuous  & / \\
Wind duration      & Number of hours with wind speed above 20, 25 and 30 m/s; KNMI & Continuous & 3 \\
Precipitation timing  & 6-hour block during which total precipitation is highest; KNMI & Categorical & 4 \\
Wind timing  & 6-hour block during which sum of cubed hourly wind speeds is highest; KNMI & Categorical & 4 \\
6-hourly wind sums  & Sum of cubed hourly wind speeds per 6-hour block; KNMI & Continuous & 4 \\
Early vs late wind timing  & Ratio of morning (0-12h) to afternoon (12-24h) wind intensity\textsuperscript{c}; KNMI & Continuous & / \\
Evening share wind  & Share of wind intensity during evening (18-24h) ; KNMI & Continuous & / \\
Evening wind indicator & Indicator denoting whether evening wind intensity is highest; KNMI & Categorical & 2\\
\bottomrule
\end{tabular}
\end{adjustbox}}
\begin{threeparttable}
\usebox{\covbox}
\begin{tablenotes}[flushleft]
\footnotesize
\item[a] For categorical variables, this column denotes the number of levels. For continuous variables, this column denotes the number of different variables, if applicable.
\item[b] Determined from the difference between the height of the 45~dBZ echo top determined by radars, and the height of the 0\textdegree C isotherm.
\item[c] Computed as the ratio of the sum of cubed hourly wind speeds before noon to the sum of cubed hourly wind speeds in the afternoon.
\end{tablenotes}
\end{threeparttable}
\caption{Covariates used in the occurrence and reporting models, grouped by covariate block. For each covariate, the table indicates the description, the variable type and, where applicable, the number of categories over which the covariate is defined. The object-specific covariates are constructed by aggregating the object-level information to the modelling level, yielding one proportion per category, as described in Section~\ref{sec:feature}. The day-specific covariates are included for the occurrence day $\text{day}(i)$ as well as for each of the $d$ potential reporting days $\text{day}(i), \ldots, \text{day}(i)+d-1$, as these are informative for the delay at which a claim is reported.}
\label{tab:covariates}
\end{table}
\subsection{Parameter tuning}

Tables~\ref{tab:optimalregular} and~\ref{tab:optimalregimeswitch} show the grid values and the chosen values for the XGBoost-specific parameters for the single-regime and the two-regime models, respectively. From the full grid, 50 combinations were randomly sampled. Additionally, we fix the maximum number of EM iterations at $K=100$, and the number of EM iterations for early stopping at $\texttt{EM\_patience}=10$.

\begin{table}[H]
\centering
\begin{adjustbox}{width=0.75\textwidth}
    \begin{tabular}{llll}
 \toprule
 && \multicolumn{2}{c}{Optimal setting}\\
 \cmidrule(lr){3-4} 
 &Tuning values&National&Regional\\
 \midrule
 \textbf{Occurrence model}&&&\\
 \quad Learning rate &$[0.10,0.15,0.20]$&0.1&0.1\\
 \quad Initial rounds &$[300,400,500]$&500&500\\
 \quad Additional rounds &$[75,100,125]$&100&100\\
 \quad Tree depth &$[2,3,4]$&4&4\\
 \quad Patience &$[15,30,45]$&30&30\\[0.5em]

 \textbf{Reporting model}&&&\\
 \quad Learning rate &$[0.05,0.075,0.10]$&0.075&0.075\\
 \quad Initial rounds &$[10,15,20]$&20&20\\
 \quad Additional rounds &$[5,10,15]$&10&10\\
 \quad Tree depth &$[3,4,5]$&5&5\\
 \quad Patience &$[15,30,45]$&15&15\\

\bottomrule
 
\end{tabular}
\end{adjustbox}

\caption{Optimal tuning parameter values for the single-regime EM-XGBoost setting. The first column enumerates the tuning parameters. The second column indicates the tuning grid values for each tuning parameter. The third and fourth column show the values that maximise the observed log-likelihood $LL(\lambda_i, p_{i,j};D_{\text{val}})$ at the national and regional level, respectively.}
\label{tab:optimalregular}
\end{table}

\begin{table}[H]
\centering
\begin{adjustbox}{width=0.8\textwidth}
    \begin{tabular}{llll}
 \toprule
 && \multicolumn{2}{c}{Optimal setting}\\
 \cmidrule(lr){3-4} 
 &Tuning values&National&Regional\\
 \midrule
 \textbf{Occurrence model (clear regime)}&&&\\
 \quad Learning rate &$[0.10,0.15,0.20]$&0.2&0.1\\
 \quad Initial rounds &$[300,400,500]$&500&500\\
 \quad Additional rounds &$[75,100,125]$&75&75\\
 \quad Tree depth &$[2,3,4]$&3&4\\
 \quad Patience &$[15,30,45]$&30&15\\[0.5em]

 \textbf{Occurrence model (severe regime)}&&&\\
 \quad Learning rate &$[0.10,0.15,0.20]$&0.2&0.15\\
 \quad Initial rounds &$[300,400,500]$&400&300\\
 \quad Additional rounds &$[75,100,125]$&100&75\\
 \quad Tree depth &$[2,3,4]$&4&4\\
 \quad Patience &$[15,30,45]$&30&45\\[0.5em]

 \textbf{Reporting model}&&&\\
 \quad Learning rate &$[0.05,0.075,0.10]$&0.1&0.05\\
 \quad Initial rounds &$[10,15,20]$&20&15\\
 \quad Additional rounds &$[5,10,15]$&5&15\\
 \quad Tree depth &$[3,4,5]$&5&4\\
 \quad Patience &$[15,30,45]$&45&30\\

\bottomrule
 
\end{tabular}
\end{adjustbox}

\caption{Optimal tuning parameter values for the two-regime EM-XGBoost setting. The first column enumerates the tuning parameters. The second column indicates the tuning grid values for each tuning parameter. The third and fourth column show the values that maximise the observed log-likelihood $LL(\lambda_i, p_{i,j};D_{\text{val}})$ at the national and regional level, respectively.}
\label{tab:optimalregimeswitch}
\end{table}

\begin{figure}[H] 
\centering
\begin{adjustbox}{width=0.95\textwidth}
    \subfloat[Empirical]{\includegraphics[width = 0.33\textwidth]{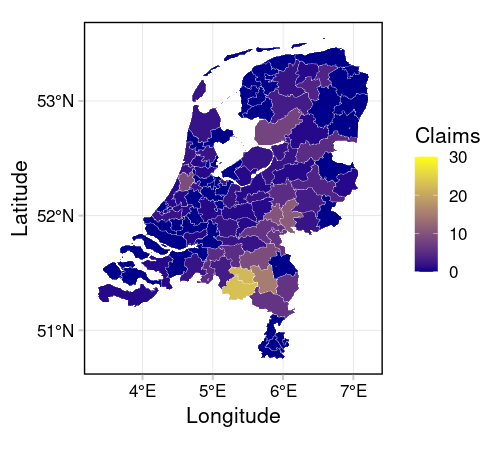}} 
\subfloat[Regional single-regime]{\includegraphics[width = 0.33\textwidth]{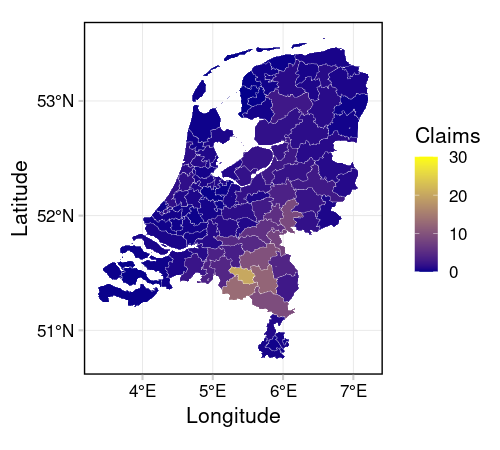}}
\subfloat[Regional two-regime]{\includegraphics[width = 0.33\textwidth]{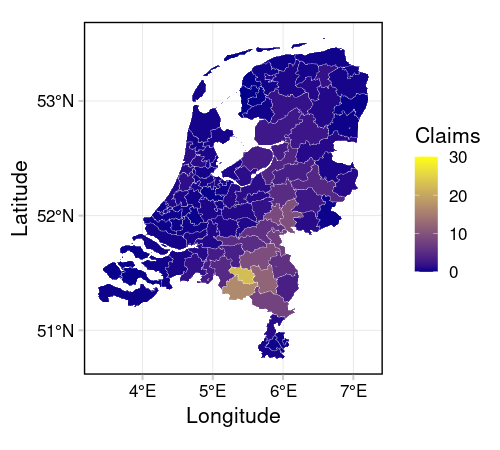}}
\end{adjustbox}
\caption{Visualisation of the observed and predicted number of claims per PC2 
    postal code area on 4 June 2019. The left panel 
    shows the observed claim counts, the middle and right panel the out-of-sample predicted 
    claim counts of the regional single-regime and two-regime model, respectively, computed on $\mathcal{D}_{\text{test}}$. } 
\label{fig:occfit_4june2019}
\end{figure}


@misc{copernicus2020eobs,
  author       = {{Copernicus Climate Change Service (C3S)}},
  title        = {{E-OBS} daily gridded meteorological data for {E}urope from 1950 to present derived from in-situ observations},
  year         = {2020},
  publisher    = {Copernicus Climate Change Service (C3S) Climate Data Store (CDS)},
  doi          = {10.24381/cds.151d3ec6}
}

@article{cornes2018ensemble,
  title={An ensemble version of the {E-OBS} temperature and precipitation data sets},
  author={Cornes, Richard C and Van Der Schrier, Gerard and Van Den Besselaar, Else JM and Jones, Philip D},
  journal={Journal of Geophysical Research: Atmospheres},
  volume={123},
  number={17},
  pages={9391--9409},
  year={2018},
  publisher={Wiley Online Library},
  doi={10.1029/2017JD028200}
}

@misc{copernicus2024era5daily,
  author       = {{Copernicus Climate Change Service (C3S)}},
  title        = {{ERA5} post-processed daily-statistics on single levels from 1940 to present},
  year         = {2024},
  publisher    = {Copernicus Climate Change Service (C3S) Climate Data Store (CDS)},
  doi          = {10.24381/cds.4991cf48}
}

@article{hersbach2020era5,
  title={The {ERA5} global reanalysis},
  author={Hersbach, Hans and Bell, Bill and Berrisford, Paul and Hirahara, Shoji and Hor{\'a}nyi, Andr{\'a}s and Mu{\~n}oz-Sabater, Joaqu{\'\i}n and Nicolas, Julien and Peubey, Carole and Radu, Raluca and Schepers, Dinand and others},
  journal={Quarterly Journal of the Royal Meteorological Society},
  volume={146},
  number={730},
  pages={1999--2049},
  year={2020},
  publisher={Wiley Online Library},
  doi={10.1002/qj.3803}
}

@misc{knmi_rad_nl25_rac_mfbs_1h_2_0,
  author       = {{Royal Netherlands Meteorological Institute (KNMI)}},
  title        = {{RAD-NL25-RAC-MFBS-24H NetCDF4} dataset (version 2.0): 1 hour precipitation accumulations from gauge-adjusted radar data for the {Netherlands} (1 km)},
  year         = {2025},
  publisher    = {KNMI Data Platform},
  note          = {url: \url{https://dataplatform.knmi.nl/dataset/rad-nl25-rac-mfbs-01h-netcdf4-2-0}}
}

@misc{knmi_radar_hail_warning_24h_v1_0,
  author       = {{Royal Netherlands Meteorological Institute (KNMI)}},
  title        = {Precipitation - radar 24 hour probability of hail over the {Netherlands} (radar\_hail\_warning\_24h), version 1.0},
  year         = {2025},
  publisher    = {KNMI Data Platform},
  note          = {url: \url{https://dataplatform.knmi.nl/dataset/radar-hail-warning-24h-1-0}}
}

@misc{knmi_stationgegevens,
  author       = {{Royal Netherlands Meteorological Institute (KNMI)}},
  title        = {Dagwaarnemingen},
  year         = {2026},
  publisher    = {KNMI},
  note          = {url: \url{https://www.daggegevens.knmi.nl/klimatologie/daggegevens}}
}

@article{wilsens2025machine,
  title={Machine learning in an expectation-maximisation framework for nowcasting},
  author={Wilsens, Paul and Antonio, Katrien and Claeskens, Gerda},
  journal={arXiv preprint arXiv:2512.07335},
  year={2025},
  doi={https://doi.org/10.48550/arXiv.2512.07335}
}

@article{karney2013algorithms,
  title={Algorithms for geodesics},
  author={Karney, Charles FF},
  journal={Journal of geodesy},
  volume={87},
  number={1},
  pages={43--55},
  year={2013},
  publisher={Springer},
  doi={10.1007/s00190-012-0578-z}
}

@article{lloyd1982least,
  title={Least squares quantization in {PCM}},
  author={Lloyd, Stuart},
  journal={IEEE transactions on information theory},
  volume={28},
  number={2},
  pages={129--137},
  year={1982},
  publisher={IEEE},
  doi={10.1109/TIT.1982.1056489}
}

@techreport{SwissReInstitute2024Sigma1,
  author      = {{Swiss Re Institute}},
  title       = {{N}atural {C}atastrophes in 2023: {G}earing {U}p for {T}oday's and {T}omorrow's {W}eather {R}isks},
  institution = {Swiss Re Institute},
  type        = {Sigma},
  number      = {1/2024},
  year        = {2024},
  month       = mar,
  address     = {Zurich, Switzerland},
  note = {url: \url{https://www.swissre.com/institute/research/sigma-research/sigma-2024-01.html}, accessed on 2026-06-03.}
}

@techreport{GallagherRe2025NatCat2024,
  author      = {{Gallagher Re}},
  title       = {Natural {C}atastrophe and {C}limate {R}eport 2024},
  institution = {Gallagher Re},
  year        = {2025},
  month       = jan,
  note = {url: \url{https://www.ajg.com/gallagherre/news-and-insights/gallagherre-natural-catastrophe-and-climate-report-2024/}, accessed on 2026-06-03.}
}

@article{shi2022assessing,
  title={Assessing hail risk for property insurers with a dependent marked point process},
  author={Shi, Peng and Fung, Glenn M and Dickinson, Daniel},
  journal={Journal of the Royal Statistical Society Series A: Statistics in Society},
  volume={185},
  number={1},
  pages={302--328},
  year={2022},
  publisher={Oxford University Press},
  doi={10.1111/rssa.12754}
}

@article{gao2022leveraging,
  title={Leveraging high-resolution weather information to predict hail damage claims: {A} spatial point process for replicated point patterns},
  author={Gao, Lisa and Shi, Peng},
  journal={Insurance: Mathematics and Economics},
  volume={107},
  pages={161--179},
  year={2022},
  publisher={Elsevier},
  doi={10.1016/j.insmatheco.2022.08.006}
}

@article{gao2025risk,
  title={Risk modeling of property insurance claims from weather events},
  author={Gao, Lisa and Shi, Peng},
  journal={ASTIN Bulletin: The Journal of the IAA},
  volume={55},
  number={2},
  pages={242--262},
  year={2025},
  publisher={Cambridge University Press},
  doi={10.1017/asb.2025.7}
}

@article{shi2024leveraging,
  title={Leveraging weather dynamics in insurance claims triage using deep learning},
  author={Shi, Peng and Zhang, Wei and Shi, Kun},
  journal={Journal of the American Statistical Association},
  volume={119},
  number={546},
  pages={825--838},
  year={2024},
  publisher={Taylor \& Francis},
  doi={10.1080/01621459.2024.2308314}
}

@article{bourget2024data,
  title={A data science approach to climate change risk assessment applied to pluvial flood occurrences for the {U}nited {S}tates and {C}anada},
  author={Bourget, Mathilde and Boudreault, Mathieu and Carozza, David A and Boudreault, J{\'e}r{\'e}mie and Raymond, S{\'e}bastien},
  journal={ASTIN Bulletin: The Journal of the IAA},
  volume={54},
  number={3},
  pages={495--517},
  year={2024},
  publisher={Cambridge University Press},
  doi={10.1017/asb.2024.19}
}

@article{fung2025investigating,
  title={Investigating the effect of climate-related hazards on claim frequency prediction in motor insurance with incomplete data},
  author={Fung, Tsz Chai and Jeong, Himchan and Tzougas, George},
  journal={Journal of the Royal Statistical Society Series C: Applied Statistics},
  volume={74},
  number={5},
  pages={1444--1475},
  year={2025},
  publisher={Oxford University Press UK},
  doi={10.1093/jrsssc/qlaf030}
}

@article{robben2026short,
  title={The short-term association between environmental variables and mortality: evidence from {E}urope},
  author={Robben, Jens and Antonio, Katrien and Kleinow, Torsten},
  journal={Journal of the Royal Statistical Society Series A: Statistics in Society},
  volume={189},
  number={2},
  pages={1131--1153},
  year={2026},
  publisher={Oxford University Press UK},
  doi={10.1093/jrsssa/qnaf052}
}

@article{robben2025granular,
  title={Granular mortality modeling with temperature and epidemic shocks: a three-state regime-switching approach},
  author={Robben, Jens and Barigou, Karim and Kleinow, Torsten},
  journal={Insurance: Mathematics and Economics},
  pages={103250},
  year={2026},
  publisher={Elsevier},
  doi = {10.1016/j.insmatheco.2026.103250}
}

@article{crevecoeur2023bridging,
  title={Bridging the gap between pricing and reserving with an occurrence and development model for non-life insurance claims},
  author={Crevecoeur, Jonas and Antonio, Katrien and Desmedt, Stijn and Masquelein, Alexandre},
  journal={ASTIN Bulletin: The Journal of the IAA},
  volume={53},
  number={2},
  pages={185--212},
  year={2023},
  publisher={Cambridge University Press},
  doi={10.1017/asb.2023.14}
}

@article{verbelen2022modeling,
  title={Modeling the occurrence of events subject to a reporting delay via an {EM} algorithm},
  author={Verbelen, Roel and Antonio, Katrien and Claeskens, Gerda and Crevecoeur, Jonas},
  journal={Statistical Science},
  volume={37},
  number={3},
  pages={394--410},
  year={2022},
  publisher={Institute of Mathematical Statistics},
  doi={10.1214/21-STS831}
}

@article{mack1993distribution,
  title={Distribution-free calculation of the standard error of chain ladder reserve estimates},
  author={Mack, Thomas},
  journal={ASTIN Bulletin: The Journal of the IAA},
  volume={23},
  number={2},
  pages={213--225},
  year={1993},
  publisher={Cambridge University Press},
  doi={10.2143/AST.23.2.2005092}
}

@article{bergstra2012random,
  title={Random search for hyper-parameter optimization},
  author={Bergstra, James and Bengio, Yoshua},
  journal={The Journal of Machine Learning Research},
  volume={13},
  number={1},
  pages={281--305},
  year={2012},
  publisher={JMLR. org},
  note = {url: \url{https://jmlr.org/papers/v13/bergstra12a.html}}
}

@inproceedings{chen2016xgboost,
  title={{XGB}oost: {A} scalable tree boosting system},
  author={Chen, Tianqi and Guestrin, Carlos},
  booktitle={Proceedings of the 22nd {ACM SIGKDD} International Conference on Knowledge Discovery and Data Mining},
  pages={785--794},
  year={2016},
  doi={10.1145/2939672.2939785}
}

@article{gao2026zero,
  title={A zero-inflated mixed-effects spatial point process for grouped storm loss data},
  author={Gao, Lisa and Jessup, S{\'e}bastien and Yan, Tianxing},
  journal={arXiv preprint arXiv:2607.03852},
  year={2026},
  doi={10.48550/arXiv.2607.03852}
}

@article{fonseca2025empirical,
  title={Empirical evaluation of windstorm losses and meteorological variables over the Netherlands},
  author={Fonseca Cerda, M d S and De Moel, H and Van Ederen, D and Aerts, JCJH and Botzen, WJW and Haer, T},
  journal={Natural Hazards},
  volume={121},
  number={5},
  pages={6085--6106},
  year={2025},
  publisher={Springer},
  doi={10.1007/s11069-024-07024-y}
}

@article{bonazzi2012spatial,
  title={The spatial structure of {E}uropean wind storms as characterized by bivariate extreme-value Copulas},
  author={Bonazzi, A and Cusack, S and Mitas, C and Jewson, S},
  journal={Natural Hazards and Earth System Sciences},
  volume={12},
  number={5},
  pages={1769--1782},
  year={2012},
  publisher={Copernicus Publications G{\"o}ttingen, Germany},
  doi={10.5194/nhess-12-1769-2012}
}

@article{tofallis2015better,
  title={A better measure of relative prediction accuracy for model selection and model estimation},
  author={Tofallis, Chris},
  journal={Journal of the Operational Research Society},
  volume={66},
  number={8},
  pages={1352--1362},
  year={2015},
  publisher={Taylor \& Francis},
  doi={10.1057/jors.2014.103}
}
\end{document}